\documentclass[
showpacs,floatfix, aps,prb,amsmath,
twocolumn,
groupaddress,eqsecnum]{revtex4}

\usepackage{epsfig}
\usepackage{epsf}
\usepackage{array}

\newcommand{\con}{\!\circ\!}
\newcommand{\beq}{\begin{equation}}
\newcommand{\eeq}{\end{equation}}
\newcommand{\be}{\begin{equation}}
\newcommand{\ee}{\end{equation}}
\newcommand{\beqa}{\begin{eqnarray}}
\newcommand{\eeqa}{\end{eqnarray}}
\newcommand{\bea}{\begin{eqnarray}}
\newcommand{\eea}{\end{eqnarray}}
\newcommand{\w}{\omega}
\newcommand{\W}{\Omega}
\newcommand{\e}{\epsilon}
\newcommand{\nn}{\boldsymbol{n}}

\renewcommand{\r}{\boldsymbol{r}}
\renewcommand{\P}{\boldsymbol{P}}
\newcommand{\p}{\mbox{\boldmath $p$}}
\newcommand{\R}{\mbox{\boldmath $R$}}

\renewcommand{\v}{\mbox{\boldmath $v$}}
\newcommand{\s}{\boldsymbol{s}}
\newcommand{\E}{\boldsymbol{E}}
\newcommand{\B}{\boldsymbol{B}}
\newcommand{\A}{\mathbf{A}}
\renewcommand{\j}{\boldsymbol{j}}
\newcommand{\nnabla}{\boldsymbol{\nabla}}

\newcommand{\q}{\boldsymbol{q}}

\newcommand{\sgn}{\mbox{sgn}}
\renewcommand{\>}{\rangle}
\newcommand{\<}{\langle}
\newcommand{\rda}{\rangle\!\rangle}
\newcommand{\lda}{\langle\!\langle}

\newcommand{\pt}{\partial_t}

\newcommand{\St}{\widehat{\mathrm{St}}}

\renewcommand{\Re}{\mathrm{Re}}
\renewcommand{\Im}{\mathrm{Im}}
\newcommand{\tr}{\mathrm{tr}}
\newcommand{\Tr}{\mathrm{Tr}}
\newcommand{\mylabel}[1]{\label{#1}
}
\newcommand{\req}[1]{Eq.~(\ref{#1})}
\newcommand{\reqs}[1]{Eqs.~(\ref{#1})}
\newcommand{\rref}[1]{(\ref{#1})}

\newcommand{\an}[1]{\langle #1 \rangle_{\boldsymbol{n}}}

\newcommand{\vgrad}{\v \! \cdot \! \nnabla}

\newcommand{\dnnn}{\delta\left(\widehat{\nn_{1}\nn_{2}}\right)}
\newcommand{\tta}{\delta t}
\newcommand{\K}{{\cal K}}
\newcommand{\cA}{\boldsymbol{\cal A}}
\renewcommand{\L}{{\cal L}}
\newcommand{\sqF}{\hat{\cal F}}  
\newcommand{\eh}{\textrm{{\scriptsize e}{\tiny -}{\scriptsize h}}}
\newcommand{\da}[2]{\textrm{{\scriptsize #1-#2}}}
 
\newcommand{\concomma}{\stackrel{\con}{,}}

\newcommand{\sd}{\sigma_{\scriptscriptstyle D}}
\newcommand{\kwf}{\hat{\kappa}_{\scriptscriptstyle W\!F}}
\newcommand{\cV}{{\scriptstyle C}_{\scriptscriptstyle V}}
\newcommand{\vertex}[4]{\gamma_{{#1}}\left(
\begin{matrix}\nn_{#4}\\
\nn_{#2};\nn_{#3}
\end{matrix}
\right)}
\newcommand{\donetwo}[2]{\delta\left(\widehat{\nn_{#1}\nn_{#2}}\right)}

\begin{document}

\title{Interaction corrections to the thermal transport coefficients in
disordered metals: quantum kinetic equation approach.}

\author{G. Catelani and I.L. Aleiner}
 \affiliation{ Physics Department, Columbia University, New
  York, NY 10027 }

\begin{abstract}
We consider the singular
electron-electron interaction corrections to the transport
coefficients in disordered metals
to test the validity of the Wiedemann-Franz law. We develop
a local, quantum kinetic equation approach in which the charge and energy
conservation laws are explicitly obeyed. To obtain the local description,
we introduce bosonic distribution functions for the neutral, low-energy
collective modes (electron-hole pairs). The resulting system of kinetic 
equations enables us to distinguish between the different physical processes
involved in the charge and energy transport: elastic electron
scattering affects both, while the inelastic processes influence only
the latter. Moreover, the neutral bosons, though incapable of
transporting charge, 
contribute significantly to the energy
transport. In our approach we  calculate on equal footing
the electrical and thermal conductivities and the specific heat 
in any dimension. We found that the Wiedemann-Franz law is always
violated
by the interaction corrections;
the violation is larger for one- and 
two-dimensional systems in the diffusive regime $T\tau \ll \hbar$ and it is
due to the energy transported by the neutral bosons.
For two-dimensional systems in the quasi-ballistic regime $T\tau \gg \hbar$
the inelastic scattering of the electron on the bosons also contributes to
the violation.
\end{abstract}
\date{\today}
\maketitle

\section{Introduction}
\label{intro}

It is well known that the measurement of the thermal transport
coefficient may provide additional information about the
scattering processes in disordered metals. 
In particular, the Wiedemann-Franz \cite{WF} law holds
as long as elastic scattering\footnote{It was shown by 
G.V. Chester and A. Thellung,
that \req{WF} remains valid for arbitrary scattering strength
as long as the scattering rates and the density of states are smooth
(${\cal C}_2$)
function of energy near the Fermi level.} 
dominates in the system:
\be
\mylabel{WF}
\mathrm{L}=\frac{\kappa}{\sigma T}=\frac{\pi^2}{3e^2},
\ee
where $\kappa$ and $\sigma$ are respectively the thermal and electric 
conductivities in the system, $T$ is the temperature in energy
units ($k_B=1$) and $e$ is the electron charge.
On the other hand for the deeply inelastic forward scattering the
Wiedemann-Franz law is violated\cite{textbook}, 
so that the Lorentz number $\mathrm{L}$
is smaller than the universal value, $\mathrm{L} < \frac{\pi^2}{3e^2}$.

Historically, the transport 
(in particular thermal transport)
coefficients were first being calculated using
the Boltzmann equation (BE)\cite{BE}. The advantage of this approach is that
it allows for the clear separation of the scale in the problem:
a particle moves freely most of the time and rarely scatters on
other particles or impurities. The BE is applicable on the time scale
much larger than the time it takes for the scattering to happen,
so all the scattering events are encoded into the local collision
integral. All the quantum mechanical part of the calculation is
reduced then into the solution of the scattering problems for the
relevant physical processes. This gives
the precise form of the collision integral but it does not affect the
general structure of the BE.
The great advantage of the BE is that its structure illuminates the
relevant conservation laws.

In the late 50s an alternative approach became popular -- the so
called Kubo formulas\cite{Kubo}. 
In this approach the transport equation is not
derived but rather the connection of the transport coefficient to the
equilibrium correlation function of certain current operators is used.
(The Kubo approach to the thermal transport was claimed to be put on rigorous
footing by Luttinger\cite{Luttinger} based
on the assumption that there exists some spatial scale in the system
so that for perturbations smooth on that scale the gradient expansion
is possible). Being exact, the Kubo formulas
are formally applicable even in the regime where the transport
equation can not be justified (separation of the evolution into
free motion and rare collisions is not possible).

However, in practice, the possibility of explicit calculation within the Kubo
formula is somewhat limited. The most spectacular results of the Kubo-formula 
calculations -- such as Maki-Thompson\cite{Maki,Thompson}, 
Aslamazov-Larkin\cite{AL} and weak localization\cite{Gorkov79} 
corrections to the electrical conductivity -- require a small
parameter which is the same parameter that determines the applicability of
the Boltzmann equation. It means that all those effects can be
also described in terms of quantum corrections to the collision
integral (for weak localization it was done in 
Ref.~\onlinecite{AleinerLarkin}). 
The most relevant for this paper effect -- the 
Altshuler-Aronov\cite{AA79} interaction correction to the electrical
conductivity in two dimensions\cite{AAL,Finkelstein}
\be
\mylabel{AAL}
\delta \sigma_{AA} =-\frac{e^2}{2\pi^2\hbar}\ln\left(\frac{\hbar}{T\tau}\right)
\left[ 1+ 3 \left( 1- \frac{1}{F_0^\sigma} \ln (1+F_0^\sigma) \right)\right]
\ee
originates from the elastic scattering of the electrons on the self-consistent
potential (Friedel oscillation)\cite{Rudin,AAG}, and it can be once again
obtained from the correction to the collision integral\cite{ZNA}.

The success of the Kubo formulas in the description of the
quantum and interaction effects in thermal transport is
by far more modest and controversial. Particularly, despite
a 20-years history there is no consensus on the answer to a
natural question:
how does the logarithmic correction to the conductivity \rref{AAL} 
translate into a correction to the Wiedemann-Franz law \rref{WF}? 

A first attempt at answering this question was made by Castellani at
al. \cite{Castellani} by analyzing Ward identities for a disordered
Fermi liquid; they found that the Wiedemann-Franz law should
hold for interacting disordered electrons. Their claim was later
disputed by Livanov et al.\cite{Livanov}: in a ``quantum kinetic
equation'' approach,
\footnote{The quantum kinetic equation with the
necessary conservation laws was not actually derived in
Ref.~\onlinecite{Livanov}, so we will not be able to compare their approach
with ours.} 
a logarithmic divergence for the thermal
conductivity in two dimensions was found to have even the sign
contrary to the Wiedemann-Franz law. 
More recently Niven and Smith\cite{Smith} 
applied Kubo formula and
found again a logarithmically divergent contribution
(for the Coulomb but not the short-range interaction)
in addition to the one that follows from the Wiedemann-Franz law.

The reason for this confusion in the literature is twofold.
Technically, the identification of the correct form of the current
operator is complicated by the presence of the electron-electron 
interaction (the energy current operator in the form defined by 
Luttinger\cite{Luttinger} is cumbersome to use due to the
presence of the additional disorder and interaction potentials in it,
whereas the superficially more elegant 
expression in the Matsubara frequency representation
in fact does not correspond to any conservation law for the
interacting system and violates gauge invariance -- see Appendix 
\ref{microcurr}). Physically, the use of the diagrammatic calculation 
within the Kubo formula prevents one from a clear identification of the 
relevant scattering processes, because each diagram taken separately 
describes some mixture of such processes
and does not have its own physical meaning.

This situation calls for the development of the kinetic equation
description, which takes into account the interaction correction
of the Altshuler-Aronov type both for the electric and thermal
transport.
The advantage of this approach is that it makes possible to keep track of the
conservation laws explicitly and thus excludes any ambiguity in the
definition of the currents. This paper is devoted to the development
and application of this machinery.

We will use units with $\hbar =1$ throughout the paper and we will restore
the Planck constant in the final results only.
The remainder of the paper is organized as follows: in  Sec.~\ref{Sec.2} 
we discuss some general features of the kinetic equation approach using a 
simple ``toy model''.
In Sec.~\ref{final}, we present our final expression for the kinetic equation
describing interacting electrons in disordered metals. Section \ref{sec:sum}
 summarizes
the results for the thermal conductivity and the specific heat obtained by
solving the kinetic equation. The derivation of the kinetic equation
is presented in Sec.~\ref{sec:derivation}, while the calculation of the 
transport coefficients and the specific heat is given in Sec.~\ref{resder}.
Some mathematical details are relegated into Appendices.



\section{Structure of the kinetic equation: currents and specific heat}
\label{Sec.2}

The purpose of this section is to show how the
structure of the kinetic equation permits the proper identification
of the relevant currents. We will remind how to
calculate the specific heat from the kinetic equation
once the conservation laws are obtained (this enables  
a direct check against the much simpler thermodynamic calculation). 
We will discuss the locality requirement for a proper kinetic
equation. The latter requirement will dictate the
number of necessary degrees of freedom 
(i.e. independent distribution functions) which should be introduced
into the kinetic description.

\subsection{Kinetic equation and conservation laws}
\label{sec:cons}

As a concrete example, we consider here electron-like and
hole-like excitations coupled to neutral bosons in the presence of an external
electric field $\E$. (As we will see later, a system of 
interacting electron can be effectively described at low temperatures by such a
coupled system for the scattering at small momentum transfer in a 
particle-hole channel.) The kinetic equations for electrons and bosons have
the  form:
\begin{subequations}
\label{kineq}
\be 
\mylabel{kineq1}
\left[ \frac{\partial}{\partial t} + 
v_F \nn\cdot \nnabla + ev_F
  \nn\cdot \E\frac{\partial}{\partial \epsilon} \right] 
f = St_e \{ f, N
\}
\ee 
\be
\mylabel{kineq2}
 \left[\frac{\partial}{\partial t} + v (\omega)\nn
  \cdot \nnabla \right] N = St_b \{ f, N \} 
\ee 
where $f=f(\e,\nn;t,\r)$
is the distribution function for the
electrons with charge $e$, $v_F$ is the Fermi velocity, and $\nn$ is
the direction of the momentum.  
Energy $\e$ is counted from the Fermi level so that
$f(\e > 0)$ describes the electron-like excitations and 
$1-f(-\e),\ \e > 0$ corresponds to the hole-like excitations.
Having in mind only singular in $T$ corrections,
we neglect the dependence of the electron velocity on energy
(electron-hole
asymmetry) as
it does not introduce anything but small correction regular in powers
of $T^2$.
\end{subequations}

The bosonic function $N=N(\omega,\nn;t,\r)$ 
is the distribution function for the bosons with velocity 
$\v (\omega )$. All the interaction
effects are included into the collision integrals $St_e$ and $St_b$;
for example, an electron-like excitation can decay into a less
energetic electron and a neutral boson, or an electron and a hole can
annihilate into a neutral bosons, etc. By locality, the collision
integrals depend on the same variables as the distribution functions,
i.e. $St_e = St_e (\e ,\nn ; t,\r)$ and $St_b = St_b (\w ,\nn ;t,\r)$.

In thermodynamic equilibrium and $\E=0$ the Fermi function
for fermions and the Planck function for the neutral bosons
\be
\label{equil}
f_F(\e)=\frac{1}{\exp(\e/T)+1}; \quad N_P(\w) =\frac{1}{\exp(\w/T)-1}; 
\ee
solve the kinetic equation. The temperature $T$ here is a constant
determined by the initial conditions for the kinetic equation.

Being the effective description for the slow dynamics of the
original quantum mechanical system, the kinetic equation must respect
the conservation laws of the original system: (i) total charge
conservation and  (ii) total energy conservation.
Those two conditions are enforced by the requirements for the
collision integrals
\begin{subequations}
\label{conservation}
\be
\mylabel{number}
\int d\epsilon \ \nu \an{ St_e \{ f,N \}} = 0,
\ee
and
\be
\mylabel{energy}
\int d\epsilon
\ \e \ \nu 
\an{
St_e \{ f,N \}}
 + \int d\w\ \w \ b(\w )  \an{St_b \{ f,N \}}
 =  0.
\ee
Here $\nu$ is the density of states (DoS) of the electrons (we will neglect
its energy dependence)and $b(\omega)$ is the density of states of the
bosons. We also introduced the short-hand notation for the angular
integral
\end{subequations}
\be
 \an{\dots}
 \equiv \int \frac{d\nn}{\Omega_d} \dots\ , 
\mylabel{angle}
\ee
where $\Omega_d$ is the total solid angle in $d$ dimensions.

Let the electron density $\rho$  be given by:
\be
\mylabel{rho}
\rho(t,\r) = e \nu \int d\e 
\an{f (\epsilon,\nn;t,\r )}.
\ee
Integrating \req{kineq1} over the energy and the direction of the
momentum and using \req{number} we arrive at the continuity equation
\be
\mylabel{rhocont}
\frac{\partial\rho}{\partial t}+ \nnabla \cdot \j = 0,
\ee
with the electron current density defined as
\be
\mylabel{j}
\j (t,\r) = e\nu v_F \int d\epsilon 
\an{ \nn f (\epsilon, \nn ;t, \r)}.
\ee 
(Strictly speaking \req{rhocont} fixes only the longitudinal component
of the current -- an 
arbitrary curl may be added to \req{j}. We will not consider the
effect of the magnetic field here, so we will disregard such
magnetization currents.)

We turn now to the analysis of the energy conservation.
We multiply \req{kineq1} by $\nu\e$ 
and integrate over $\e$ and $\nn$. Next, we multiply \req{kineq2} 
by $\w b(\w)$ and
integrate over the $\w$ and $\nn$. Adding the two results together we
obtain, with the help of \reqs{energy} and \rref{j}
\be\mylabel{encon}
\frac{\partial u_{tot} }{\partial t} + 
\nnabla \cdot \j^\e_{tot} = \j\cdot \E
\ee
where
\begin{subequations}
\label{utot}
\bea
&&u_{tot}=u_e(t,\r) + u_b (t,\r);\\
&&u_e(t,\r) = \nu \int d\e \ \e \ \an{f (\e ,\nn ;t ,\r )};\label{ue} \\
&&u_b (t,\r) = \int d\w \ \w \ b(\w ) 
\an{ N (\w ,\nn ;t,\r )},
\label{ub}
\eea
\end{subequations}
and
{\setlength\arraycolsep{0pt}
\begin{subequations}
\label{jtot}
\bea
&&\j^\e_{tot} = \j^\e_e + \j^\e_b;\\
&&\j^\e_e(t,\r) = v_F\nu \int d\e \ 
 \e \ \an{\nn f (\e ,\nn ; t ,\r )}; \label{jee} \\
&&\j^\e_b (t,\r) = \!\int\!d\w \, \w \, b (\w ) v(\omega) 
\an{\nn N (\w ,\nn ;t,\r )} \label{jeb}
\eea
\end{subequations} }

The right hand side of \req{encon} is nothing but the Joule heat.
For the homogeneous system the gradient term in the left hand side
vanishes, and by virtue of the energy conservation the expression
\rref{utot}
must be identified with the total energy density of the system. On the other
hand,
for $\E=0$ \req{encon} has the form of the continuity equation for the
energy density; therefore  \reqs{jtot} has to be identified with the
total energy current density.
This statement is not entirely trivial. One could imagine that for
an interacting system the DoS entering into the expressions
for the charge \rref{rho} and for the energy density \rref{utot}
are renormalized differently. The energy conservation equation
\rref{encon} excludes such a possibility. 

The conservation of energy \rref{encon} is valid for any rate of the
energy flow in and out of the system. On the other hand, the collision
integrals in \reqs{kineq} define a certain time scale $\tau_{in}$
-- the dynamics slow on the scale of $\tau_{in}$ can be characterized
by the distribution functions \rref{equil} with the temperature
dependent on time, $T(t)$ [corrections to such an adiabatic
description
are of the order of $\tau_{in}\pt \ln T$]. Substituting such form of
the distribution function in \reqs{utot}, and
the result in \req{encon}, we find for the homogeneous system
\be
\mylabel{econ1}
\cV(T)\frac{\partial T}{\partial t} = \j\cdot\E,
\ee
where 
\be
\cV=  \frac{\partial}{\partial T} \left[\nu \int d\e \ \e f_F (\e)+
\int d\w \ \w \ b (\w ) N_P(\w)\right] 
\mylabel{cV}
\ee    
is nothing but the specific heat of the system.
The latter quantity may be calculated independently by applying the standard
diagrammatic technique for the equilibrium systems. The agreement of
such a calculation with the structure of the kinetic equation result (\ref{cV})
will be the most important check of the consistency of our
description of the thermal transport.

\subsection{Locality of the kinetic equation and the number of degrees
of freedom}
\label{sec:local}

The local in space and time form of the collision integrals is clearly a
simplified description. Actually the collision integral may be nonlocal
on the time scale of the order of $\hbar/T$ and on the spatial scale
of the order of $\hbar v_F/T$. We will call such a description {\em local}, and
the description where the non-locality is involved on larger spatial
and time scales -- {\em non-local}.

The number of distribution functions to be introduced into the
description is governed by the locality of the kinetic equation.
Let us use the model of \reqs{kineq} to illustrate the point. We had a
local description in terms of the fermionic and bosonic distribution
functions. One can, however, try to eliminate the bosonic distribution
function and obtain a description in terms of the electronic degrees
of freedom only.

Assuming that the deviation of the distribution function from its
equilibrium value is small, one can linearize the bosonic collision
integral to the form
\be
St_b\{f,N\}= - \hat{I} \left(N - \tilde{N}\{f\}\right) 
\mylabel{stblinear}
\ee 
where $\hat{I}$ is some positive definite integral operator and 
$\tilde{N}\{f\}$ is the functional of the fermionic distribution
function $f(\e)$, such that for $f(\e)=f_F(\e)$, $\tilde{N}(\w)=N_P(\w)$. 
Using \req{stblinear} one can formally solve \req{kineq2} 
\be
N = \frac{1}{\frac{\partial}{\partial t} + v (\omega)\nn
  \cdot \nnabla + \hat{I}}
\hat{I}\tilde{N}\{f\} 
\mylabel{sol2}
\ee
Substituting \req{sol2} into \req{kineq1} we apparently obtain the
kinetic equation in terms of the electron distribution function only
\be\label{kineq11} \begin{split}
& \left[ \frac{\partial}{\partial t} + 
v_F \nn\cdot \nnabla + ev_F
  \nn\cdot \E\frac{\partial}{\partial \epsilon} \right] 
f = St_e^{?} \{ f \} \\
& St_e^{?} \{ f \}\equiv 
St_{e}\left\{ f, \frac{1}{\frac{\partial}{\partial t} + v (\omega)\nn
  \cdot \nnabla + \hat{I}}
\hat{I}\tilde{N}\{f\}  \right\}
\end{split} \ee 

If we are interested in the linear response to a weak and smooth
external perturbation, the description in terms of this single kinetic
equation is completely equivalent to the original coupled system
\rref{kineq}. However, there are clearly drawbacks: the presence of the
integral operator $\hat{I}$ in the collision integral makes it
non-local on the scale determined by the kinetic equation
itself rather than by the temperature.
Moreover, though it is still easy to identify the continuity equation
for the electron charge
using \req{number}, there is no longer a relation similar to \req{energy}.
This is why
the analysis of the energy conservation law becomes cumbersome:
the calculation of specific heat and energy current requires
the time expansion of the collision integral, which in turns seems to require
the knowledge of the concrete form of the inelastic collision
integral.


The example we have just considered is somewhat trivial because
the separation of the system into the fermionic and bosonic modes was
given from the very beginning. The problem we are dealing with in
this paper is how to include the collective modes of the interacting
electron system into the kinetic equation. Indeed, in this case any
calculation will give the result in terms of the electronic
distribution function only, and it is not clear a priori how to
introduce the occupation numbers for the collective modes into the
description.

As we will show, it may be possible to reverse our previous argument.
We will
consider a system of interacting electrons and we will
find that the interactions are described by a {\em non-local} 
collision integral.  Therefore we will introduce
bosonic degrees of freedom that will enable us to rewrite the non-local
kinetic equation in terms of coupled, local kinetic equations. Then
we will be able to identify the energy density and energy current
density as sums of fermionic and bosonic contributions.
The concrete example will be discussed briefly in the following subsection.

\subsection{Degrees of freedom for the kinetics of
disordered Fermi liquid}

We now focus  on the disordered, interacting Fermi liquid. For simplicity, 
we consider the interaction in the singlet channel only.
Our goal is to show that the thermodynamic result for the 
interaction correction to the specific heat has indeed the kinetic 
equation structure \rref{cV}. As a result, we will be able to determine
the necessary number of bosonic degrees of freedom for the 
local kinetic equation.
For the paper to be self-contained, we briefly 
review the thermodynamic approach, referring the reader to the 
literature \cite{ring} for further details.

The thermodynamic calculation of the specific heat $\cV$ is based on 
the relation between $\cV$ and the thermodynamic potential $\W$:
\be
\label{cv2}
\cV = -T\frac{\partial^2 \W}{\partial T^2}
\ee
The thermodynamic potential can be written as the sum of the thermodynamic 
potential for non-interacting quasiparticle $\W_0$ 
and a correction $\delta\W$ associated with the soft modes in the
system. Keeping such a correction is legitimate because it turns
out to be a more singular function of temperature than the $T^3$ correction
due to the electron-hole asymmetry. 

\begin{figure}[tbh]
\includegraphics[width=0.42\textwidth]{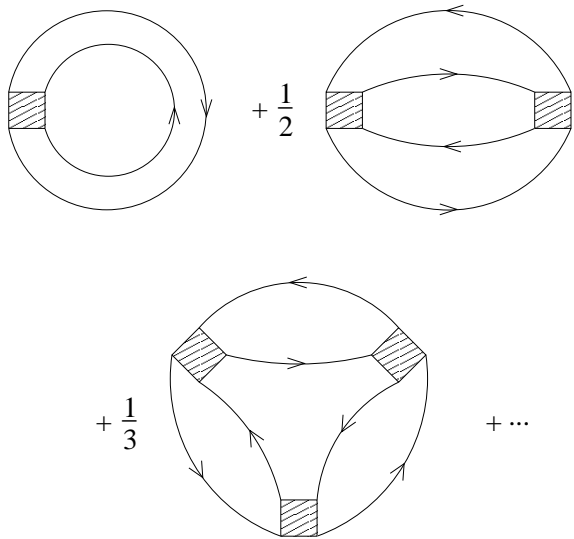}
\caption{Leading singular contribution to the thermodynamic potential for the
clean system. The
shaded box corresponds to $F/\nu$, defined through the two particle
vertex $\Gamma^\w$, see Ref.~\onlinecite{FL}; 
the solid lines are coherent parts of the electron Green's 
functions. For the disordered system, the polarization bubbles should be
dressed by the impurity scattering \cite{AA85}.}
\label{fig1}
\end{figure}

The correction $\delta\W$ is given by the sum of the so-called ring 
diagrams,  see Fig.~\ref{fig1}.
The Matsubara representation for this diagram is 
\be 
\label{dom}
\delta \W = \frac{T}{2} \sum_{\w_n} \int\!\frac{d^d q}{(2\pi)^d} \ln
\left( 1 + \frac{F}{\nu} \Pi (i |\w_n| , \q) \right) 
\ee 
Here $F$ is the
interaction constant, $\w_n = 2\pi T n$ are the bosonic Matsubara
frequencies and $\Pi$ is the polarization operator. The
explicit expression for this operator is not important for the present
discussion and will be given later, see \req{pira}.

A straightforward calculation, relegated to Appendix \ref{app1}, 
enables us to rewrite \req{dom} as
\be \label{dop} 
\delta \W = -\int\!\frac{d\w}{2\pi} 
\left(\frac{1}{2}\coth \frac{\w}{2T}\right) 
\int\!\frac{d^d q}{(2\pi)^d} 
\Im \, \Tr \Big[ \ln \hat{\L}^\rho - \ln \hat{\L}^g \Big], 
\ee
The explicit expressions for the bosonic propagators
$\L^\rho$, $\L^g$ are not relevant [they can be found in \req{Ls} -- the
trace should be understood as a sum or integration over all variables other 
than  $\w, \, \q$]; 
we just mention here that $\L^\rho =\L^g$
in the absence of interaction, $F=0$. 
Substituting \req{dop} into \req{cv2} and performing an
integration over $\w$ by parts, we find  
\begin{subequations}
\label{cVDoS}
\be
\label{dcV}
\delta\cV = \frac{\partial}{\partial T} \int_0^{\infty}\!\!\! d\w\, 
\w N_P(\w ) \left[b^{\rho}(\omega) - b^{g}(\omega)\right],
\ee
where the densities of states are defined as
\bea
&&b^{\rho}(\omega)= \frac{1}{\pi}\Im 
 \int\frac{d^d q}{(2\pi)^d}
\partial_\w \Tr  \ln \L^\rho; \\
&& b^{g}(\omega)= \frac{1}{\pi}\Im 
 \int\frac{d^d q}{(2\pi)^d}
\partial_\w \Tr \ln \L^g.
\eea
\end{subequations}
The function $b^{\rho}(\w)$ has the physical meaning of the density of
states (DoS) of the bosonic degrees of freedom in the system (soft 
electron-hole pairs).  The function $b^{g}(\w)$ 
has the meaning of the density of states of fictitious bosons
(we will call them ``ghosts'') which describe the soft electron-hole
pairs in the absence of interaction. The physical meaning of
the minus sign in front of $b^g(\w)$ is that with 
the formation of the collective modes
some degrees of freedom are removed from the description of the
non-interacting system; the introduction of the ghost
bosons in the last term in \req{dcV} takes this
reduction into account. 

Comparison of \req{dcV} with \reqs{cV} and \rref{utot} hints the
following expression for the contribution of the collective modes to
the energy density in the non-equilibrium case
\be
u_b = \int_0^{\infty}\!\!\! d\w \, \w 
\left[N^{\rho}(\omega) b^{\rho}(\omega) - 
N^{g}(\omega)
b^{g}(\omega)\right],
\label{utot2}
\ee
where $N^{\rho}=N^g=N_P$ in the equilibrium and they have to be found from
some kinetic equation otherwise [this definition 
requires that \req{encon} holds for an arbitrary
distribution function]. A similar expression can be obtained for the
contribution due to the interaction in the triplet channel by introducing
the additional propagator $\L^\sigma$ and distribution function $N^\sigma$.
This means that the proper local
kinetic equation must include four distribution functions:
one for fermions, $f(\e)$, and three for bosons,
$N^{\rho,\sigma,g}(\omega)$.
In the subsequent sections we will be able to derive such a description.

\section{Final form of the kinetic equation and scattering processes}
\label{final}

This section summarizes the final form of the quantum kinetic
equation, the conservation laws and the corresponding currents.
The derivation of these results is presented in
Sec.~\ref{sec:derivation}.

In accord with the previous section,
the kinetics of the system is described by the electronic
distribution function $f(\e,\nn; t, \r)$,
the ``distribution functions'' of the bosonic singlet and triplet excitations,
$\hat{N}^\rho$ and $\hat{N}^\sigma$, and the ``distribution function''
of the ghost excitation, $\hat{N}^g$.

The electron distribution function $f(\e,\nn;t,\r)$ is
diagonal in the space of the momentum directions.
On the contrary, the bosonic excitations are
characterized by the density matrices  ${N}^{\alpha}(\w,\q; \nn_i,\nn_j;
t, \r)$ [$\alpha = \rho, \sigma , g$] 
which may be not diagonal in the space of the momentum direction
$\nn$. Only in the thermal equilibrium 
\be\label{thermeq}\begin{split}
& f_{eq} (\e,\nn; t,\r) = f_F(\e) \, , \\
& {N}^{\alpha}_{eq}(\w,\q; \nn_i,\nn_j; t, \r)=
\W_d\donetwo{i}{j} {N}_P(\w) \, ,
\end{split}\ee
with the Fermi and Plank distribution functions given by \req{equil},
the matrices  
${N}^{\alpha}(\w,\q; \nn_i,\nn_j; t, \r)$ acquire the diagonal form
\footnote{As given in \req{thermeq}, the equilibrium distribution 
functions $N_{eq}^\alpha(\w)$
are defined only for $\w > 0$; for $\w < 0$ they are found by using the 
property in \req{symN2}.}. 
However even out of equilibrium these matrices have the property
\be
\label{symN2}
\begin{split}
N^{\alpha}&\left(\w,\q; \nn_i,\nn_j\right) 
\\ &= - \left[N^{\alpha}\left(
-\w,-\q; \nn_j,\nn_i \right)+\W_d\donetwo{i}{j} \right].
\end{split}
\ee
(hereafter, the spectator $t,\, \r$ variables might be suppressed.)

Strictly speaking, $f(\e,\nn;t,\r)$ is the $2\times 2$ density
matrix in the spin space 
and ${N}^{\sigma}$ is the  $3\times 3$  density
matrix in the angular momentum $L=1$ space;
 however this will not be important
in further calculations and we will write the equations
for the diagonal components only.
To account for the three folded degeneracy of the triplet mode 
we will explicitly introduce factors of $3$
in the corresponding collision integrals and currents.

For the sake of compactness, we will use an operator notation for 
matrices in the space of momentum direction, so that for example
$\hat{N}$ should be understood as an operator acting on a function
$a(\nn_i)$ as follows:
\be
\left[ \hat{N} a \right] (\nn_i) \equiv 
\int\frac{d\nn_j}{\W_d} N(\nn_i,\nn_j) a (\nn_j). 
\label{operator}
\ee

The kinetic equation for the electrons in the electric field $\E$ (we
will not consider the magnetic field effects) has the canonical form:
\be\begin{split}
\bigg[ & \pt + \vgrad + e\v\!\cdot\!\E \frac{\partial}{\partial \e} \bigg]
f(\e,\nn ;t,\r) =\St_e(\e,\nn;t,\r);\\
&\St_e=\St_\tau f   +\St^{\da{$e$}{$\rho$}} \{ f,N^\rho\} + 
3 \St^{\da{$e$}{$\sigma$}} \{f,N^\sigma \}
 \\ & \qquad - 4 \St^{\da{$e$}{$g$}} \{ f, N^g \} +\St^{\da{$e$}{$e$}} \{f \},
\end{split}
\label{eqf}
\ee
where the first term on the right hand side is the ``bare'' collision integral
\be
{\rm St}_\tau(\nn_i,\nn_j) =
\frac{1}{\tau(\theta_{ij})}-
\donetwo{i}{j}\int  \frac{d\nn_k}{\tau(\theta_{ik})} ,
\label{sttau}
\ee
with $\theta_{ij} = \widehat{\nn_i\nn_j}$, 
and the other terms, which will be written shortly, take into account
the interaction effects.

The bosonic distributions, for $\alpha=\rho,\sigma, g$,
are governed by:
{\setlength\arraycolsep{0pt}
\bea\label{boseq}
&& \w \biggl[ \left\{ \frac{1}{1+\hat{F}^\alpha} ; \pt {\hat N}^\alpha \right\}
 + \left\{ \hat{\s}^\alpha(\w,\q) ; \nnabla  {\hat N}^\alpha \right\} 
 \\ && \ + i \left[\hat{H}^\alpha_\eh(\w,\q); \hat{N}^\alpha \right]\biggr]
= \St^{\da{$\alpha$}{$e$}}\left\{N^\alpha, f\right\}(\w,\q;\nn_i,\nn_j; t,\r)
\nonumber
\eea
where the commutator and anticommutator are defined as
\be 
\left\{\hat{A};\hat{B}\right\}\equiv \frac{1}{2}
(\hat{A}\hat{B}+\hat{B}\hat{A}); \quad
\left[\hat{A};\hat{B}\right] \equiv \hat{A}\hat{B}-\hat{B}\hat{A}.
\mylabel{compm1}
\ee}

The operators $\hat{H}^\alpha_\eh$ acting in the angular (momentum
direction) space are defined as
\be
 \hat{H}^\alpha_\eh(\w,\q)= \v\cdot\q-\frac{\w}{1+\hat{F}^\alpha}
\ee
and the velocity operator is
\be
\hat{\s}^\alpha(\w,\q) = 
\frac{\partial\hat{H}^\alpha_\eh(\w,\q) }{\partial \q}=
\v +\w  \frac{\partial }{\partial \q}
\left(\frac{\hat{F}^\alpha}{1+\hat{F}^\alpha}\right).
\ee
The action of the operators $\hat{F}^\alpha$ in the angular space
is the same as in \req{operator} and
they have the following forms:
\be
\begin{split}
&\hat{F}^g=0\\
& [\hat{F}^\sigma](\nn_i,\nn_j) = F^\sigma(\theta_{ij})\\
& [\hat{F}^\rho](\nn_i,\nn_j) = \nu V(\q)+ F^\rho(\theta_{ij}),
\end{split}
\ee
where  $F^{\rho,\sigma}(\theta)$ are the Landau Fermi-liquid
interaction parameters. The  angular independent $\nu V(\q)$
term takes into account the long range part of the Coulomb 
density-density interaction.

To characterize the density of states for the bosonic excitations
we introduce the propagators 
$\hat{\L}^\alpha (\w,\q;\nn_i,\nn_j)$, $\alpha=\rho,\sigma,g$ as
\be
\label{Ls}
\left[
i\hat{H}^\alpha_\eh(\w,\q)
- \St_\tau   \right] \hat{\L}^\alpha = \hat{1},
\ee
They describe the propagation of
the electron-hole pair which is scattered by the disorder
potential. This propagation is affected by the corresponding
interactions for $\alpha=\rho,\sigma$, whereas it reduces to
the usual diffusion for the ghosts.

We are now prepared to write down the conservation laws
which must be satisfied by the collision integrals
independently of their explicit form or the particular
shape of the distribution functions.
The conservation of the number of particles is guarded
by the condition
\begin{subequations}
\label{cons}
\bea
&&\int\St^{\da{$e$}{$\alpha$}} \{ f,N^\alpha\}\left(\e,
\nn;t,\r\right) d\nn\, d\e  = 0.
\quad \alpha=g,\rho,\sigma;
\nonumber\\
&&\int\St^{\da{$e$}{$e$}} \{ f\}\left(\e,
\nn;t,\r\right) d\nn\, d\e =0,
\label{cons1}
\eea
and the impurity collision integral \rref{sttau} 
conserves the number of particles on each energy shell
\be
\int \St_\tau f (\e,\nn;t,\r) \, d\nn =0.
\label{cons2}
\ee
The conservation of energy during purely electron-elec\-tron collisions
is dictated by
\be
\int\e\,
\St^{\da{$e$}{$e$}} \{ f\}\left(\e,
\nn;t,\r\right) d\nn\, d\e =0.
\label{cons3}
\ee
Finally, the conservation of energy during the electron-boson
collision is guaranteed by the conditions
\bea
&&\nu \int\epsilon \,
\St^{\da{$e$}{$\alpha$}}\{ f, N^\alpha\}\left(\e, \nn;t,\r\right) 
\frac{d\nn\, d\e}{\W_d}
\label{cons4}
\\
&&+ \int
\Tr \left[
\hat{\L}^\alpha (\w) 
\St^{\da{$\alpha$}{$e$}} \{ f, N^\alpha\}\left(\w ;t,\r\right)
\right] \frac{d\w}{2\pi} \nonumber
\\ &&
= -i \int   \Tr \left[ 
\left[ \hat{H}^\alpha_{\eh}(\w) ; \hat{\L}^\alpha (\w) \right] 
\hat{N}^\alpha(\w; t ,\r ) \right] \frac{\w d\w}{2\pi} ;
\nonumber
\eea
for $\alpha=g,\rho,\sigma$, where the trace is defined as
\end{subequations}
\be\label{trace1}
\Tr \hat{A}\hat{B} = \int \frac{d\nn_1 d\nn_2}{\W_d^2}
\int\frac{d^dq}{(2\pi)^d} A(\q;\nn_1,\nn_2) B(\q;\nn_2,\nn_1).
\ee

The existence of the conservation laws \rref{cons}
immediately enables us to establish the expressions for the
conserved currents in the spirit of Sec.~\ref{sec:cons}.
We find by integrating both sides of
\req{eqf} over $\epsilon$ and $\nn$:
\bea
&& \frac{\partial\rho}{\partial t}+ \nnabla \cdot \j = 0,
\label{consrho} \\
&& \rho(t,\r) = e \nu \int f (\epsilon,\nn;t,\r ) \frac{d\e\,d\nn}{\W_d}
\nonumber\\
&&
\j (t,\r )= e\nu v_F \int \nn f (\epsilon,\nn;t, \r) \frac{d\e\,d\nn}{\W_d}
\nonumber
\eea
which express the conservation of charge in terms for the usual charge density
and electric current density -- cf. \reqs{rho}-\rref{j}.

Turning to the energy conservation, we multiply \req{eqf} by $\e$ and
then integrate over $\nn, \, \e$. 
Similarly, we pre-multiply \req{boseq} by $\hat{\L}^\alpha$, take the 
trace \rref{trace1} and integrate over $\w$. Adding the results
together, one finds:
\begin{subequations}
\label{consen}
\bea
&& \frac{\partial u_{tot}}{\partial t}+ \nnabla \cdot
\j^\e_{tot}=\j\cdot\E 
 \\
&& u_{tot} = u_e + u_\rho +3 u_\sigma - 4 u_g \nonumber \\
&& \j^\e_{tot} = \j^\e_e + \j^\e_\rho + 3 \j^\e_\sigma - 4 \j^\e_g .\nonumber
\eea
The electronic contributions to the energy density and current density are
given by
\bea\label{elen}
&& u_e (t,\r) = \nu \int \frac{d\e\,d\nn}{\W_d}
\e \, f (\epsilon, \nn ;t,\r ) 
\nonumber\\
&& \j^\e_e (t,\r )= \nu v_F \int\frac{d\e\,d\nn}{\W_d} \e \,
\nn f (\epsilon, \nn;t, \r).
\eea
The contributions of the bosonic neutral excitations are 
\bea\label{bosen}
&& u_\alpha(t,\r) = \int\Tr\left\{ \frac{1}{1+\hat{F}^\alpha}  
\hat{\L}^\alpha (\w)
\hat{N}^\alpha (\w;t,\r) \right\} \frac{\w \, d\w}{2\pi}
\nonumber \\
&& j_\alpha^\e(t,\r) = \int\Tr\left\{ \hat{\s}^\alpha(\w)  \hat{\L}^\alpha(\w)
\hat{N}^\alpha(\w;t,\r) \right\} \frac{\w \, d\w}{2\pi}
\nonumber\\
\eea
for $\alpha=g,\rho,\sigma$.
\end{subequations}

Equations 
\rref{consrho}-\rref{consen} constitute our main results: the conserved 
currents are defined in terms of the distribution functions of the 
quasiparticles which describe the low energy excitations of the interacting
electron gas for interaction in the particle-hole channel. 
In contrast with previous calculations\cite{Livanov,Langer,Smith,Raimondi}, 
we explicitly show the 
validity of the continuity equation for the energy transport; no such
proof has been presented before in the quantum kinetic 
equation approach\footnote{The current operator
used in Ref.~\onlinecite{Langer} does not satisfy the continuity equation
for the long range interaction potential.}. 
Moreover we believe that the form of the energy current in those references 
is not correct, since it is not gauge invariant -- see Appendix \ref{microcurr}
for more details.
As an additional benefit, our approach enables us
to clearly identify the contributions of the collective modes and
the scattering processes involved -- this last task is accomplished
by analysing the explicit form of the collision integrals, which is
is also needed to calculate the transport coefficients.
The detailed derivation of the collision integrals
is presented in Sec.~\ref{sec:derivation}; here we
summarize the results and give them physical interpretation.
 
To shorten the  formulas we introduce the following combinations
of the distribution functions: 
\begin{subequations}
\bea
\label{ups}
&&\Upsilon^{\alpha}_{ij;kl}
\left(\e,\w, \q ; t,\r \right)
\equiv N^{\alpha}\left(\w,\q;\nn_i,\nn_j;t,\r\right)
\nonumber \\
&& \quad \times
\Big\{f(\e,\nn_k;t ,\r)-f(\e-\w,\nn_k ;t,\r)
\Big\}
\\ && \quad+
\W_d\delta(\widehat{\nn_i\nn_j})\Big\{f(\e,\nn_l;t,\r)
\left[1-f(\e-\w,\nn_k;t,\r )\right]
\Big\} \nonumber
\eea
which, as it follows from \req{symN2}, has the property
\be
\label{upsilon1}
\int\!d\e \, \Upsilon^{\alpha}_{ij;kl} (\e,\w ,\q) =  
\int\!d\e \, \Upsilon^{\alpha}_{ji;lk} (\e,-\w ,-\q)
\ee
and
\be
\label{psi}
\begin{split}
&\Psi_{ij;kl}\left(\e,\e_1;\w\right) \\
& \equiv f(\e-\w,\nn_i)\left[1-f(\e,\nn_j)\right]f(\e_1,\nn_k)
\left[1-f(\e_1-\w,\nn_l)\right] \\
& - f(\e,\nn_i)\left[1-f(\e-\w,\nn_j)\right]f(\e_1-\w,\nn_k)
\left[1-f(\e_1,\nn_l)\right]\! . 
\end{split}
\ee
It is easy to check that $\Upsilon=\Psi=0$ in the thermal equilibrium
\rref{thermeq}.
The last combination enters the collision integral in the symmetric form
\be\label{psis}
\begin{split}
\Psi^s_{ij}\equiv \frac{1}{4}
\Big[{\Psi}_{ij;\,ij} +{\Psi}_{ji;\,ij}
+{\Psi}_{ij;\,ji} +{\Psi}_{ji;\,ji}\Big]
\end{split}
\ee
It is worth noticing that the terms involving four distribution
functions $f$ are in fact cancelled from $\Psi^s$; moreover it has the
properties:
\be\label{psip}
\begin{split}
& \int \Psi^s_{ij} \left(\e,\e_1;\w\right) \, d\e \, d\e_1 = 0 \\
\int a(\w) & \int \e \, \Psi^s_{ij} \left(\e,\e_1;\w\right) \, d\e \, d\e_1 
\, d\w = 0
\end{split} \ee
\end{subequations}
for any even function $a(\w)$. 
Finally we introduce the vertex $\gamma$ for the impurity scattering
\be\label{vert}
\gamma_{ij}^k \equiv 
\frac{1}{\tau(\theta_{ij})}\W_d\left[\donetwo{j}{k}-\donetwo{i}{k}\right]
\ee
and the shorthand notation
\[
\L^\alpha_{ij} \equiv \L^\alpha(\w,\q; \nn_i,\nn_j)
\]

The explicit expression for the boson-electron collision integral is then
\be
\begin{split}
&\St^{\da{$\alpha$}{$e$}}
\left(\w,\q; \nn_1,\nn_2 ;  t,\r \right)= 
- \int\!d\e\int\!\frac{d\nn_3d\nn_4}{\W_d^2} 
\\ & \ \times 
\Big\{\gamma_{14}^{3}
\Upsilon^{\alpha}_{32;41}\left(\e,\w,\q;t,\r \right)
+ \gamma_{34}^{2}
\Upsilon^{\alpha}_{13;14}\left( \e,\w,\q;t,\r \right)
\Big\},
\end{split}
\label{boselci}
\ee
for $\alpha=g,\rho,\sigma$.
The formula for the electron-boson collision integral can be conveniently 
decomposed into {\em local} $(l)$ and {\em non-local} $(n)$ 
[in the sense of Sec. \ref{sec:local}]
parts
\begin{subequations}
\label{elbosci}
\be
\St^{\da{$e$}{$\alpha$}} = \St_l^{\da{$e$}{$\alpha$}}
+ \St_n^{\da{$e$}{$\alpha$}}.
\ee
The local part of the collision integral is
\be
\label{stlocal}
\begin{split}
& \St_l^{\da{$e$}{$\alpha$}} (\e,\nn_1) 
=  \frac{1}{\nu} \int\!\frac{d\w}{2\pi} \frac{1}{\w}
\int\!\frac{d^dq}{(2\pi)^d} \int\!\frac{d\nn_2 d\nn_3 d\nn_4}{\W_d^3}
 \\
& \times \big\{ \gamma_{12}^3 \big[ \L^\alpha_{34}
\Upsilon^\alpha_{41;21}(\e,\w,\q) + 
\Upsilon^\alpha_{34;21}(\e,\w,\q) \bar{\L}^\alpha_{41} \big] \\
& +\gamma_{21}^{3} \big[
\L^\alpha_{34} \Upsilon^\alpha_{42;21}(\e,\w,\q)
+ \Upsilon^\alpha_{34;21}(\e,\w,\q) \bar{\L}^\alpha_{42} 
\big] \big\},
\end{split}
\ee
where the bar indicates Hermitian conjugation
\be\label{barL}
\bar{\L}^\alpha (\w,\q;\nn_i,\nn_j) = \L^\alpha (-\w,-\q;\nn_j,\nn_i)
\ee
Using \req{Ls} and the definitions \rref{trace1},
\rref{vert} we can verify that the pair
\reqs{boselci}-\rref{stlocal} satisfies the energy
conservation law \req{cons4} on its own. 

Equation \rref{stlocal}
satisfies the particle number conservation law \rref{cons1} as well.
To check this, we make the change of variables $(\w,\q) \to (-\w,-\q)$ in
the terms containing $\bar{\L}^\alpha$ and then use
 \reqs{upsilon1} and \rref{barL} to rewrite the integral of \req{stlocal}
 in terms of $\L^\alpha$ only:
\[
\begin{split}
& \int \St_l^{\da{$e$}{$\alpha$}} (\e,\nn_1) d\e d\nn_1
=\dots \int\!\frac{d\nn_1 \ldots d\nn_4}{\W_d^4}
 \\
& \times \big\{ \gamma_{12}^3 \big[ \L^\alpha_{34}
\Upsilon^\alpha_{41;21}(\e,\w,\q) - 
\L^\alpha_{14} \Upsilon^\alpha_{43;12}(\e,\w,\q)  \big] \\
& +\gamma_{21}^{3} \big[
\L^\alpha_{34} \Upsilon^\alpha_{42;21}(\e,\w,\q)
- \L^\alpha_{24} \Upsilon^\alpha_{43;12}(\e,\w,\q)  
\big] \big\}
\end{split}
\]
We perform the $\nn_3$ integration
using the delta functions in \req{vert} and we obtain the
result antisymmetric with respect to the $\nn_1 \leftrightarrow \nn_2$
permutation. Hence the above expression vanishes after 
the $\nn_{1,2}$ integrations.

\begin{figure}
\includegraphics[width=0.47\textwidth]{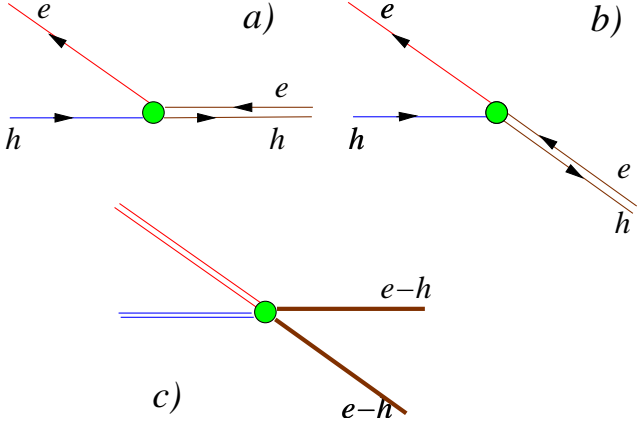}
\caption{
The scattering amplitudes leading to the creation of the same
electron and hole out of different electron hole pairs (double
lines) (a-b) and their interference contribution (c). The impurity
is denoted by filled circle.}
\label{figscat}
\end{figure}
The physical meaning of the collision integrals
\rref{boselci} and \rref{stlocal} is the following. In the absence
of the disorder the electron-hole pair propagates for an
infinitely long time. Due to the impurity potential the decay
of the pair into the electron and hole moving in different directions
as shown in Fig.~\ref{figscat} is allowed. Equations
\rref{boselci} and \rref{stlocal} are the probabilities for such a decay.
[See also  Sec.~\ref{abf}
after \req{st0clq} for further discussion.]

The non-local contribution to the collision integral
\bea
&&\St_n^{\da{$e$}{$\alpha$}} (\e,\nn_1) 
= \frac{2}{\nu} \int\!\frac{d\w}{2\pi} \frac{1}{\w^2} 
\int\!\frac{d^dq}{(2\pi)^d} \int\!\frac{d\nn_2\ldots d\nn_7}{\W_d^6} 
\nonumber \\
&&\times \gamma_{13}^2 \gamma_{46}^5 
\left[ \L^g_{14}  - \L^g_{34} \right] 
\sinh^2\left(\frac{\w\partial}{
2\partial\epsilon}\right)
\big[ f(\e,\nn_6)-f(\e,\nn_4) \big] 
\nonumber \\
&&\times \int\!d\e_1 \big\{ 
 \left[ \Upsilon^\alpha_{57;64}(\e_1,\w,\q) 
+ \Upsilon^\alpha_{57;46}(\e_1,\w,\q)  \right] \bar{\L}^\alpha_{72}
\nonumber \\
&& \qquad +\L^\alpha_{57} \left[ \Upsilon^\alpha_{72;13} (\e_1,\w,\q)
+ \Upsilon^\alpha_{72;31}(\e_1,\w,\q) \right] \big\}
\nonumber \\
\label{stnonloc}
\eea
by construction obeys its own conservation law
\be\label{ap}
\int \e^m \, \St_n^{\da{$e$}{$\alpha$}} (\e,\nn_1) \, d\e =0 , \quad m=0,1,
\ee
i.e. it conserves the energy and the number of the electrons moving
for a given momentum direction $\nn$. Moreover, one
can see that the collision integral \rref{stnonloc} does not
contribute to the linear response at all because in the thermodynamic
equilibrium $\Upsilon=0$ and $f$ does not depend on the angle.
The non-locality of this collision integral indicates that the
task formulated in  Sec.~\ref{sec:local} has not been quite
accomplished. Technically, this non-locality can be 
decoupled by introducing a density matrix non-diagonal
in the boson-ghost space. We choose not to pursue this line
because the term \rref{stnonloc} does not contribute to any observable
quantity we are interested in and does not affect any conservation laws.
\end{subequations}

The electron-electron collision integral can be split into 
elastic, {\em non-local} and {\em local} parts:
\begin{subequations}
\label{elelci}
\be
\St^{\da{$e$}{$e$}} (\e,\nn_1)= \St_{el}^{\da{$e$}{$e$}}(\e,\nn_1)
+ \St_{n}^{\da{$e$}{$e$}} (\e,\nn_1) + \St_{l}^{\da{$e$}{$e$}} (\e,\nn_1).
\ee

The elastic term describes the scattering of the electron
on the static self-consistent potential created by all the other
electrons
\bea
&& \St_{el}^{\da{$e$}{$e$}} = 
\frac{2}{\nu}\Re \int\!\frac{d\w}{2\pi}\frac{1}{\w}
\int\!\frac{d^dq}{(2\pi)^d}
\int\!\frac{d\nn_2\dots d\nn_6}{\W_d^5}
\gamma_{13}^{2}\gamma_{46}^{5}
\nonumber \\
&& \quad \times
\left[\L^\rho+3\L^\sigma-4\L^g\right]_{52}
\left[ f(\e-\w, \nn_6) - f(\e-\w, \nn_4) \right]
\nonumber \\ \label{steeel}
&& \quad \times \Big[\L^g_{14} f(\e, \nn_3)
+\L^g_{34} f(\e, \nn_1) \Big].
\eea
Its physical origin is discussed in details 
in Ref.~\onlinecite{ZNA}. 
Being elastic, it conserves the number of particle for
each energy shell:
\be
\int \St_{el}^{\da{$e$}{$e$}} (\e,\nn_1) \, d\nn_1 =0,
\ee
as one can see from the property $\gamma_{13}^2=-\gamma_{31}^2$
of the vertex \rref{vert}.

The nonlocal term 
{\setlength\arraycolsep{0pt}
\bea
&&\St_n^{\da{$e$}{$e$}}  =
-\frac{4}{\nu}\Re\int\!\frac{d\w}{2\pi}\frac{1}{\w^2}
\int\!\frac{d^dq}{(2\pi)^d} \int\!\frac{d\nn_2\ldots d\nn_6}{\W_d^5} 
\gamma_{13}^2 \gamma_{46}^5 \nonumber \\
&& \ \times   \int\!d\e_1 \big\{ 
\left[ \L^g_{14} - \L^g_{34} \right] 
\left[ \bar{\L}^\rho+3\bar{\L}^\sigma
-4\bar{\L}^g \right]_{52} \Psi^s_{46} (\e,\e_1,;\w) 
\nonumber \\ \label{steen}
\eea
describes the inelastic electron-electron collisions
during which bosons and ghosts act as virtual states.
[The function $\Psi^s$ were introduced in \req{psis}]. 
The real part being an even function, we can use 
\req{psip} to verify that \req{steen} obeys the conservation law
\be\label{ap2}
\int \e^m \, \St_{n,l}^{\da{$e$}{$e$}} (\e,\nn_1) \, d\e =0 , \quad m=0,1.
\ee
As indicated, the same law is satisfied by the local (and elastic) term:
\bea
&&\St_l^{\da{$e$}{$e$}}  =
\frac{2}{\nu}\int\!\frac{d\w}{2\pi}\frac{1}{\w^2}
\int\!\frac{d^dq}{(2\pi)^d} \int\!\frac{d\nn_2\ldots d\nn_6}{\W_d^5} 
\int\!d\e_1 \gamma_{13}^2 \gamma_{46}^5 \nonumber \\
&& \ \times \left[ \L^g_{14} + \L^g_{34} \right] 
\left[ \L^\rho+3\L^\sigma - 4\L^g \right]_{52} 
\nonumber \\
&& \ \times \sinh^2\left(\frac{\w\partial}{
2\partial\epsilon}\right) \left[ f(\e,\nn_6 ) - f(\e,\nn_4 )\right]
\nonumber \\
&& \ \times \left[ f(\e_1,\nn_1) [ 1-f(\e_1-\w,\nn_3)] + 
(\nn_1 \leftrightarrow \nn_3)  \right].
\label{steel}
\eea
Therefore equation \rref{ap2} enables us to conclude that both collision
integrals \rref{steen} and \rref{steel} do not affect the transport 
coefficients [when they can be considered as perturbations in comparison
to the bare impurity collision integral]. }

We note that, although it might not be evident, the present form
of the kinetic equation permits the proper identification of the inelastic
kernel that determines the phase relaxation time -- further details can 
be found in Appendix \ref{deph}.

\end{subequations}

\section{Summary of the results for thermal transport and specific heat}
\label{sec:sum}

In this section we present our final answers for the interaction corrections
to the thermal conductivity and the specific heat. They are obtained by
solving the kinetic equations and then substituting the solutions into 
the definitions \rref{consen} for the energy and energy current densities.
The explicit calculations are performed in Sec. \ref{resder}.
We will consider the short range impurities $\tau (\theta ) = \tau$.
We report our results for quasi one-dimensional and three-dimensional systems
in the diffusive limit $T\tau \ll \hbar$; for two-dimensional systems, we will
not put such a restriction on the temperature range. However common to all
dimensionality is the zeroth harmonic approximation for the Fermi liquid
constants [see \req{f0}].

\subsection{Thermal conductivity}

In the absence of the magnetic field the thermal conductivity tensor is 
diagonal, $\kappa_{\mu\nu} = \kappa \delta_{\mu\nu}$, and
we will write the expression for the diagonal components only as:
\be\label{wfvio}
\kappa = \kappa_{\scriptscriptstyle W\!F} + \Delta \kappa \, .
\ee
The first term is given by the Wiedemann-Franz law 
$\kappa_{\scriptscriptstyle W\!F} = L \sigma T$ with the inclusion of the 
interaction corrections to the conductivity and the Lorentz number given in
\req{WF}. The second term causes a violation of the Wiedemann-Franz
law. In the diffusive limit and for low dimensionality the main contribution
to $\Delta\kappa$ is due to the long range nature of the bosonic energy 
transport, which originates from the long range part of the interaction in
the singlet channel. In the quasi ballistic case a large contribution 
comes from the inelastic scattering of the electron on the bosons
as well. Smaller
corrections arise due to the triplet channel bosonic transport and
to the energy dependence of the elastic scattering. 

For quasi one-dimensional and three-dimensional systems in the diffusive
limit we write
\[
\Delta \kappa = \delta \kappa^\rho + 3 \delta\kappa^\sigma 
+ \delta\kappa_{el}  
\]
where the bosonic corrections include the ghost contributions
\[
\delta \kappa^\alpha = \kappa^\alpha - \kappa^g , \ \alpha=\rho ,\sigma
\]
[see \req{thcona} for the definition of $\kappa^\alpha$] and
we neglect the inelastic contributions $\delta\kappa_{in}$ which are smaller
by the parameter $T\tau / \hbar$.

\begin{subequations}\label{k1d}
For quasi one-dimensional systems the explicit expressions are:
\bea
\delta\kappa_{el} &=& \frac{1}{8\sqrt{2\pi}} \zeta \left( \frac{3}{2}
\right) \sqrt{\frac{DT}{\hbar}} \\
&&\times \left\{ - 1 +3 \left[ 1 - \frac{2}{F_0^\sigma}
\left( 1+F_0^\sigma - \sqrt{1+F_0^\sigma} \right)
\right] \right\} \nonumber \\
\delta\kappa^\rho &=& \frac{3}{8\sqrt{2\pi}} \zeta \left( \frac{3}{2}
\right) \sqrt{\frac{DT}{\hbar}} ak \ln^{\frac{1}{2}} 
\left( \frac{\hbar D k^2}{T} \right) \\
\delta\kappa^\sigma &=& \frac{3}{8\sqrt{2\pi}} \zeta \left( \frac{3}{2}
\right) \sqrt{\frac{DT}{\hbar}} \left[ \sqrt{1+F_0^\sigma} - 1 \right]
\eea
\end{subequations}
where $a$ is a length of the order of the wire width and 
$k = \sqrt{4\pi e^2 \nu}$ is the inverse screening length in the bulk.

\begin{subequations}\label{k3d}
For three-dimensional systems the results are:
\bea
\delta\kappa_{el}&=&\frac{5}{48\sqrt{2\pi^3}} \zeta \left( \frac{5}{2}
\right) \sqrt{\frac{T^3}{\hbar^3 D}} \\
&& \times \left\{ 1 + 3 \left[ 1 - \frac{2}{F_0^\sigma}
\left( 1- \frac{1}{\sqrt{1+F_0^\sigma}} \right) 
\right] \right\} \quad \nonumber \\
\delta\kappa^\rho&=&\frac{15}{32\sqrt{2\pi^3}} \zeta \left( \frac{5}{2}
\right) \sqrt{\frac{T^3}{\hbar^3 D}} \\
\delta\kappa^\sigma&=&\frac{15}{32\sqrt{2\pi^3}} \zeta \left( \frac{5}{2}
\right) \sqrt{\frac{T^3}{\hbar^3 D}} 
\left[ 1- \frac{1}{\sqrt{1+F_0^\sigma}} \right]
\eea
\end{subequations}
In these expressions, $\zeta (x)$ is the Riemann zeta function and 
$\zeta (3/2) \approx 2.612$ , $\zeta (5/2) \approx 1.341$.

For two-dimensional systems, we separate the corrections due to the singlet
and the triplet channel interactions:
\begin{subequations}\label{k2d}
\be
\Delta\kappa = \Delta\kappa_s + 3\Delta\kappa_t
\ee
The singlet channel contribution is, with logarithmic accuracy:
\be\label{k2ds}\begin{split}
\Delta\kappa_s = & \
\frac{T}{6\hbar} g_1 \left(2\pi \frac{T\tau}{\hbar} \right) 
\ln \left( \frac{ \hbar v_F k}{T} \right) \\
& - \frac{T}{24\hbar} g_2\left(\pi \frac{T\tau}{\hbar} \right) 
\ln \left( 1+ \frac{\hbar^2}{(T\tau)^2} \right) \\
& -\frac{\pi^2}{15}\frac{T}{\hbar}
\left(\frac{T\tau}{\hbar} \right)^2 \ln\left(\frac{E_F}{T}\right)
\end{split}\ee
where $k=2\pi e^2 \nu$ is the two-dimensional inverse screening length.
The cross-over functions $g_1$ and $g_2$ are given in \req{gsdef}. Here we note
that $g_1(x),g_2(x) \simeq 1$ for $x\ll 1$ and $g_1(x) \simeq 3/x$, 
$g_2(x)\simeq 14x^2/15$ for $x\gg 1$.

For the triplet channel we have:
\be\label{k2dt}
\Delta\kappa_t = \left\{
\begin{matrix}
-\frac{T}{18\hbar} \left[ 1- \frac{1}{F_0^\sigma} 
\ln \left( 1+F_0^\sigma \right) \right] &  \\ + \frac{T}{12\hbar} \ln \left( 
1+F_0^\sigma \right) & , \ T\tau \ll 1 \\ & \\
 -\frac{\pi^2}{15}\frac{T}{\hbar} \left( \frac{T\tau}{\hbar} \right)^2 
\ln\left(\frac{E_F}{T}\right) 
\left( \frac{F_0^\sigma}{1+F_0^\sigma}\right)^2 & , \ T\tau \gg 1
\end{matrix} \right.
\ee
\end{subequations}

In the diffusive limit $T\tau\ll \hbar$ our results are consistent
with those of Ref.~\onlinecite{Smith}, even though the form of the 
energy current operator used in this reference is, to our opinion, 
incorrect -- see Appendix~\ref{microcurr}.

\subsection{Specific heat}
\label{cvsumm}

The specific heat is given by
\be
\cV = \frac{\pi^2}{3} \nu T + \delta \cV
\ee
where the first term is the usual non-interacting electronic contribution 
and the second term is the bosonic interaction correction one.

For quasi one-dimensional and three-dimensional systems 
\be\label{dcvdiff}
\delta \cV = \left( 1+ 3\Lambda_d \right) 
\left( \frac{T}{\hbar D}\right)^{\frac{d}{2}} a_d
\ee
The two terms in the first bracket are respectively the singlet and triplet 
channel contributions. The singlet
channel term is considered in the unitary limit and is therefore independent
of any interaction parameter. On the other hand,
the Fermi liquid parameter for the interaction in the 
triplet channel enter into \req{dcvdiff} as
\be
\Lambda_d = 1 - \frac{1}{\left(1+F^\sigma_0\right)^{d/2}}
\ee
and the numerical factors $a_{1,3}$ are:
\[ \begin{split}
a_3 = \frac{15}{32\pi \sqrt{2\pi}}\zeta \left( \frac{5}{2}\right) \\
a_1 = -\frac{3}{8\sqrt{2\pi}} \zeta \left( \frac{3}{2}\right)
\end{split}\]

For two-dimensional systems the result is, with logarithmic accuracy:
\be\label{dcv2d}\begin{split}
\delta \cV = - \left( \frac{T}{\hbar D} \right) \bigg[ &
\left(1+3\Lambda_2\right) \frac{1}{12} \ln  \left( \frac{E_F}{T}\right) \\
 & + \left( 1+3\Lambda_2^2 \right)  \frac{3}{4\pi} \zeta(3) T\tau \bigg]
\end{split}\ee
where $\zeta (3) \approx 1.202$.
The first term on the right hand side extends to higher temperatures the 
logarithmic behaviour known in the diffusive limit (the upper cutoff 
is of the order of the Fermi energy $E_F$ and not $\hbar/\tau$); 
the second term becomes relevant in the quasi 
ballistic limit. In the diffusive limit our results are the same
as those obtained in Ref.~\onlinecite{AA85} by explicit thermodynamic
calculation.


\section{Derivation of the kinetic equation}
\label{sec:derivation}

This section is devoted to the derivation of the local kinetic equation. 
We first introduce the Eilenberger equation and some basic notation. 
Then we perform a (generalized) gauge transformation: this is the 
crucial step that enable us to obtain the local description. Then we 
introduce the bosonic degrees of freedom and derive the collision integrals.

\subsection{Eilenberger equation}

Our starting point for the derivation of the kinetic equation is the
same as in Ref.~\onlinecite{ZNA} and we briefly summarize it here.

The interaction with small momentum and energy transfer in the
singlet channel [the triplet channel will be discussed in 
section \ref{triplet}] is decoupled using the two
Hubbard-Stratonovich fields ${\phi}_{\pm}(t,\r,\nn)$.
For the purpose of the one loop approximation we will employ,
these fields can be considered as Gaussian with the propagators
\beqa
&&
\lda \phi_+(1) \phi_+(2) \rda = 
-\frac{i}{2} D^K(1,2), 
\nonumber\\
&&
\lda \phi_+(1) \phi_-(2) \rda = 
-\frac{i}{2} D^R(1,2), 
\nonumber\\
&&
\lda \phi_- (1)\phi_+(2) \rda =
-\frac{i}{2} D^A(1,2), 
\nonumber\\
&&
\lda \phi_- (1)\phi_-(2) \rda = 0.
\label{Ds}
\eeqa
where $\lda\dots\rda$ means averaging over the fields $\phi_\pm$.
We used the short hand notation
\be
\label{notation}
\begin{split}
&(i) \equiv (t_i,\r_i,\nn_i), 
\ \int d i \equiv \int dt_i d\r_i\int \frac{d \nn_i}{\Omega_d},
\\
& (i^*) \equiv (\r_i,\nn_i), 
\ \int d i^* \equiv \int  d\r_i\int \frac{d \nn_i}{\Omega_d}
\end{split}
\ee
where
$i=1,2,\dots$ and  $\Omega_d$ is the total solid angle.

We introduce the disorder averaged Green's function of the electron
in the field $\phi_{\pm}$ in its matrix form in Keldysh space:
\beq
\widehat G(1, 2 | \phi) = 
\left(
\begin{matrix}
G^R(1, 2 | \phi) & G^K(1, 2 | \phi) \cr
G^Z(1, 2 | \phi) & G^A(1, 2 | \phi) 
\end{matrix}
\right)_K,
\label{nosense}
\eeq
such that its average over the fluctuating field $\phi_{\pm}$ gives the
usual expressions for the physical propagators:
\begin{eqnarray} 
&&\lda G^R(1, 2) \rda = 
-i\theta(t_1-t_2)\langle \psi (1) \psi^\dagger(2)+
\psi^\dagger(2) \psi (1) \rangle,
\nonumber
\\
&&\lda G^A(1, 2) \rda = 
i\theta(t_2-t_1)\langle \psi (1) \psi^\dagger(2)+
\psi^\dagger(2) \psi (1) \rangle,
\nonumber
\\
&&\lda G^K(1, 2) \rda = -i \langle \psi (1) \psi^\dagger(2)-
\psi^\dagger(2) \psi (1) \rangle,
\nonumber
\\
&&\lda G^Z(1,2) \rda=0 \, .
\label{GFs}
\end{eqnarray} 
Here $\theta(t)$ is the Heaviside step function, $\psi^\dagger,\ \psi$
are the fermionic creation/annihilation operators in the Heisenberg
representation and quantum mechanical averaging $\<\dots\>$ is
performed with an arbitrary distribution function to be found from the
solution of the kinetic equation.

For the disorder averaged Green's function, the semi-classical approximation 
is obtained by integrating the Wigner transform of $\widehat G(1,2 | \phi)$
over the distance from the Fermi surface:
\be
\widehat G(t_1,t_2,\p ,\R ) =
\int\! d^2r \ e^{i\P \cdot \r} \widehat G(1,2 | \phi),
\ee
\bea
\r = \r_1 - \r_2 ; \quad 
\R = \frac{1}{2}(\r_1 + \r_2); \nonumber
\\
\P = \p - \frac{1}{2}
\left[\A \left(t_1, \R\right)+
\A \left(t_2, \R\right)\right]; \nonumber
\eea
\be
\label{intg}
\hat{g}(t_1, t_2, \nn, \r ) = \frac{i}{\pi} 
\int\limits_{-\infty}^{\infty} \! d\xi \ \widehat G\left(t_1,
t_2, \nn \left[p_F + \frac{\xi}{v_F}\right], \r\right),
\ee
where $\A$ is the vector potential of an external electromagnetic field, 
$p_F$ the Fermi momentum and $v_F$ the Fermi velocity.
The dynamics of the semiclassical Green's function 
$\hat g$ in the matrix form is governed by the 
Eilenberger equation\cite{Eilenberger}:
\be \label{eil1} 
\left[
{\tilde\partial_{t}} + \v \cdot \tilde{\nnabla}  
+ \boldsymbol{\w}_c\cdot
\left(\nn \times 
\frac{\partial}{\partial \nn} \right)
\right] \hat{g} + i\left[ \hat{\phi}; \hat{g} \right] = 
\frac{\left[\hat{g} \concomma \St_\tau \hat{g}\right]}{2},  
\ee
where $\v = v_F \nn$, the action of the
``bare'' collision integral on any function
$a(\nn)$ is defined as
\be
\begin{split}
& \left[\St_{\tau}a\right](\nn) = 
\int \frac{d\nn_1}{\W_d}
{\rm St}_\tau(\nn,\nn_1)
a(\nn_1) \, ,\\
& {\rm St}_\tau(\nn_1,\nn_2) =
\frac{1}{\tau(\theta_{12})}-
\donetwo{1}{2}\int  \frac{d\nn_2}{\tau(\theta_{12})}
\end{split}
\mylabel{Sttau}
\ee
and $\theta_{12}=\widehat{\nn_1\nn_2}$ [for the short-range
impurity $\tau(\theta)$ does not depend on $\theta$; however, the
formulas derived here will be valid for the arbitrary impurity
scattering]. 
The time convolution of two 
matrices $\hat{a}(t_1,t_2)$ and $\hat{b}(t_1,t_2)$ is:
\be\label{tcon}
\hat{a}\, \con\, \hat{b} = \int \! dt_3 \ \hat{a}(t_1,t_3)\hat{b}(t_3,t_2);
\qquad
\left[\hat{a} \,\concomma\, \hat{b}\right]= \hat{a} \circ \hat{b} -
\hat{b} \circ \hat{a}. 
\ee
Defining the commutator between a matrix $\hat{c}(t,\r,\nn )$ and $\hat{g}$ as
\be
\label{com1}
\left[ \hat{c};\hat{g} \right] = 
\hat{c}(t_1,\r,\nn ) \hat{g}(t_1,t_2,\nn,\r) - 
 \hat{g}(t_1,t_2,\nn,\r) \hat{c}(t_2,\r ,\nn)
\ee
the covariant derivatives in \req{eil1} are given by:
\begin{subequations}
\label{covariant}
\bea
\tilde{\partial_t} \hat{g} & = & 
\partial_{t_1} \hat{g}  + \partial_{t_2} \hat{g} + 
i \left[\hat{\varphi};\hat{g}\right]
\\
\tilde{\nnabla}\hat{g} & = & \nnabla \hat{g} + 
i\left[\hat{\A};\hat{g}\right]
\eea
\end{subequations}
with $\hat{\A} =\A \hat{\openone}_K$ and 
$\hat{\varphi} = \varphi \hat{\openone}_K$. 
Here $\hat{\openone}_K$ denotes the unit matrix in Keldysh space and $\varphi$ 
is the scalar potential for an external electromagnetic field such that:
\be
e\E = -\nnabla \varphi + \pt \A \quad , \quad e\B =-c 
\nnabla \times \A \nonumber
\ee
The vector $\boldsymbol{\w}_c=e\B/(mc)$ has the magnitude of the cyclotron 
frequency and the direction of the magnetic field $\B$. Finally, 
$\hat{\phi}$ is the  matrix in the Keldysh space:
\be
\hat{\phi} = 
\left(
\begin{matrix}
\phi_+ & \phi_- \cr
\phi_- & \phi_+ 
\end{matrix}
\right)_K
\ee

The matrix Green's function $\hat{g}$ is subject to the following 
constraints:
\begin{subequations}
\label{constr}
\bea
 && \hat{g}(\nn ,\r )\con \hat{g}(\nn ,\r ) = \delta (t_1-t_2) \hat{\openone}_K
\\
\hspace{-1cm} && \int\! dt \ \Tr\  \hat{g}(t,t,\nn ,\r ) =0.
\eea
\end{subequations}
In thermal equilibrium, the relation
\be
\label{gkeq}
\begin{split}
&g^K(t_1,t_2)= \left[g^R \con\, n- n \con\,g^A \right](t_1,t_2)
\\
&
n(t_1,t_2) =\int
\frac{d\e}{2\pi}e^{i\e(t_2-t_1)}n(\e);
\\
&
n(\e)=1-2f_F(\e)=2\tanh \frac{\e}{2T}
\end{split}
\ee
must hold independently of the form of the spectral functions $g^{R,A}$.

In what follows we will assume that there is no magnetic field,  
$\B =0$ and $\boldsymbol{\w}_c =0$, but no gauge choice is made: 
although one could set $\A=0$ by a gauge transformation,
both the scalar and vector external potentials are left arbitrary
in order to keep track of the gauge invariance of the equations.

As for the propagators defined by \req{Ds}, they satisfy 
matrix Dyson's equation:
\be
\begin{split}
&\hat{D}(1,2) = \hat{D}_0(1,2) + 
\int\!\! d3 \!\int\!\! d4 \ \hat{D}_0 (1,3) \hat{\Pi}(3,4) \hat{D}(4,2);
\\
&
\hat{D}_0 (1,2) = - \left[ V(\r_{12}) + 
\frac{F^{\rho}\!\left(\theta_{12}\right) \delta (\r_{12})}
{\nu}\right] \delta (t_{12}) \hat{\openone}_K;
\end{split}
\label{dys}
\ee
where $V(\r)$ is the long range part of the interaction
(for the Coulomb interaction $V(\r)=e^2/|\r|$),
$\theta_{12}=\widehat{\nn_1\nn_2}$, $\r_{12}=\r_1\! -\! \r_2$,
$t_{12}=t_1-t_2$.
 The matrix propagator is denoted by $\hat{D}$
and $\hat{\Pi}$ is the matrix polarization operator. They have a structure
similar to the Green's function one:
\be
\hat{D} =  \left( \begin{matrix}
D^R & D^K \cr
0 & D^A 
\end{matrix}
\right)_K,
\quad
\hat{\Pi} =  \left( \begin{matrix}
\Pi^R & \Pi^K \cr
0 & \Pi^A 
\end{matrix}
\right)_K
\ee

The polarization operators are given by variational derivatives of the 
solutions to the Eilenberger equation \rref{eil1}: 
\begin{subequations}
\label{polar}
\be
\Pi^R (1,2) = \Pi^A(2,1) = \nu \left[ \delta_{12} + \frac{\pi}{2}
\frac{\delta {g^K (t_1,t_1,\nn_1,\r_1 )}}{\delta \phi_+ (t_2, \r_2, \nn_2 )}
\right],
\ee 
\be
\Pi^K (1,2) = \frac{\pi \nu}{2} 
\left[
\frac{\delta {( g^K +g^Z )(t_1,t_1,\nn_1,\r_1 )}}
{\delta \phi_-(t_2, \r_2,\nn_2 )}
\right],
\ee
\end{subequations}
where 
\be
\delta_{12}\equiv \W_d \dnnn\delta(\r_1-\r_2)\delta(t_1-t_2),
\label{d12}
\ee
 with $\Omega_d$ being the total solid angle.

\subsection{The gauge transformation}
\label{sec:gaugetr}

With the Eilenberger equation \rref{eil1} at hand, one could proceed as 
in Ref.~\onlinecite{ZNA} in order to derive an equation for the 
distribution function. However the resulting
inelastic part of the collision integral, expressed in terms of 
the electron distribution function only, is {\em non-local} and 
the evaluation of e.g. the thermal conductivity 
would require the time and spatial gradient expansion of this term in
the spirit of \req{kineq11}. As we already discussed, such a route makes
the energy conservation in the kinetic equation obscure.
 Here we follow a different approach,
inspired by the following considerations\cite{AK}: if the fluctuating fields
were uniform, they could be eliminated from \req {eil1} 
by a gauge transformation
\be
\hat{g} \to e^{-i\int^{t_{1}} \hat{\phi}(t) dt}\hat{g}
e^{i\int^{t_{2}} \hat{\phi}(t) dt }.
\ee 
In other words,  the position-independent fluctuations 
of the $\phi$ fields define the time dependent position of the energy
levels but the occupation numbers for such levels do not change.
Therefore such fluctuations affect neither the electric transport 
nor the electron contribution  to the thermal transport in the system. 
Moreover, if the path of the electron were a straight line,
all the smooth fluctuating fields could still be eliminated in the eikonal
approximation and once again they should not affect the 
electronic contribution to the transport. 
To get rid of such spurious contributions
we will employ the gauge transformation described below. 

We introduce a new matrix 
field $\hat{K}(t,\nn ,\r )$:
\be
\hat{K} = 
\left(
\begin{matrix}
K_+ & K_- \cr
K_- & K_+ 
\end{matrix}
\right),
\ee
which is a functional of the field $\hat{\phi}$ and is used to perform 
the ``generalized'' gauge transformation
\be\label{gaugetr}
\hat{g} \to e^{-i\hat{K}(t_1,\nn ,\r )}\hat{g} \,
e^{i\hat{K}(t_2,\nn ,\r )}.
\ee
This transformation is unitary and as such preserves
the constraints \rref{constr}.  As we will see, 
it will lead us to the 
{\em local} kinetic equations. Applying the transformation 
to the Eilenberger equation \rref{eil1}, we obtain
\be
\left[ {\tilde\partial_{t}} + \v \cdot \tilde{\nnabla}  
\right] \hat{g} 
-i \left[ \left( \partial_{t}+\vgrad \right) 
\hat{K} -\hat{\phi}, \hat{g} \right] 
=\frac{1}{2}
\left[ \hat{g} \concomma 
{\St}_\tau^\phi\hat{g}
 \right],
\label{eil2}
\ee
where
\[
\begin{split}
&\left[{\St}_\tau^\phi\hat{g}\right](t_1,t_2,\nn )
\equiv\int \frac{d\nn_1}{\W_d}{\rm St}_\tau(\nn,\nn_1) 
\\
&
\times
 e^{i\hat{K}(t_1,\nn )}
e^{-i\hat{K}(t_1,{\nn}_1 )} 
\hat{g}(t_1,t_2,{\nn}_1 )  e^{i\hat{K}(t_2, \nn_1 )}  
e^{-i\hat{K}(t_2, {\nn} )}.
\end{split}
\]
The ``bare'' impurity collision integral and the derivatives
are defined respectively in \req{Sttau} and \rref{covariant}. 
We suppressed the $\r$ argument which is the same in
all functions.

We will look for a perturbative solution to the Eilenberger equation
in the form \rref{eil2} in the first  loop approximation; 
to do so it will suffice to  retain only the terms at most 
quadratic in the $K$ fields in the collision integral. 
At the lowest order, $\hat{g}$ has the form:
\be
\label{gzero}
\hat{g} = \left( \begin{matrix}
\delta (t_1-t_2) & g^K \cr
0 & -\delta (t_1 - t_2) \end{matrix} \right).
\ee
We will require that such form is preserved even in the first and
second order in $K$, i.e. the corrections to the spectrum (described by
$g^{R,A}$) are indeed eliminated by the gauge transformation.

At the linear order, the retarded, advanced and `Z' components 
 of \req{eil2} vanish if  $K_{-}$ satisfies the equation
\begin{equation}
\mylabel{eqKminus}
\Big( \partial_t + \vgrad \Big) K_{-} +  \St_{\tau}
K_{-} = \phi_{-}.
\end{equation}
The solution to the integro-differential equation \rref{eqKminus} 
can be written in terms of the diffuson propagator 
$\L^g(t_1,t_2;\nn_1,\nn_2;\r_1, \r_2)$ --
the retarded solution to the classical kinetic equation
\be
\mylabel{Diffuson}
\left( \partial_{t_1} + \v_{1} \cdot \nnabla_{\r_1} - \St_{\tau} \right) 
\L^g =\delta_{12},
\ee
where $\delta_{12}$ is defined in \req{d12}.
Using \req{Diffuson}, we find
\be
\mylabel{km}
K_{-}(1) = -\int\!\! d 2\,\bar{\L}^g(1,2) \phi_{-}(2),
\quad \bar{\L}^g(1,2)=\L^g(2,1),
\ee
where we use the shorthand notation \rref{notation}.
In an operator notation, \req{km} can be rewritten as
\be
\label{kmprime}
\tag{\ref{km}$^\prime$}
K_{-} = -\hat{\bar{\L}}^g \phi_{-}.
\ee

To simplify further manipulations,
we introduce the following function of three angular variables
\be
\vertex{}{i}{j}{k} \equiv 
\frac{1}{\tau(\theta_{ij})}\W_d\left[\donetwo{j}{k}-\donetwo{i}{k}\right]
\equiv \gamma_{ij}^k.
\label{gamma1}
\ee
This function is related to the impurity collision integral
\rref{Sttau}
by
\be
\begin{split}
&\int \frac{d\nn_2}{\W_d} \vertex{}{1}{2}{3} = 
\left[{\rm St}_\tau\right](\nn_1,\nn_3);\\
&\int \frac{d\nn_1}{\W_d} \vertex{}{1}{2}{3} = 
-\left[{\rm St}_\tau\right](\nn_2,\nn_3);
\\
&\int \frac{d\nn_3}{\W_d} \vertex{}{1}{2}{3}=0.
\end{split}
\label{gamma2}
\ee

Denoting with $\delta g^K$ the first order correction to $g^K$, 
the Keldysh component of the
Eilenberger equations at the linear order is:
\bea
\label{keldcom}
&&i \Big( \tilde{\pt} + \v\cdot\tilde{\nnabla} -\St_{\tau} \Big) \delta g^{K}
\\ 
&& +\left[ \left( \partial_{t} + \vgrad - \St_{\tau}\right) K_{+}
-\phi_+ ; g^K \right] = \hat{Q} K_- 
\nonumber
\\
&& + \int\frac{d\nn_2 d\nn_3}{\W_d^2}\vertex{}{}{2}{3}
\left[K_+(\nn_3); g^K(\nn_2)-g^K(\nn) \right]
\nonumber 
\eea
with [cf. \req{covariant}]
\[ \begin{split}
\tilde{\pt} \delta g^K & = 
\Big( \partial_{t_{1}} + \partial_{t_{2}} \Big) \delta g^K 
+ i [ \varphi ; \delta g^K ] \\
\tilde{\nnabla} \delta g^K & = \nnabla \delta g^K + i [ \A ; \delta g^K ].
\end{split} \]
Here $\hat{Q}$ is a local in space operator
\[
\hat{Q} K_-   =  \int\! dt_3 
\int \frac{d\nn_1}{\W_d}
{Q}(t_1,t_2,\nn; t_3,{\nn}_1; \r ) K_-(t_3,{\nn}_1,\r )
\]
with the kernel 
\be
\mylabel{Q}
\begin{split}
&Q(t_1,t_2,\nn_1; t_3,{\nn}_2)=\frac{1}{2} \int
\frac{d\nn_3}{\W_d}
\vertex{}{1}{3}{2}
\\
& \times\left[
 P g^K(t_1,t_3,\nn_1) g^K(t_3,t_2,{\nn}_3)
+(\nn_1\leftrightarrow \nn_3)\right].
\end{split}
\ee
We suppressed the spectator argument $\r$, which is the same in
each term of the equation, and the last term means that one must add terms 
like the ones shown but with the angular arguments of the Green's function 
switched.
The principal value sign $P$ in \req{Q} means that the divergent at $t_1
\to t_2$ part of the product of the Green's functions 
\be\label{gklim}
g^K(t_1,t_2,\nn ,\r )\Big|_{t_1 \to t_2} = -\frac{2i}{\pi (t_1-t_2)}
+ \mbox{regular} 
\ee
must be excluded:
\[
\begin{split}
Pg^K(t_1,t_3&)g^K(t_3,t_2) 
\\
&\equiv
 g^K(t_1,t_3)g^K(t_3,t_2) -4 \delta (t_1-t_3) \delta (t_3-t_2),
\end{split}
\]
or, equivalently, 
\[
\begin{split}
&Pg^K(t_1,t_3)g^K(t_3,t_2) 
\\
&
\equiv \frac{1}{2}\sum_{\sigma=\pm 1}
g^K(t_1,t_3+\sigma i 0)g^K(t_3+\sigma i0, t_2).
\end{split}
\]
It is worth noticing that all non-equilibrium effects contribute to
the regular part in \req{gklim} but not to the singular part -- 
the states deep into the Fermi sea, which are not 
perturbed, contribute to it.

In order to solve \req{keldcom} we define a new field $\tilde{K}_{-}$
by the relation 
\be
\mylabel{Ktildem}
\tilde{K}_{-} (t,\nn,\r) = 
\left( i\pt
\right)^{-1} 
\hat{M} K_{-} \, ,
\ee
where the operator $\hat{M}$ will be shown to be
related to certain products of the Green's
functions $g^K$, see \req{mg}. The operator $\hat{M}$ is Hermitian and 
local in space but not in the momentum direction and time. We used again 
an operator notation
\be
\mylabel{hatM}
\hat{M} \, K_{-}\equiv
\int\! dt_1 \int \frac{d\nn_1}{\W_d}{M(t, \nn; t_1,{\nn}_1;\r ) 
K_{-}(t_1,{\nn}_1,\r )}.
\ee

We require $K_+$ to satisfy:
\be
\label{kpeq}
\Big( \partial_t + \vgrad -\St_{\tau} \Big) K_{+} = 
\phi_{+} - {2} \tilde{K}_{-},
\ee
whose solution is:
\be\mylabel{kp}
K_+ = \hat\L^g  \phi_+ -{2}\hat\L^g \tilde{K}_-.
\ee
The operator notation here is the same as in \req{km}.

\begin{subequations}
The next task is to choose the ``best'' form for the operator $\hat{M}$
to maximally simplify any further perturbative expansion.
Writing $\delta g^K = \delta g^K_+ +\delta g^K_-$ , we 
obtain the following equations:
\label{dgpmeq}
\bea
\label{dgmeq}
i \hat{L} \delta g^K_{-} 
&=& \hat{Q} K_- + {2} \Big[ \tilde{K}_-, g^K \Big] 
\\
\label{dgpeq} 
i \hat{L} \delta g^K_{+} 
 &=& \int\frac{d\nn_2 d\nn_3}{\W_d^2}\vertex{}{}{2}{3}
\\
&& 
\times \left[K_+(\nn_3),\left({g^K}(\nn_2)-{g^K}(\nn)\right)\right],
\nonumber
\eea
\end{subequations}
where
$\hat{L} \equiv \Big(
 \tilde{\pt} + \v\cdot\tilde{\nnabla}  - \St_{\tau} \Big) $.

We note that the right hand side of \req{dgpeq} vanishes 
in equilibrium, since $g^K =\an{g^K}$. 
Therefore $\delta g^K_+$ also vanishes in equilibrium and cannot
contribute to equilibrium properties such as the specific heat.
Moreover, even in non-equilibrium
$\delta g^K_+ (t_1, t_1, \nn ,\r ) = 0$, because the right hand side 
of \req{dgpeq} vanishes, see the remark after \req{gklim}. 
It means that $\delta g^K_+$ does not contribute 
to the electron density or current.

We are now ready to choose the operator $\hat{M}$.
We require that $\delta g^K_-(t_1, t_1, \nn ,\r ) = 0$,
i.e. $\delta g^K_-$ also does not contribute 
to the electron density or current. It means the the
right-hand side of \req{dgmeq} must vanish for $t_1=t_2$
for any field $K_-$.
Imposing this requirement, we obtain 
\be
\hat{M}(t_1,t_2,\nn ,\tilde{\nn}, \r )
 = \frac{\pi}{4} \hat{Q}(t_1,t_1,\nn; t_2,\tilde{\nn};\r ).
\label{MQ}
\ee
Together with \req{Q}, it yields
\be
\begin{split}
& {M}(t_1,\nn_1;t_2,{\nn}_2;\r ) = \frac{\pi}{8} 
 \int \frac{d\nn_3}{\W_d}\vertex{}{1}{3}{2}
\\
& 
\times\left[P g^K(t_1,t_2,\nn_1) g^K(t_2,t_1,{\nn}_3)
 + (\nn_1 \leftrightarrow \nn_3)\right].
\end{split}
\label{mg}
\ee
Expression \rref{gamma1} for the vertex $\gamma$ enables us
to establish the following properties of kernel \req{mg}
\be
{M}(t_1,\nn_1;t_2,{\nn}_2) = {M}(t_1,\nn_2; t_2,{\nn}_1)
= {M}(t_2,\nn_2; t_1,{\nn}_1),
\tag{\ref{mg}$^\prime$}
\label{mgsymm}
\ee
i.e. the operator $\hat M$ is Hermitian.

It is instructive to find $\hat{M}$ in the thermal equilibrium.
Using \req{gkeq} and the fact that due to the choice \rref{eqKminus}
the retarded and advanced components of $\hat{g}$ are still given 
by \req{gzero}, we find from \reqs{gamma2} and \rref{mg}:
{\setlength\arraycolsep{0pt}
\bea
&&{M}_{eq}(t_1,\nn_1; t_2,\nn_2; \r)
=\int \frac{d\w}{2\pi}e^{i\w(t_2-t_1)}\hat{M}_{eq}(\w;\nn_1,\nn_2) 
\nonumber\\
\label{meq}
&&{M}_{eq} (\w;\nn_1,\nn_2)=
-\omega \coth \left( \frac{\omega }{2T} \right)
 \Big[ \hat{St}_\tau \Big](\nn_1,\nn_2).
\eea
Equation \rref{meq} will be useful for checking the 
fluctuation dissipation theorem.}

\subsection{Polarization operators and propagators}
\label{ehpairs}

The knowledge of the linear order corrections to the Green's function permits 
the calculation of the polarization operators as variational 
derivatives of the original Green's functions (i.e. before the gauge 
transformation) in the limit $t_2 \to t_1$, see \req{polar}.
At linear order the corrections to the original Green's functions are given 
by the relations [cf. \req{gaugetr}]:
\begin{subequations}
\bea
\hspace{-.5cm} \delta g^K &\to& \delta g^K - i \left[ K_+ , g^K \right]
-2K_-\delta (t_1-t_2 ) ,
\\
\hspace{-0.5cm} \delta g^Z &\to& 2K_- \delta (t_1-t_2).
\eea
\end{subequations}
By construction of the previous subsection:
\be 
\lim_{t_2\to t_1} \delta g^K(t_1,t_2 ,\nn,\r) =0 
\ee
and using \req{gklim}:
\be
\lim_{t_2\to t_1} -i\left[ K_+ , g^K \right] = -\frac{2}{\pi} 
\pt K_+ (t,\nn,\r ).
\ee
Substituting these results into \req{polar} and using
\reqs{km}, \rref{Ktildem} and \rref{kp}, we obtain:
\begin{subequations}
\label{polar2}
\be
\begin{split}
&\Pi^R (1,2) = \nu \left[ \delta_{12} - \partial_{t_{1}} 
\L^g(1,2 ) \right];\\
&\Pi^A (2,1) = \nu \left[ \delta_{12} - \partial_{t_{2}} 
\bar{\L}^g(1,2 ) \right],
\end{split}
\label{pira}
\ee 
and we use the notation \rref{notation} throughout this subsection.
The result for the Keldysh component is 
\be\label{pik}
\Pi^K (1,2) = {2i\nu} 
\left[\hat\L^g\hat{M} \hat{\bar{\L^g}} \right](1,2).
\ee
\end{subequations}
The actions of the operators $\hat{M}$ and $\hat{D}$ are defined
in \reqs{km} and \rref{hatM}.

It is easy to check that the 
fluctuation-dissipation relation between the polarization operators
holds in the thermal equilibrium. 
As follows from \reqs{Diffuson} and \rref{km}
\be\label{diffid}
-2 \hat\L^g\hat{St}_\tau \hat{\bar{\L}}^g   = 
\hat\L^g + \hat{\bar{\L^g}}.
\ee
We perform the time Fourier transform for all the propagators
and the polarization operators in thermodynamic equilibrium
\be
A(1, 2)= \int \frac{d\w}{2\pi}e^{i\w(t_2-t_1)}A(\w; 1^*, 2^*),
\ee
where we use the short hand notation \rref{notation}.
Substituting \reqs{meq} and \rref{diffid} into \req{pik}, 
we obtain that in equilibrium:
\be\label{poleq}
\Pi^K_{eq} (\w;1^*\!,2^* )= \Big[ \Pi^R (\w;1^*\!,2^*) - \Pi^A(\w;1^*\!,2^*
  ) \Big]
\coth \frac{\w}{2T}.
\ee
and with the help of \req{dys}, we derive the fluctuation-dissipation
relation
\be
D^K_{eq} (\w; 1^*\!,2^* )= \Big[D^R (\w; 1^*\!,2^*) - D^A(\w; 1^*\!,
2^*) \Big] 
\coth\frac{\w}{2T}.
\label{FDTD}
\ee

Given the expressions for the polarization operators obtained above, we  
can solve the Dyson equation \rref{dys} and obtain the explicit expressions
for the interaction propagators. In the operator notation:
\begin{subequations}
\label{Dresult}
\bea
&&\nu \hat{D}^R = -  \frac{1}{1+\hat{F}-\partial_t\hat{F} 
\hat\L^g  } \hat{F}
\\
&&\nu \hat{D}^A =-\hat{F} \frac{1}{1+\hat{F}+\partial_t
\hat{\bar{\L}}^g \hat{F}  } 
\\
\label{dkeq}
&&\hat{D}^K = 2 i\nu  \hat{D}^R \hat \L^g\hat{M} 
\hat{\bar{\L^g}}\hat{D}^A,
\eea
\end{subequations}
where 
the action of the operator $\hat{F}$ on any function $a(t,\nn,\r)$ is
defined by
\be
\begin{split}
\big[\hat{F} a\big]&(t;\nn,\r)
\equiv \int\!\frac{d\nn_1}{\W_d} 
\bigg[F^\rho (\widehat{\nn\nn_1})a(t,\nn_1,\r)
\\ & +
\int d\r_1 \nu V(\r -\! \r_1) a(t,\nn_1,\r_1) \bigg]
,
\end{split}
\label{Fdef}
\ee
see also the text after \req{dys}.

To find the propagators
for the  fields $K_\pm$ given in \reqs{kmprime} and  \rref{kp}, defined as:
\beqa
&&
\lda K_+(1) K_+(2) \rda = 
\frac{i}{2} {\cal K}^K(1,2), 
\nonumber\\
&&
\lda K_+(1) K_-(2) \rda = 
\frac{i}{2} {\cal K}^R(1,2), 
\nonumber\\
&&
\lda K_- (1)K_+(2) \rda =
\frac{i}{2} {\cal K}^A(1,2), 
\nonumber\\
&&
\lda K_- (1)K_-(2) \rda = 0 
\label{KKs}
\eeqa
we use \reqs{Ds} and \rref{Dresult} and obtain  
for the retarded and advanced propagators
\begin{subequations}
\label{Ks}
\be \label{kra}
\hat{\cal K}^R =  \hat\L^g \hat{D}^R \hat\L^g;
\quad
\hat{\cal K}^A =  \hat{\bar{\L^g}} \hat{D}^A \hat{\bar{\L^g}},
\ee
whereas the result for the Keldysh propagator is
\be \label{kk1} \begin{split}
\hat{\cal K}^K\! = &- \hat\L^g \hat{D}^K \hat{ \bar{\L^g}} 
\\& + {2i} 
\Big[ \hat\L^g  (\pt)^{-1} \hat{M} 
 \hat{\bar{\L^g}} \hat{D}^A \hat{\bar{\L^g}}
- \hat\L^g \hat{D}^R \hat\L^g  \hat{M}  
(\pt)^{-1} \hat{\bar{\L^g}}  
 \Big]. 
\end{split}\ee
\end{subequations}
The fluctuation-dissipation relation between the $D$ pro\-pagators
in \req{FDTD}, the 
equilibrium form for $\hat{M}$ in \req{meq} and the 
identity \rref{diffid} enable us to verify the 
fluctuation-dissipation relation for the ${\cal K}$ pro\-pagators:
\be
{\cal K}^K_{eq} (\w;1^*\!,2^* )= \left[{\cal K}^R (\w;1^*\!,2^*) 
- {\cal K}^A(\w;1^*\!,2^* ) \right] 
\coth \frac{\w}{2T}.
\label{FDTK}
\ee

\subsection{Additional bosonic fields}
\label{abf}

Equation \rref{kk1} together with \reqs{kra} and \rref{mg} permits
to express the Keldysh propagator ${\cal K}^K$ in terms of the
electron distribution function. This relation, however, would be
nonlocal on the spatial scale much larger than the temperature length
\be
L_T \simeq {\rm min}\left[\frac{\hbar v_F}{T}, 
v_F \sqrt{\frac{\hbar\tau}{T}}\, \right], \label{tlength}
\ee
recall the discussion of Sec.~\ref{sec:local}.
Indeed the collision integral and all the physical quantities 
will be given by integrals of the type
\[
{\cal I}^{\alpha} =
\int d\omega f(\omega) {\cal K}^{\alpha}(\omega),\quad {\alpha=R,A,K},
\]
where the function $f(\omega)$ depends on its argument on the
characteristic scale of $T$. 
A retarded function is an analytic function of $\omega$ at $\Im \, \w >0$,
which means that for $\alpha=R$ the integral will be determined only by 
the singularities
of $f(\omega)$, i.e. ${\cal I}^{R} \simeq {\cal K}^{R}(\omega=iT)$.
This immediately restricts the spatial scales to $L_T$. The same
argument applies to the advanced case, because of analyticity at
$\Im \, \w < 0$. The function ${\cal K}^{K}(\omega)$, however, is not
analytic. Moreover, according to \req{kk1} it contains overlapping
singularities of the retarded and advanced propagators. It means
that the characteristic frequencies entering ${\cal I}^{K}$ are
determined by the poles of the propagator rather than by the width of
the function $f$, i.e. the spatial scale may by far exceed  $L_T$
and any expression of the type ${\cal I}^K$ is thus {\em non-local}.

To overcome this difficulty, the standard parametrization of
the Keldysh function $D^K= D^R \con N - N\con D^A$ is 
usually introduced and
the kinetic equation for the distribution function $N$ is then
derived. All the non-locality in the problem is then contained
in the partial solution of the kinetic equation, compare with
\req{sol2}, whereas the kinetic equation itself is {\em local}.

In what follows, we adopt this program in a slightly modified form.
We introduce a new retarded propagator $\L^\rho(1,2)$,
compare with \req{Diffuson}
\be
\label{LR}
\begin{split}
&
\left[
i\hat{H}_\eh(i\partial_{t_1},-i\nnabla_1)
- \St_\tau   \right] \L^\rho =\delta_{12},
\\
& \hat{H}_\eh(\w,\q)= \v\cdot\q-\frac{\w}{1+\hat{F}}
\end{split}
\ee
and its advanced counterpart $\bar{\L}^\rho(1,2)= \L^\rho(2,1)$.
The multiplications in \req{LR} are to be understood in the operator
sense and the action of the operator $\hat{F}$ on a function 
$a(t,\nn,\r )$ is defined by \req{Fdef}.

For $\hat{F}=0$, $\L^\rho$ ($\bar{\L}^\rho$) reduces to the
usual diffuson $\L^g$ ($\bar{\L}^g$).
Physically, $\L^\rho$ describes the  spectrum of the propagating
electron-hole pair and the operator in the left hand side of \req{LR}
corresponds to the kinetic equation for the collective mode in the
Fermi liquid theory\cite{FL}. The operator $\hat{H}_\eh(\w,\q)$
can be interpreted as a ``Hamiltonian'' [see also App. \ref{alter}] 
of the interacting electron-hole pair.

In terms of $\L^\rho$ and $\L^g$, \reqs{Ks} acquire the form
\begin{subequations}
\label{KL}
\be \label{krfin}
\begin{split}
&{\nu}\hat{\K}^R =  \left( \pt \right)^{-1}\Big[  
\hat\L^g - \hat{\L}^\rho \Big] 
\\
&{\nu}\hat{\K}^A = - \Big[  
\hat{\bar{\L^g}} - \hat{\bar{\L}}^\rho \Big] \left( \pt \right)^{-1}
\end{split}
\ee
\be
\label{kkfin}
 \nu \hat{\K}^K = -{2i}\left( \pt \right)^{-1}
\Big[ \hat\L^g \hat{M} \hat{\bar{\L^g}}  
  - \hat{\L}^\rho \hat{M} \hat{\bar{\L}}^\rho \Big]\left( \pt \right)^{-1}. 
\ee
\end{subequations}

\begin{subequations}
\label{Ns}
Let us introduce two bosonic ``distribution functions''
(the density matrices to be more precise), $\hat{\cal N}^g$ and
$\hat{\cal N}^\rho$, that satisfy the following  equations:
\bea
&& (\hat\L^g)^{-1} \hat{\cal N}^g + \hat {\cal N}^g 
(\hat{\bar{\L^g}})^{-1} = 2 \hat{M},
\\
&& (\hat{\L}^\rho)^{-1}\hat{\cal N}^\rho + \hat{\cal N}^\rho 
(\hat{\bar{\L}}^\rho)^{-1}= 2 \hat{M} . 
\eea
The operator $\hat{M}$ is defined in \reqs{hatM} and \rref{mg} and,
in a more explicit notation, the action of the 
operators $\hat{{\cal N}}^{\rho,g}$ 
on any function $a(t,\nn,\r)$ is to be understood
as
\end{subequations}
\[
\left[\hat{\cal N}^{\rho,g}a\right](1) 
=\int \! d 2 \ {\cal N}^{\rho,g}\left(1,2\right)a(2),
\]
where the shorthand notation \rref{notation} is used. Note that from the above
equations it follows that the bosonic
functions ${\cal N}^{\rho,g}$ are symmetric
\be
\label{symN}
{\cal N}^{\rho,g}(1,2) = {\cal N}^{\rho,g}(2,1).
\ee

Equations \rref{Ns} enable us to rewrite \req{kkfin} as
\be\begin{split}
\label{kkn}
\nu \hat{\K}^K =& -i \left( \pt \right)^{-1} 
\Big[ \hat\L^g \hat{\cal N}^g +  \hat{\cal N}^g
\hat{\bar{\L^g}} \Big] 
\left( \pt \right)^{-1} \\
& +i \left( \pt \right)^{-1}
\Big[ \hat\L^\rho  \hat{\cal N}^\rho   + 
\hat{\cal N}^\rho \hat{\bar{\L}}^\rho \Big]
\left( \pt \right)^{-1}.
\end{split}\ee
This expression is local in the above discussed sense and
will be used in the construction of the conserved
energy current.
Obtaining the local expression, however, required the
introduction of two additional bosonic distribution functions:
${\cal N}^\rho$, describing the interacting electron-hole pairs
and the ghost field distribution ${\cal N}^g$, subtracting the contribution 
of the electron-hole pairs in the absence of interactions.

Closing this subsection, let us rewrite \reqs{Ns} in a form
resembling the kinetic equation of Sec.~\ref{Sec.2}.
We substitute \reqs{Diffuson} and \rref{LR} in \reqs{Ns}
and obtain
{\setlength\arraycolsep{0pt}  
\begin{subequations}
\label{Ns2}
\bea
\Big[\pt + \v\cdot\nnabla ;\ \hat{\cal N}^g\Big]
 & = &\, \St^{b}\left\{{\cal N}^g, g^K\right\}
\label{Ns2a}\\
\left[i\hat{H}_\eh(i\partial_{t_1},-i\nnabla_1)
 ;\ \hat{\cal N}^\rho \right]
& =& \, \St^{b}\left\{{\cal N}^\rho, g^K\right\}
\label{Ns2b},
\eea
where the collision integrals are
\be
\St^{b}\left\{{\cal N}^\alpha, g^K\right\}\equiv
2\Big\{\St_\tau; \hat{\cal N}^\alpha\Big\} +2
\hat{M};
\label{Ns2c}
\ee
for $\alpha=g,\rho$.
\end{subequations}
They depend on $g^K$ via $\hat{M}$ and we used the notation
\be 
\left\{\hat{A};\hat{B}\right\}\equiv \frac{1}{2}
(\hat{A}\hat{B}+\hat{B}\hat{A}); \quad
\left[\hat{A};\hat{B}\right] \equiv \hat{A}\hat{B}-\hat{B}\hat{A}.
\mylabel{compm}
\ee}

We perform the time and space Wigner transforms of \reqs{Ns2} to
introduce the bosonic distribution functions $N^{g,\rho}$
\be
\label{Nb}
\begin{split}
&\hat{\cal N}^{g,\rho}(1,2) =
\int\!\frac{d\w}{2\pi}\,
e^{-i\w\left(t_1-t_2\right)}
\int\!\frac{d^dq}{\left(2\pi\right)^d}\,
e^{i\q\cdot\left(\r_1-\r_2\right)}
\\
&\times{\w}
\big[2 N^{g,\rho}\left(\w,\q; 
\nn_1,\nn_2 ; t,\r \right) + \W_d\dnnn\big],
\end{split}
\ee
where $t=\frac{t_1+t_2}{2}$, $\r=\frac{\r_1+\r_2}{2}$.
The symmetry relation \rref{symN} translates into the condition
\be
\label{symN1}
\begin{split}
N^{g,\rho}&\left(\w,\q; \nn_1,\nn_2\right) 
\\ &= 
- \left[N^{g,\rho}\left(
-\w,-\q; \nn_2,\nn_1
\right)+\W_d\dnnn\right].
\end{split}
\ee
The physical meaning of this relation is 
boson statistics: at $\w >0$, $N^{g,\rho}$
corresponds to  the occupation numbers entering  the probability
of the absorption of the bosons whereas the $\w<0$ part describes 
the boson emission.

The fermionic distribution function $f$ is obtained in two steps:\
(a) we introduce the gauge invariant Green's function $g$ (see also
the next subsection) and (b) we perform the time 
Wigner transform: 
\begin{subequations}
\label{wigtr2}
\bea
g^K (t_1,t_2,\nn,\r) &\equiv& \exp 
\left( -i \int^{t_{1}}_{t_{2}}\!\!d\tilde{t} 
\, \varphi (\tilde{t},\r)\right) \, g(t_1,t_2,\nn,\r) \nonumber \\
\label{gt2} \\
g (t_1,t_2,\nn,\r) &=& 2 \int\!\frac{d\e}{2\pi}\,
 e^{-i\e \left(t_1-t_2\right)}
\big[1-2 f\left(\e,\nn ; t,\r \right) \big]. \nonumber \\
\label{gkWT}
\eea
\end{subequations}

Performing such Wigner transforms on  \reqs{Ns} and \rref{mg}, we find
{\setlength\arraycolsep{0pt}
\begin{subequations}
\label{kineqN}
\bea
\label{kineq0}
&& \w \biggl[\pt {\hat N}^g + 
\left\{\v; \nnabla  {\hat N}^g\right\} +
i\left[\v\cdot\q ; {\hat N}^g\right]\biggr]
= \St^b\left\{N^g, f\right\}
\nonumber \\ && \\
\label{kineqrho}
&& \w \biggl[
\left\{ \frac{1}{1+\hat{F}} ; \pt {\hat N}^\rho \right\}
 + \left\{ \hat{\s}(\w,\q) ; \nnabla  {\hat N}^\rho \right\} \nonumber \\
&& \quad \quad \quad + i \left[\hat{H}_\eh(\w,\q) , {\hat N}^\rho \right]
\biggr]= \St^b\left\{N^\rho, f\right\},
\eea
\end{subequations}
where the collective mode velocity operator is
\be
\hat{\s}(\w,\q) = \frac{\partial\hat{H}_\eh(\w,\q) }{\partial \q}=
\v +\w  \frac{\partial }{\partial \q}
\left(\frac{\hat{F}}{1+\hat{F}}\right).
\label{sv}
\ee
In the left hand side of equation \rref{kineqrho} we limited ourselves to the 
leading Poisson brackets [the equation becomes exact for the short range
interaction, since $\partial_{\q} \hat{F}=0$, and in the unitary limit
$\hat{F} \to \infty$].
However no Poisson brackets arise in
the right hand sides of \reqs{kineqN} -- this is a consequence of the 
locality of the kinetic equations.}

The right hand sides of \reqs{kineqN} describe the decay of the
electron-hole pair into the electron and the hole moving into
different directions. 
To write down the expression for
this collision term, it is convenient to introduce the following object:
\bea
\label{Delta}
&&\Upsilon^{g,\rho}_{ij;kl}
\left(\e,\w, \q ; t,\r \right)
\equiv N^{g,\rho}\left(\w,\q ;\nn_i,\nn_j; t,\r\right)
\nonumber \\
&& \quad \times
\Big\{f(\e,\nn_k ;t,\r)-f(\e-\w,\nn_k ;t,\r)
\Big\}
\\ && \quad+
\W_d\delta(\widehat{\nn_i\nn_j})\Big\{f(\e,\nn_l ;t,\r)
\left[1-f(\e-\w,\nn_k ;t,\r )\right]
\Big\} \nonumber.
\eea
It is easy to see that $\Upsilon^{g,\rho}=0$
in the thermal equilibrium \req{thermeq}.

In terms of  this object and the vertex \rref{gamma1} we have 
\be
\begin{split}
\St^b&\left\{N^{g,\rho}, f\right\}
\left(\w,\q; \nn_1,\nn_2 ;t,\r \right)=-
\int\!d\e\int\!\frac{d\nn_3d\nn_4}{\W_d^2} 
\\ &\times 
\Big\{\gamma_{14}^{3}
\Upsilon^{g,\rho}_{32;41}\left(\e,\w\right)
+
\gamma_{34}^{2}
\Upsilon^{g,\rho}_{13;14}\left( \e,\w\right)
\Big\},
\end{split}
\label{st0clq}
\ee
where we  suppressed the spectator  arguments
$t,\ \r$ and $\q$ in the right-hand side of the equation.
Deriving \req{st0clq} we
used \reqs{gamma2}, \rref{mgsymm}  and the property 
\[\int\!d\e \, \big[ f(\e)-f(\e-\w)
\big] =-\w.
\]

To understand  the physical meaning of the processes described
by the collision integral \rref{st0clq}, we use the explicit
form of the vertex $\gamma$ [\req{gamma1}] for the isotropic
impurity scattering $\tau(\theta_{12})=\tau$.
Then, the collision integrals can be decomposed into the sum of two
contributions
\[
\St^{b}\big\{N^{g,\rho}, f\big\}=\St^{b}_{cl}\big\{ N^{g,\rho}, f \big\}
+\St^{b}_{q}\big\{N^{g,\rho}, f\big\}
\]
The first term in the right-hand side can be obtained
from a simple counting of the probabilities of the processes depicted
on Fig.~\ref{figscat}(a-b)
\[
\begin{split}
&\St^b_{cl}\left\{N^{g,\rho}, f\right\}
\left(\w ; \nn_1,\nn_2\right)
\\&\quad =\frac{1}{\tau}
\int\!d\e\int\!\frac{d\nn_3}{\W_d} 
\Big\{
\Upsilon^{g,\rho}_{12;32}\left( \e,\w\right)
+\Upsilon^{g,\rho}_{12;13}\left( \e,\w, \q\right)
\Big\} .
\end{split}
\]
The second term in the right-hand side  originates
from the interference of two scattering processes, see
Fig.~\ref{figscat}(c). That is why it produces contributions
to $\hat{N}$ which are not diagonal in the momentum directions:
\[
\begin{split}
&\St^b_{q}\left\{N^{g,\rho}, f\right\}
\left(\w ;\nn_1,\nn_2\right)
\\ &\quad =-\frac{1}{\tau}
\int\!d\e \int\!\frac{d\nn_3}{\W_d}
\Big\{
\Upsilon^{g,\rho}_{13;12}\left( \e,\w\right)
+\Upsilon^{g,\rho}_{32;21}\left(\e,\w\right)
\Big\} .
\end{split}
\]

\subsection{The collision integral for electrons}
\label{elcolint}

With the bosonic propagators $\K$ 
at hand, we can proceed with the calculation and 
include the second order in the fluctuating
fields $K_{\pm}$ contributions to the 
collision term of the 
Eilenberger equation \rref{eil2}. With the fluctuating fields 
$K_{\pm}$ given by \reqs{kmprime} and \rref{kp}, Eilenberger equation becomes: 
\be\label{eil30}\begin{split}
& \Big[ \tilde{\partial}_{t} + \v \cdot \tilde{\nnabla}\Big] \hat{g} \\ \
& = \left[ \hat{g} \concomma \frac{1}{2} \St_\tau^\phi \hat{g} 
- i \left( \St_\tau K_+ - 2 \tilde{K}_- \right) \hat{\openone}_K
 +i \St_\tau K_- \hat{\sigma}^x_K \right],
\end{split}\ee
where we use the notation \rref{covariant} for the derivatives, $\St_\tau^\phi$
was defined after \req{eil2} and $\tilde{K}_-$ in \req{Ktildem}, and 
$\hat{\sigma}^x_K$ is the Pauli matrix. 

\begin{figure}[th]
\includegraphics[width=0.43\textwidth]{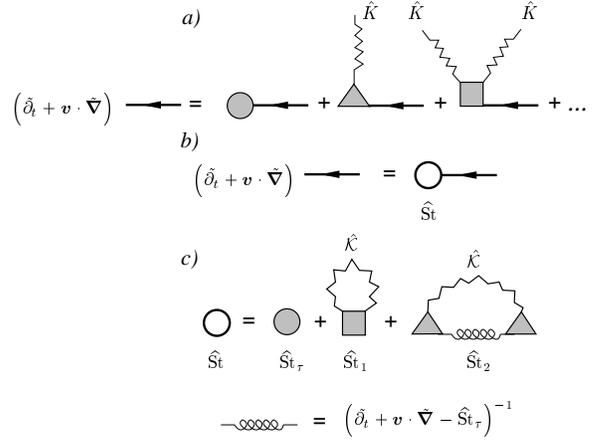}
\caption{Schematic representation of the averaging
  over the fluctuation fields $\hat{K}$. a) Expansion of the
  Eilenberger equation \rref{eil30} 
before the averaging; b) The equation
  for the Green function averaged over $\hat{K}$, see \req{eil3}; 
c) The
  contribtutions to the total collision integral in the first loop
  approximation. }
\label{fig2}
\end{figure}

We expand the right hand side of
\req{eil30} up to the second order in $\hat{K}$, see Fig.~\ref{fig2}a;
then we average it to obtain  Fig.~\ref{fig2}b.
The  resulting 
second order contributions can have two different origins:
(1) they may arise from the expansion of the exponentials 
truncated at the second order, term $\St_1$ on Fig.~\ref{fig2}c,
or (2) they 
are obtained as products between the linear correction $\delta g^K$ 
of \reqs{dgmeq} and the first order expansion of the exponentials, 
term $\St_2$  on Fig.~\ref{fig2}c. 

The Eilenberger equation for the averaged Green function 
takes the form 
\be \label{eil3}
\left[ {\tilde\partial_{t}} + \v \cdot \tilde{\nnabla}  
\right] \hat{g} = \hat{\St} \{ \hat{g},N^\rho,N^g \}
\ee
where $\hat{\St}$ contains both zero and second order contributions.
We find [see Appendix~\ref{app2} for details on the 
cancellation of the second order corrections in the R,A,Z sectors]:
\be
\hat{\St}\left\{g^R_0, g^A_0,
g^K\right\} = \left( \begin{matrix}
0 & \St\left\{ g^K, N^\rho, N^g \right\} \cr
0 & 0 
\end{matrix} \right),
\ee
where $g^R_0=-g^A_0=\delta(t_1-t_2)$. It means that the
matrix Green's function of the form \rref{gzero} is still the
solution of the Eilenberger equation -- the main gain of the gauge 
transformation \rref{gaugetr} --  provided that
 the kinetic equation for the Keldysh component is satisfied:
accordingly, we concentrate on this component only.

Performing the gauge transformation \rref{gt2} on the Keldysh component of
the Eilenberger equation, we arrive at the explicitly
gauge-invariant form of the kinetic equation:
\be \label{gikineq}
\left[ \partial_{t_{1}} + \partial_{t_{2}} + \vgrad + i\v\!\cdot \!\! 
\int^{t_{1}}_{t_{2}} \!\!d\tilde{t} \, e\E (\tilde{t},\r) \right] g = \St 
\{ g,N^\rho,N^g \}
\ee

In the collision integral $\St\{ g,N^\rho,N^g \}$, we indicate with 
$\St_1\{ g,N^\rho,N^g \}$ the contributions of the type (1) and with 
$\St_2 \{ g,N^\rho,N^g \}$ those of type (2) [this separation has 
no particular physical meaning -- it is just a matter of
practicality in the calculations; we will return on the physical aspects when
analyzing the conservation laws in the next subsection]:
\be
\St\{ g,N^\rho\!,N^g \} = \St_{\tau} g - 4\St_1 \{ g,N^\rho\!,N^g \} 
-4\St_2 \{ g,N^\rho\!,N^g\}.
\label{stsum}
\ee
where $\St_{\tau}$ is defined in \req{Sttau}.
The numerical factors in front of the last two terms are introduced to 
facilitate the transformation to the canonical form of the kinetic
equation in further sections.

The expression for $\St_1$ written in terms of the $\K$ propagators 
\rref{KKs} and the $\gamma$-vertex \rref{gamma1} is
{\setlength\arraycolsep{0pt}
\bea \label{i1}
&&\left[\St_1\right](t_1,t_2;\nn_1) 
= -\frac{i}{16} 
\int \frac{d\nn_2d\nn_3}{\W_d^2}
\, \gamma_{12}^3
\\
&&\times \Big\{ g(t_1,t_2,{\nn}_2)
\left[
\tilde{\K}^K_{32}(t_1,t_2) 
- \tilde{\K}^K_{31}(t_1,t_2) 
\right] 
\nonumber \\
&&+\!\int\!\!dt_3  g(t_1,t_3,\nn_1) 
g(t_3,t_2,{\nn}_2)
\left[
{\K}_{32}^A(t_3,t_2) 
- {\K}_{31}^A(t_3,t_2) 
\right]
\nonumber \\ 
&& - g(t_1,t_3,{\nn}_2) g(t_3,t_2,\nn_1) 
\left[
{\K}_{23}^R(t_1,t_3) 
- {\K}_{13}^R(t_1,t_3) 
\right]\Big\}.
\nonumber
\eea
We introduced the short hand notations
\be\label{knot}
{\cal K}_{ij}(t_1,t_2)\equiv {\cal K}(t_1,\nn_i; t_2, \nn_j),
\ee
and
\[
\tilde{\K}^K(t_1,t_2) \equiv
2 {\K}^K(t_1,t_2)
- {\K}^K(t_1,t_1)
- {\K}^K(t_2,t_2). 
\]
We omitted the variable $\r$, which always appears in the distribution
function as $g(t_1,t_2,\nn,\r)$ and in the propagators as 
$\K(t_1, \nn_1, \r;  t_2, \nn_2,\r)$.
The dependence on the electron distribution function $g$ is explicit, 
whereas the dependence on the bosonic distribution functions is hidden 
into the propagators -- see \req{kkn}.
For reasons which will become clear in the next subsection,
we split \req{i1} into two parts
\begin{subequations}
\label{i1split}
\bea
&& \St_1 \, = \, \St_1^{\mathrm{in}} + \St_1^{\mathrm{el}}; 
\label{i1split1}\\
&&\left[\St_1^{\mathrm{in}}\right](t_1,t_2;\nn_1) 
= -\frac{i}{32} 
\int \frac{d\nn_2d\nn_3}{\W_d^2}
\, \gamma_{12}^3  \label{i1split2} \\
&&\times \Big\{ 2 g(t_1,t_2,{\nn}_2)
\left[
\tilde{\K}^K_{32}(t_1,t_2) 
- \tilde{\K}^K_{31}(t_1,t_2) 
\right] 
\nonumber \\
&&+\!\!\int\!\!dt_3  
\left[g(t_1,t_3,\nn_1) 
g(t_3,t_2,{\nn}_2) \!
+ g(t_1,t_3,{\nn}_2) g(t_3,t_2,\nn_1)
\right]
\nonumber\\
&&
\times \left[
{\K}_{32}^A(t_3,t_2) - {\K}_{23}^R(t_1,t_3)
- {\K}_{31}^A(t_3,t_2) + {\K}_{13}^R(t_1,t_3) 
\right]\Big\}.
\nonumber\\
&&\left[\St_1^{\mathrm{el}}\right](t_1,t_2;\nn_1) 
= -\frac{i}{32} \!\int\!\!dt_3 
\int \frac{d\nn_2d\nn_3}{\W_d^2}
\, \gamma_{12}^3  \label{i1split3} \\
&&\times \Big\{ 
\left[g(t_1,t_3,\nn_1) 
g(t_3,t_2,{\nn}_2)
- g(t_1,t_3,{\nn}_2) g(t_3,t_2,\nn_1)
\right]
\nonumber\\
&&
\times \left[
{\K}_{32}^A(t_3,t_2) + {\K}_{23}^R(t_1,t_3)
- {\K}_{31}^A(t_3,t_2) - {\K}_{13}^R(t_1,t_3) 
\right]\Big\}.
\nonumber
\eea
\end{subequations} }

As for $\St_2$, it is convenient to separate it in two parts,
depending on which field, $K_+$ or $K_-$, we retain in the expansion:
\begin{subequations}
\label{st2}
\be
\St_2 = \lda \St_+ \rda + \lda \St_- \rda
\ee
where $\lda \dots \rda$ stands for the averaging over the fluctuating
fields $K_{\pm}$ with the propagators \rref{KKs}
and
{\setlength\arraycolsep{0pt}
\bea
&&\St_{-}(t_1,t_2; \nn,\r)= 
\frac{i}{4}
\int dt_3\int \frac{d\nn_1}{\W_d}K_-(t_3,\nn_1,\r)
\nonumber\\
&&\quad \times
\Big[\delta Q(t_1,t_2, \nn; t_3, \nn_1; \r)
\label{stm}\\
&&\quad -\frac{i\pi}{2}  
\delta g_+\left(t_1,t_2;\nn,\r \right)
\int\limits_{t_2}^{t_1}dt_5 
Q(t_5,t_5, \nn; t_3, \nn_1; \r)
\Big].
\nonumber\\ 
&&\St_{+}(t_1,t_2\nn)=\frac{i}{4} \int\frac{d\nn_2d\nn_3}{\W_d^2}
\vertex{}{}{2}{3}\label{stp}\\
&&\quad \quad \times
\Big[K_+(\nn_3); \big( {\delta g}(\nn_2)-{\delta g}(\nn) \big) \Big](t_1,t_2).
\nonumber
\eea
The commutator was defined in \req{com1} and
the kernel $\delta Q$ is the first variation
of the operator \rref{Q} with respect to the Keldysh component
of the electron Green's function
\bea
&&
\delta Q(t_1,t_2, \nn; t_3,\nn_1;\r) =\frac{1}{2}
\int \frac{d\nn_2}{\W_d}\vertex{}{}{2}{1}
\nonumber
\\
&& \quad \times
\bigg\{ \Big[\delta g_+(t_1,t_3,\nn,\r) g(t_3,t_2,{\nn}_2,\r)
\label{deltaQ}
\\&&
\quad\quad
+g(t_1,t_3,\nn,\r) \delta g_+(t_3,t_2,{\nn}_2,\r)
\Big] + \left(\nn_2 \leftrightarrow \nn \right)
\bigg\}.
\nonumber
\eea
}
\end{subequations}

Finally, the functions $\delta g_{\pm}$ and $\delta g=\delta
g_{+}+\delta g_-$ 
are obtained by solving \reqs{dgmeq} and \rref{dgpeq} 
[after the transformation \rref{gt2}];
with the help of \req{MQ}:
{\setlength\arraycolsep{0pt}
\bea
&&\delta g_+(t_1,t_2; \nn, \r) =
-i \int d3 \int\frac{d\nn_2d\nn_4}{\W_d^2}
\, \gamma_{32}^{4} \label{dgp} \\
&&\L^g\left(t_1,\nn,\r ;3\right)
\Big[
K_+(t_3,\nn_4,\r_3)-K_+(t_3-t_{12},\nn_4,\r_3)
\Big]
\nonumber
\\
&&\times 
\Big[
{g^K}(t_3,t_3-t_{12}; \nn_2,\r_3)
- {g^K}(t_3,t_3-t_{12}; \nn_3,\r_3)
\Big]
\nonumber
\eea
\bea
&&\delta g_-(t_1,t_2; \nn, \r)  
=-i\int\! d 3\int dt_4
\int\frac{d\nn_4}{\W_d}
\label{dgm}
\\
&&\L^g\!\left(t_1,\nn,\r ;3\right)K_-(t_4,\nn_4,\r_3)
\Big[
Q(t_3,t_3-t_{12}, \nn_3; t_4, \nn_4; \r_3 ).
\nonumber\\
&&
-\frac{i\pi}{2}  
g^{K}\left(t_3,t_3-t_{12};\nn_3,\r_3 \right)\!\!\!
\int\limits_{t_3-t_{12}}^{t_3}\!\!\! \! dt_5 
Q(t_5,t_5, \nn_3; t_4, \nn_4; \r_3)
\Big],
\nonumber
\eea
where $t_{12}=t_1-t_2$ and the notation \rref{notation} is used
\footnote{These solution are exact only in the absence of the electric field, 
since in its gauge invariant form the operator acting on $\delta g_{\pm}$ is 
the same that appears on the left hand side of \req{gikineq}. We could include 
perturbatively field dependent corrections to our expressions, which would 
be of the first order in $\E$ for $\delta g_-$ and of the second order
for $\delta g_+$ [since $\delta g_+$ vanishes in equilibrium]. 
However, as noted before, the first property 
in \req{dgtt} implies that these corrections cannot contribute to
the physical quantities we are interested in and therefore we do not include
them in our calculations.}. 
}

For future use, we note following properties
[see also the discussion following \req{dgpmeq}]:
\be
\label{dgtt}
\delta g_-(t,t)= \delta g_+(t,t)=0; \
\int\!d\nn\, \delta Q(t_1,t_2, \nn; t_3,\nn_1;\r)=0,  
\ee
as one can see from \reqs{dgp}-\rref{dgm} and from the definitions 
\rref{deltaQ} and \rref{gamma1}.

The canonical form of the kinetic equation is obtained 
 by performing the time Wigner transform 
\rref{gkWT} of both sides of
\req{gikineq}. It is clear from the structure 
of the collision integrals that this procedure will lead to the 
appearance of Poisson
brackets in the right hand side of the kinetic equation.
We choose, however, another route: we will prove
the existence of the conservation laws before the Wigner
transform. It will enable us to argue that these
Poisson brackets (in our formulation of
the kinetic equation) give only small contributions, 
which can be neglected within the accuracy of the kinetic equation. 

\subsection{Conservation laws}
\label{conlaw}

The derivation of the conservation laws is based on the following
properties of the collision integrals of the previous subsection
\begin{subequations}
\label{stel}
\bea
&& \an{\St_\tau g} = 0;\\
&& \an{\St_-} = 0;\\
&& \an{\St_1^{\mathrm{el}}} = 0.
\eea
\end{subequations}
The physical meaning of conditions \rref{stel} is that the
corresponding terms in the collision integral conserve the number
of particles within the energy shell. Equations \rref{stel} follow
immediately from definitions \rref{Sttau}, \rref{i1split2}, 
\rref{stm}, \rref{deltaQ} and \rref{dgtt}.

The two remaining contributions to the collision integral have the properties
\begin{subequations}
\label{stin1}
\bea
&&\lim_{t_1\to t_2} \St_+(t_1,t_2)=0;
\label{stin1a}
\\
&& \lim_{t_1\to t_2}\St_1^{\mathrm{in}}(t_1,t_2)=0,
\label{stin1b}
\eea
\end{subequations}
and
{\setlength\arraycolsep{0pt}
\begin{subequations}
\label{stin2}
\bea
\lim_{t_1\to t_2} && \left(\partial_{t_1}-\partial_{t_2}\right) 
\, \St_+(t_1,t_2)=0;
\label{stin2a}
\\
\label{stin2b}
\lim_{t_1\to t_2} && \left(\partial_{t_1}-\partial_{t_2}\right)
\, \St_1^{\mathrm{in}} (t_1,t_2) 
\label{stin2c}
\\
 =&& {\displaystyle \frac{1}{4\pi}
\int\!\frac{d\nn_2 d\nn_3}{\W_d^2} } \, \gamma_{12}^{3} 
\Big\{ \left[\pt\, \K^K \pt\right]_{31} (t_1,t_1) \nonumber \\
&& -  {\displaystyle \frac{i\pi}{4}\!\int\!\!dt_3 }
\left[\pt \K^R \right]_{13} (t_1,t_3)
\big[ g(t_1,t_3,\nn_1) g(t_3,t_1,\nn_2) \nonumber \\ 
&& + g(t_1,t_3,\nn_2) g(t_3,t_1,\nn_1) \big] - (\nn_1 \to \nn_2 )
\Big\},
\nonumber
\eea
\end{subequations}
where we use the notation \rref{knot} and the vertex is defined in 
\req{gamma2}. 
Equations \rref{stin1a} and \rref{stin2a} immediately follow from
the definition \rref{stp} and the condition \rref{dgtt}.
Derivations of  \reqs{stin1b} and \rref{stin2b} are presented in 
Appendix~\ref{deriv1}.

Expressions \rref{stin1} mean that while not conserving the number of
particles for a given energy shell, the terms $\St_+$ and 
$\St_1^{\mathrm{in}}$ conserve the total number of particles
for a given direction (small angle inelastic scattering).
Equation \rref{stin2a} means that the inelastic $\St_+$ term 
conserves not only the number of particles but also the energy for a
given direction. Equation \rref{stin2b} means 
$\St_1^{\mathrm{in}}$
term does not conserve the energy for a given direction, thus 
describing the energy exchange between the quasiparticles and
the electron-hole pairs discussed in Sec.~\ref{ehpairs}.

The possibility to find the conserved energy current is based on
a certain relation between \req{stin2b} and the collision
integral for the electron-hole pairs -- we now turn to the discussion of
this relation.

We substitute \reqs{krfin} and \rref{kkn} into \req{stin2c} and average the
result over $\nn$; then using the analytical property
\be\label{anprop}
\K^R(t,t) = \int\!\frac{d\w}{2\pi} \K^R(\w) =0
\ee
together with $\L^\rho(1,2)=\bar{\L}^\rho(2,1)$,
$\L^g(1,2)=\bar \L^g(2,1)$, $
{\cal N}^{\rho,g}(1,2)=  {\cal N}^{\rho,g}(2,1)$ and \req{mg}, we find:
{\setlength\arraycolsep{0pt}
\bea 
\lim_{t_1\to t_2} && \left(\partial_{t_1}-\partial_{t_2}\right)
\an{\St^{\mathrm{in}}_1} 
\label{shit1}\\
 = &&-
 \frac{i}{\pi\nu}  \Tr_{\nn} \, 
\left[ \hat\L^g 
\left(\hat{\cal N}^g \St_\tau +  \hat{M} \right)
\right] \nonumber \\
&& + \frac{i}{\pi\nu} \Tr_{\nn} \,\left[\hat{\L}^\rho
\left( \hat{\cal N}^\rho  \St_\tau  
 +  \hat{M} \, 
 \right)
\right], \nonumber
\eea
where $\St_\tau$
is defined in \req{Sttau} and $\Tr_{\nn}$ acts as
\be
\label{trace}
\left[\Tr_{\nn}\hat{A}\right](t,\r)\equiv
\int \frac{d\nn}{\W_d}A\left(t,\nn,\r; t,\nn,\r\right).
\ee 
}

The corresponding traces of the collision integrals for the
electron-hole pairs -- \req{Ns2c}\footnote{The operator $\hat{M}$
is gauge-invariant and it has the same form in terms of $g$ or $g^K$.}
-- are:
{\setlength\arraycolsep{0pt}
\begin{subequations}
\label{shit2}
\bea
\Tr_{\nn}&& \left[ \hat\L^g \, \St^g \left\{ {\cal N}^g ,g \right\} 
\right] = 
\Tr_{\nn} \bigg[
\Big[ \hat\L^g , \St_\tau \Big] \hat{\cal N}^g 
\nonumber \\&& + 2
\hat\L^g \bigg(\hat{\cal N}^g \St_\tau + 
 \hat{M} \bigg) 
 \bigg] \\
\Tr_{\nn}&& \left[ \hat{\L}^\rho \St^\rho \left\{
{\cal N}^\rho, g \right\} \right] =
\Tr_{\nn} \bigg[ 
\Big[ \hat{\L}^\rho , \St_\tau \Big] \hat{\cal N}^\rho
\nonumber \\&& + 2
\hat{\L}^\rho \bigg(\hat{\cal N}^\rho \St_\tau + 
\hat{M} \bigg)
 \bigg]. 
\eea
\end{subequations}
Comparing \reqs{shit1} and \rref{shit2} and using
once again \reqs{Diffuson} and \rref{LR}, we arrive
to the desired relation between the collision integrals:
\bea \label{rel}
&&
{2i\pi \nu} \lim_{t_1\to t_2} 
\big(\partial_{t_1}  -\partial_{t_2}\big)
\an{\St \left\{ g,{\cal N}^\rho,{\cal N}^g \right\} }  \\
&& + \Tr_{\nn} \left[ \hat{\L}^\rho \St^\rho \left\{
{\cal N}^\rho, g \right\} \right]
-\Tr_{\nn} \Big[ \hat\L^g \, \St^g  \left\{ {\cal N}^g ,g \right\} 
\Big]
\nonumber \\
&& 
= \Tr_{\nn} \bigg[
\Big[ \vgrad , \hat\L^g \Big] \hat{\cal N}^g
- i\bigg[
\hat{H}_\eh(i\partial_{t},-i\nnabla)
, \hat{\L}^\rho \bigg] \hat{\cal N}^\rho 
\bigg]. \nonumber
\eea}

The left hand side of \req{rel} is the quantum counterpart of the
relation (\ref{energy}) 
and \reqs{stel}-\rref{stin1} are related to \req{number}; 
we will derive the expressions for
the electric and energy currents in the spirit of our discussion in 
Sec.~\ref{sec:cons}.

We begin with the conservation of electric charge.
According to \reqs{GFs} and \rref{intg} the charge density is given by
\be
\rho(t,\r)= -\frac{e \nu\pi}{2}
\lim_{t_1\to t_2\to t}\an{g(t_1,t_2, \nn,\r)}.
\mylabel{rho1}
\ee
Taking the limit $t_1\to t_2\to t$ in both sides of \req{gikineq}
and using \reqs{stsum}, \rref{stel} and \rref{stin1}, we obtain
 the continuity equation 
\be
\pt \rho + \nnabla \cdot \j =0,
\label{rhocont1}
\ee
where
\be
\j(t,\r)= -\frac{e v_F\nu\pi}{2}
\lim_{t_1\to t_2\to t}\an{\nn g(t_1,t_2, \nn,\r)},
\mylabel{j1}
\ee 
compare with \reqs{rho} -- \rref{j}.

Having found the usual equation for the electric currents, we turn to
the energy conservation. Acting with the operator
$(\partial_{t_1}-\partial_{t_2})$ on both sides of \req{gikineq}
and introducing the quantities
{\setlength\arraycolsep{1pt}
\bea
\label{jel}
&&u_e(t,\r) = - \frac{i\pi\nu}{4} \lim_{t_{1} \to {t_2} \to t} 
\big( \partial_{t_{1}} - \partial_{t_{2}} \big)
\an{ g(t_1,t_2,\nn,\r)}
\nonumber\\
&& \j_e^\e(t,\r) = - \frac{i\pi\nu v_F}{4} \lim_{t_{1} \to {t_2} \to t} 
\big( \partial_{t_{1}} - \partial_{t_{2}} \big)
\an{ \nn g(t_1,t_2,\nn,\r)},
\nonumber\\
\eea
we find
\bea
\pt u_e &+& \nnabla \cdot \j_e^\e =\j\cdot \E
\label{jEcon1} \\
&+&i\nu \pi \lim_{t_{1} \to {t_2} \to t}
\big(\partial_{t_1}  -\partial_{t_2}\big)
\an{\St \left\{ g,{\cal N}^\rho,{\cal N}^g \right\} }.
\nonumber
\eea}
The expression in the left hand side of \req{jEcon1} has the form
of the continuity equation for the energy current of the electrons
and the first term in the right hand side is the Joule heat acting
as an energy source.
The last term in the right-hand side indicates that the electron
system by itself is open, due to the energy exchange with the 
electron-hole pairs. As we discussed in Sec.~\ref{sec:cons} this means that
in the definition of the conserved energy and energy current 
densities the contribution
of these degrees of freedom must be taken into account. To accomplish
this task, we premultiply \reqs{Ns2a} and \rref{Ns2b} by $\hat\L^g$ and
$\hat{\L}^\rho$ respectively. Using \reqs{Diffuson}, \rref{LR} 
and \rref{Ns2c} and taking the trace $\Tr_{\nn}$ 
[see \req{trace}] of both sides, we obtain:
\begin{subequations}
\label{bos1}
\bea
\partial_t u_g + \nnabla \cdot \j^{\e}_g 
&-& \frac{1}{2}\Tr_{\nn}
\Big[ \vgrad , \hat\L^g \Big] \hat{\cal N}^g
\nonumber \\ & = &
\frac{1}{2}\Tr_{\nn}  \hat\L^g \; \St^g \, ,
\label{bos1a}
\\ 
\partial_t u_\rho + \nnabla \cdot \j^{\e}_\rho 
&-& \frac{i}{2}\Tr_{\nn} \bigg[
\hat{H}_\eh(i\partial_{t},-i\nnabla), \hat{\L}^\rho \bigg] 
\hat{\cal N}^\rho \nonumber \\ &=& 
\frac{1}{2}\Tr_{\nn} \hat{\L}^\rho \St^\rho \, ,
\label{bos1b}
\eea 
where the energy densities $u_{\rho,g}$ and currents $\j_{\rho,g}^\e$
are defined
 as
{\setlength\arraycolsep{0pt}
\bea
& u_g(t,\r) = \frac{1}{2} \Tr_{\nn} \hat\L^g \ \hat{\cal
  N}^g; \;
u_\rho(t,\r) = \frac{1}{2} \Tr_{\nn} \Big\{ \frac{1}{1+\hat{F}};
\hat{\L}^\rho  \hat{\cal N}^\rho\Big\}\nonumber\\
\label{bos1c}\\
& \j^{\e}_g =  \frac{1}{2} \Tr_{\nn} \v \hat\L^g \; \hat{\cal
  N}^g;\quad
\j^{\e}_\rho  =  \frac{1}{2} \Tr_{\nn} 
\Big\{\hat{\s}\,;\, \hat{\L}^\rho \hat{\cal
  N}^\rho\Big\},
\nonumber\\
\label{bos1d}
\eea
The velocity operator $\hat{\s}$ is defined in \req{sv} and the notation
\rref{compm} is used.
} \end{subequations}

We now add \req{bos1a} to \req{jEcon1} and subtract \req{bos1b}.
According to \req{rel} all the collision terms as well as
the commutators cancel and we obtain the energy balance equation
[compare with \req{encon}]:
\begin{subequations}
\label{bal}
{\setlength\arraycolsep{0pt}
\bea
&&\pt u_{tot} +\nnabla\cdot\j^\e_{tot}=\j\cdot\E ;
\label{bal1}\\
&&u_{tot}(t,\r)=u_e(t,\r) + u_\rho(t,\r) - u_g(t,\r);
\label{bal2}\\
&&\j^\e_{tot}(t,\r)=\j^\e_e(t,\r) + \j^\e_\rho(t,\r) - \j^\e_g(t,\r).
\label{bal3}
\eea 
}
\end{subequations}

Equations \rref{bal}, \rref{jel}, \rref{bos1c} and \rref{bos1d}
constitute the main result of this subsection. They
define the conserved currents in terms of quantities to
be found from the kinetic equations. 
It is important to emphasize
that the  conservation laws thus found are exact (at one loop) 
in the sense that 
no approximation has been made beyond the usual Fermi-liquid theory: in
detail, no gradient or harmonic expansion has been made and no time 
or space (except those suppressed by a factor of $q/p_F$) Poisson
brackets have been neglected yet.

Within the same accuracy with which the kinetic equations \rref{kineqN} were
derived, performing the Wigner transforms \rref{Nb}-\rref{wigtr2} of
\req{j1}, \rref{jel} and \rref{bos1d} we find 
\be\label{elcurr}
 \j^{\phantom \e} = e v_F \nu \int\!d\e \, \an{\nn f(\e,\nn ; t,\r)}
\ee
for the electric current density and
\begin{subequations}
\label{encurrfin}
\bea
\label{encurrel}
\j^\e_e &=& v_F \nu \int\!d\e \, \e \an{\nn f(\e,\nn ; t,\r)}
\\
\j^\e_\rho &=& \!\int\!\frac{d\w}{2\pi}\,\w \!
\int\!\!\frac{d^d q}{(2\pi)^d}
\an{ \Big\{ \hat{\s}(\w,\q) ;
\hat{\L}^\rho (\w,\q) \hat{N}^\rho (\w,\q ;t,\r ) \Big\} }
\nonumber \\ \label{encurrb} \\
\j^\e_g &=& \!\int\!\frac{d\w}{2\pi} \, \w \int\!\frac{d^d q}{(2\pi)^d}
\an{\v \hat\L^g (\w,\q) \hat{N}^g (\w,\q ; t,\r ) } 
\label{encurrg}
\eea
\end{subequations}
for the energy current densities, in agreement with 
\reqs{consrho}-\rref{consen}.

What remains to be done is to reduce the found expressions to the usual
form of the quantum Boltzmann equation. This is the subject of the
next subsection.

\subsection{The quantum kinetic equation}

After the Wigner transforms \rref{Nb}-\rref{gkWT}, \req{gikineq} becomes
\be\label{fke}
\left[ \pt + \vgrad + e\v\!\cdot\!\E \frac{\partial}{\partial \e} \right]
f(\e,\nn ; t,\r) = \St \{ f, N^\rho, N^g \}.
\ee
The collision integral is the sum of the inelastic and elastic parts
\be
\St \{f, N^\rho, N^g \} = \St_{\mathrm{in}} \{ f, N^\rho, N^g \} + 
\St_{\mathrm{el}} \{ f \}.
\ee

The elastic part is obtained by adding the transform of the 
``bare'' collision integral [the first term on the right
hand side of \req{stsum}] to the transforms of 
$\St_1^{\mathrm{el}}$, \req{i1split3}, and $\St_-$, \req{stm}.
The inelastic part is given by the transform of $\St_1^{\mathrm{in}}$, 
\req{i1split2}, plus the transform of $\St_+$, \req{stp}.
However for the sake of compactness we will not make such a distinction
between elastic and inelastic contributions and, using a notation
resembling that of section \ref{elcolint}, we write the collision integral in 
the following form:
\be\label{colint}
\begin{split}
\St \{f, N^\rho, N^g \} = & \, \St_\tau f + \St_1\{f, N^\rho, N^g \} 
\\ & + \St_- \{f \} + \St_+ \{f, N^\rho, N^g \},
\end{split}\ee
where the first term on the right hand side is the transform of the 
``bare'' collision integral and the other terms are given below.

Keeping the elastic and inelastic
parts of the collision integral \rref{i1} in a single formula,
the corresponding contribution is obtained by first substituting \reqs{kkn}
and then performing the Wigner transforms \rref{Nb}-\rref{wigtr2} 
[and using their property \rref{anprop}].
We decompose the result into distinct contributions due
to the two bosonic degrees of freedom:
\be
\label{dc1}
\St_{1}(\e,\nn ; t,\r) = \St^{e-\rho}_{1}(\e,\nn ; t,\r) - 
\St^{e-g}_{1}(\e,\nn ; t,\r).
\ee
As usual the collisions with the ``ghost'' particles enter with the
opposite sign.
In terms of the combination \rref{Delta} of distribution
functions that we denoted by $\Upsilon$ and the vertex
\rref{gamma2}, these 
contributions are (we suppress the spectator arguments $t$ and $\r$ in
both sides of the equations):
\begin{subequations}
\bea
&& \St_1^{\da{$e$}{$\rho$}} (\e,\nn_1) 
=  \frac{1}{\nu} \int\!\frac{d\w}{2\pi} \frac{1}{\w}
\int\!\frac{d^dq}{(2\pi)^d} \int\!\frac{d\nn_2 d\nn_3 d\nn_4}{\W_d^3}
 \nonumber \\
&& \times \big\{ \gamma_{12}^3 \big[ \L^\rho_{34}
\Upsilon^\rho_{41;21}(\e,\w,\q) + 
\Upsilon^\rho_{34;21}(\e,\w,\q) \bar{\L}^\rho_{41} \big] \nonumber \\
&& +\gamma_{21}^{3} \big[
\L^\rho_{34} \Upsilon^\rho_{42;21}(\e,\w,\q)
+ \Upsilon^\rho_{34;21}(\e,\w,\q) \bar{\L}^\rho_{42} 
\big] \big\}
\nonumber \\ && \\
&& \St_1^{\da{$e$}{$g$}} (\e,\nn_1) 
=  \frac{1}{\nu} \int\!\frac{d\w}{2\pi} \frac{1}{\w}
\int\!\frac{d^dq}{(2\pi)^d} \int\!\frac{d\nn_2 d\nn_3 d\nn_4}{\W_d^3}
 \nonumber \\
&& \times \big\{ \gamma_{12}^3 \big[ \L^g_{34}
\Upsilon^g_{41;21}(\e,\w,\q) + 
\Upsilon^g_{34;21}(\e,\w,\q) \bar{\L}^g_{41} \big] \nonumber \\
&& +\gamma_{21}^{3} \big[
\L^\rho_{34} \Upsilon^g_{42;21}(\e,\w,\q)
+ \Upsilon^g_{34;21}(\e,\w,\q) \bar{\L}^g_{42} 
\big] \big\}
\nonumber \\ 
\eea
\end{subequations}
Here and below the short hand notation
\[
\L^\alpha_{ij} = \L^\alpha(\w,\q; \nn_i,\nn_j),
\, \alpha= g,\rho
\]
is used. It is readily seen that this contributions coincide with the
local electron-boson collision integral of \req{stlocal}.

Proceeding as above, the transform of \req{stm} is
\begin{subequations}
\bea
&&\St_- (\e,\nn_1) = \St_{el} (\e,\nn_1) + \St_{-,l} (\e,\nn_1)
\\ && \St_{el} = \frac{2}{\nu}\Re \int\!\frac{d\w}{2\pi}\frac{1}{\w}
\int\!\frac{d^dq}{(2\pi)^d}
\int\!\frac{d\nn_2\dots d\nn_6}{\W_d^5}
\gamma_{13}^{2}\gamma_{46}^{5}
\nonumber \\
&& \quad \times \left[\L^\rho-\L^g\right]_{52}
\left[ f(\e-\w, \nn_6) - f(\e-\w, \nn_4) \right]  \nonumber \\ 
&& \quad  \times \Big[\L^g_{14} f(\e, \nn_3)
+\L^g_{34} f(\e, \nn_1) \Big]
\label{stmel} \\
&& \St_{-,l} = \frac{1}{\nu} \int\!\frac{d\w}{2\pi}\frac{1}{\w^2}
\int\!\frac{d^dq}{(2\pi)^d} \int\!\frac{d\nn_2\dots d\nn_6}{\W_d^5}
 \gamma_{13}^{2}\gamma_{46}^{5} \nonumber \\
&& \times \Big[ 2f(\e,\nn_4) - f(\e-\w,\nn_4) - f(\e+\w,\nn_4) 
- (\nn_4 \to \nn_6) \Big] \nonumber \\
&& \times \int\!d\e_1 \Big[ f(\e_1,\nn_1) [ 1- f(\e_1-\w,\nn_3)] + 
(\nn_1 \leftrightarrow \nn_3) \Big] \nonumber \\
&& \times \left[\L^\rho - \L^g\right]_{52} \L^g_{14} + 
\frac{1}{2} \left[\L^\rho - \L^g\right]_{52} \left[ \L^g_{14} -\L^g_{34}\right]
\label{stml}
\eea
\end{subequations}
 Equation \rref{stmel} is the (singlet part of the) elastic 
electron-electron collision integral, \req{steeel}. To obtain \req{stml} 
in the given form, we used the analytic properties of the propagators and 
performed the change of variable $\e_1 \to \e_1 +\w$ in some of the terms.

Finally we perform the transform of \req{stp} and obtain:
\begin{subequations}
\be
\St_{+} (\e,\nn_1) = \St^\da{$e$}{$e$}_n (\e,\nn_1) + 
\St_{+,n} (\e,\nn_1) + \St_{+,l} (\e,\nn_1)
\ee
The first term is given by (the singlet part of) \req{steen}.
The second term is 
\be
\St_{+,n} = \St^\da{$e$}{$\rho$}_{+,n} - \St^\da{$e$}{$g$}_{+,n}
\ee
with $\St^\da{$e$}{$\alpha$}_{+,n}$ given by \req{stnonloc} excluding the
last line. The third term is:
\bea
&& \St_{+,l} =\frac{1}{2\nu} \int\!\frac{d\w}{2\pi}\frac{1}{\w}
\int\!\frac{d^dq}{(2\pi)^d} \int\!\frac{d\nn_2\dots d\nn_6}{\W_d^5}
 \gamma_{13}^{2}\gamma_{46}^{5} \nonumber \\ 
&& \times \Big[ 2f(\e,\nn_4) - f(\e-\w,\nn_4) - f(\e+\w,\nn_4) 
- (\nn_4 \to \nn_6) \Big]  \nonumber \\
&& \times \left[ \L^\rho \left( 2N^\rho + 1 \right) -
\L^g \left( 2N^g +1 \right) \right]_{52} 
\left[ \L^g _{34} -\L^g_{14} \right] \label{stpl}
\eea
\end{subequations}
Adding \reqs{stpl} and \rref{stml} we recover the last line of \req{stnonloc} 
and the (the singlet part of the) local electron-electron collision integral, 
\req{steel}. This concludes the derivation of the quantum kinetic equation 
for the singlet channel. 

\subsection{The triplet channel}
\label{triplet}

The inclusion of the interaction in the triplet channel is 
straightforward; in the Eilenberger equation \rref{eil1}
we add to the left hand side the term:
\be\label{tripinter}
i \left[ \hat{\boldsymbol{\phi}} \cdot \boldsymbol{\sigma}; \hat{g} \right],
\ee
where $\sigma_i$ are the Pauli matrices [$i = x,y,z$] and the fluctuating 
field $\hat{\phi}_i$ is a 3 component vector in the $L=1$ angular 
momentum space. Therefore all the triplet channel propagators, the 
polarization operators and the density matrices should be treated 
as $3\times 3$ matrices; for example we have
\[
\left[ D_0 \right]_{ij} (1,2) = - \frac{F^{\sigma}\!\left(\theta_{12}\right) 
\delta (\r_{12})} {\nu}  \delta (t_{12}) \delta_{ij}
\]
[cf. \req{dys}] and the retarded polarization operator is given by
\[
\Pi^R_{ij} (1,2)= \nu \left\{ \delta_{12} + \frac{\pi}{4} \tr \left[ \sigma_i
\frac{\delta {g^K (t_1,t_1,\nn_1,\r_1 )}}{\delta \phi_{+j}(t_2, \r_2, \nn_2 )}
\right] \right\},
\]
where the trace is over spin indices. 

The trace of triplet channel operators includes the sum over the 
indices $i,j$. In the absence of the magnetic field, all the operators 
are diagonal, e.g.:
\[
\left[ \L^\sigma \right]_{ij} = \L^\sigma \delta_{ij} 
\]
and the trace results in extra factors of 3 in comparison to the singlet 
channel. The derivation 
can therefore be repeated with simple modifications 
and it gives the quantum kinetic equation presented in Sec.~\ref{final}. 
We note only one main difference in the derivation for the triplet channel:
the gauge transformation, which has the form:
\[
\hat{g} \to e^{-i \hat{\boldsymbol{K}} \cdot \boldsymbol{\sigma}}
\hat{g} \,  e^{ i \hat{\boldsymbol{K}} \cdot \boldsymbol{\sigma}} ,
\]
does not commute with the interaction term \rref{tripinter}. 
Additional second order terms arise due to commutators of the Pauli matrices;
however these terms vanish in the one loop approximation 
and we can neglect them.

In the next section we will use the quantum kinetic equation to calculate
the interaction corrections to transport coefficients and specific heat.

\section{Derivation of transport coefficients and specific heat}
\label{resder}

In this section the calculation of the transport coefficients
for quasi one-dimensional, two-dimensional and three-dimensional systems is
presented; the evaluation of the interaction correction
to the specific heat is in the final subsection.

In order to calculate the currents in the presence of an external
field (electric field $\E$ or temperature gradient $\nnabla T$), we 
need to solve the kinetic equations. We assume that the external
fields are weak, i.e.
\[
e E L_T \ll T , \quad \nabla T L_T \ll T
\]
and the temperature length was defined in \req{tlength}.
These conditions ensure that the deviations from the
equilibrium distribution functions are small and we can solve the
equations by iteration. 

In the lowest order, the distribution functions should null the
collision integrals in \reqs{eqf} and \rref{boseq}, and the sought corrections 
$\delta f, \delta N^{\alpha}$ are linear in  the electric field or in the 
gradients of the distribution functions. In other words, we look for a
solution to those equations of the form:
\bea\label{itsol}
&&f(\e,\nn ;\r) = f_F(\e;\r) + \delta f (\e,\nn; \r) \, ;\\
&&\hat{N}^\alpha(\w,\q;\r)=N_P(\w;\r)\hat{\openone} + 
\delta \hat{N}^\alpha(\w,\q; \r) \, ; \nonumber
\eea
where the Fermi and Plank distribution functions \rref{equil}
depend on the spatial coordinate only through the temperature $T(\r)$.
For compactness, we will consider explicitly the singlet channel only and
we will indicate how to include the triplet channel.

Let us start from the electronic part of the kinetic equation. 
The bare impurity collision part $\St_\tau$ is larger than the other
terms, thus it suffices to calculate the latter in the first order of
perturbation theory. Considering short range impurities, so that $\tau$ is 
independent of the scattering angle, we find:
\bea
\label{df}
&&\delta f = \delta f_0 + \delta f_1; \\
&&\delta f_0= \tau\v\cdot\left(e \E  -  \frac{\e \nnabla T}{T} 
\right)
\left(-\frac{\partial
  f_F(\e)}{\partial \e}\right)
\nonumber
\\ 
&&\delta f_1 = 
\tau \delta \St 
\left\{ f_F + \delta f_0, N_P +
\delta N^\rho, N_P + 
\delta N^{g} \right\}
\nonumber
\eea
where $\delta \St$ is the linearized collision integral. Note that according
to the discussion of the conservation laws in Sec.~\ref{final}, we need to
consider only the local electron-boson contribution, \req{stlocal}, and 
the elastic electron-electron one, \req{steeel}.

Expression \rref{df} is to be substituted in \reqs{consrho} and 
\rref{elen} to find the electric current and the electron component of 
the energy current. Integration of the $\delta f_0$ term is straightforward.
Due to the structure of the collision integrals \rref{elbosci}-\rref{elelci}, 
the integration over $\e$ can be performed before the $\w$ and $\q$ 
integrations in the $\delta f_1$ term. For the combination of 
distribution functions entering into \req{ups}, we find
\be
\begin{split} 
\delta & \Upsilon_{ij,kl}(\e,\w)  = 
\delta N (\w,\q;\nn_i,\nn_j ) \left\{ f_F(\e) - f_F(\e-\w) \right\} \\ 
& + \W_d\delta (\widehat{\nn_i\nn_j}) N_P (\w)
\left\{ \delta f(\e,\nn_k) - \delta f(\e-\w,\nn_k) \right\} \\ 
& + \W_d\delta (\widehat{\nn_i\nn_j}) 
\Big[ \delta f(\e,\nn_l) \big( 1-f_F(\e-\w)\big)  \\
& \quad \quad - f_F(\e)\delta f(\e-\w,\nn_k) \Big], 
\end{split}
\ee
which with the help of the identities
\begin{subequations}
\be
\begin{split}
&\int\!d\e \, f_F(\e) \frac{\partial f_F (\e-\w)}{\partial \e}
= \frac{\partial}{\partial \w} \Big[ \w N_P (\w) \Big]
\\
&\int\!d\e \frac{\partial f_F(\e)}{\partial \e}\big[ 1 - f_F (\e-\w) \big]
= \frac{\partial}{\partial \w} \Big[ \w N_P (\w) \Big]
\end{split} \ee
\be \begin{split}
&\int\!d\e \, \e \frac{\partial f_F(\e)}{\partial \e} 
\left[ 1 - f_F(\e-\w) \right]  = \frac{1}{2} \w^2 
\frac{\partial N_P(\w)}{\partial \w}; \\
&\int\!d\e \, (\e-\w) \frac{\partial f_F(\e-\w)}{\partial \e} 
f_F(\e) = -\frac{1}{2} \w^2 \frac{\partial N_P(\w)}{\partial \w};
\\
&\int\!d\e \, \e^2 \frac{\partial f_F(\e)}{\partial \e}  \big[ 
1 - f_F(\e-\w) \big] \\ 
&\quad\quad= \frac{\pi^2}{3} T^2 
\frac{\partial}{\partial \w} \Big[ \w N_P (\w) \Big] + 
\frac{1}{3} \w^3 \frac{\partial N_P(\w)}{\partial \w} ;\\
&\int\!d\e \, \e (\e-\w) f_F(\e)  
\frac{\partial f_F(\e-\w)}{\partial \e} \\ 
&\quad\quad = \frac{\pi^2}{3} T^2 
\frac{\partial}{\partial \w} \Big[ \w N_P (\w) \Big] - 
\frac{1}{6} \w^3 \frac{\partial N_P(\w)}{\partial \w};
\end{split}
\ee \end{subequations}
yields
{\setlength\arraycolsep{0pt} \begin{subequations}
\bea
&&\frac{1}{2}\int\!d\e
\big[\delta \Upsilon_{ij,kl}(\e,\w,\q)- 
\delta \Upsilon_{ij,kl}(\e,-\w,-\q) \big]
\label{idu}\\ &&
= e v_F\tau \E \cdot (\nn_k-\nn_l) 
 \frac{\partial}{\partial \w} \left( \w N_P^{ij} \right);
\nonumber\\
&& \frac{1}{2} \int\!d\e \, \e 
\big[\delta\Upsilon_{ij,kl}(\e,\w,\q)- \delta
 \Upsilon_{ij,kl}(\e,-\w,-\q)\big] \\
&&   =-\frac{\w^2}{2}\delta N(\w,\q; \nn_i,\nn_j)
- \frac{v_F\tau \nnabla T}{4T} \cdot
\left( \nn_l+\nn_k\right)\w^3 \frac{\partial N_P^{ij}}{\partial\w}
\nonumber
\\&& \quad +
\frac{v_F\tau \nnabla T}{T} \cdot \left( \nn_l-\nn_k\right) 
\left[\frac{\pi^2T^2}{3}
 \frac{\partial  \big(\w N_P^{ij}\big)}{\partial \w}
+ \frac{\w^3}{12} \frac{\partial {N}_P^{ij}}{\partial\w}
\right],
\nonumber
\eea\end{subequations}
where 
\be\label{npij}
{N}_P^{ij}\equiv \left[ \hat{N}_P \right] (\nn_i,\nn_j) \equiv 
N_P(\w)\W_d\delta (\widehat{\nn_i\nn_j}) \, .
\ee
Here we retained only the odd in $\w$ contribution because the even part
vanishes after the $\w$ integration in the relevant
collision integral, see \req{stlocal}.

The combination of the distribution functions entering 
the elastic collision part, \req{steeel}, gives 
\begin{subequations}
\bea
&&\frac{1}{2}\int d\epsilon f_F(\epsilon)
\big[\delta f(\epsilon-\w,\nn)-\delta f(\epsilon+\w,\nn)  \big] 
\\
&& \quad =- e v_F\tau \E\cdot\nn 
 \frac{\partial}{\partial \w} \Big[ \w N_P(\w) \Big].
\nonumber\\
&&\frac{1}{2}\int d\epsilon \, \epsilon
f_F(\epsilon)
\big[\delta f(\epsilon-\w,\nn)-\delta f(\epsilon+\w,\nn)  \big]
\\ &&
\quad =\frac{v_F\tau \nn\cdot\nnabla T}{T}
\left[\frac{\pi^2T^2}{3}
\frac{\partial \big( \w N_P(\w) \big)}{\partial \w} 
-\frac{\w^3}{6}\frac{\partial  N_P(\w) }{\partial \w} 
\right],
\nonumber
\eea\end{subequations}
where once again we retained only the odd in $\w$ part, 
non-vanishing after the $\w$ integration in \req{steeel}. }

Using \req{consrho}, we find the electric current $\j=\hat{\sigma}\E$
with the conductivity tensor $\hat{\sigma}$ given by:
\be\label{elcon}
\hat{\sigma} = \sd \bigg\{ \hat{\openone} + \int\!d\w
\left[ \hat{\cal S}^{el}(\w) + \hat{\cal E}(\w)
\right]\frac{\partial}{\partial \w} \Big[ \w N_P (\w) \Big] \bigg\}
\ee
where $\sd = \tau v_F^2 e^2 \nu /d$ is the Drude conductivity. Indicating 
spatial indices with $\mu,\nu =1, \ldots,d$,
the elastic kernels ${\cal S}^{el}_{\mu\nu}$ 
and ${\cal E}_{\mu\nu}$, which originate from \req{stlocal} and 
\req{steeel} respectively, are:
\be\label{kernels}
\begin{split}
{\cal S}^{el}_{\mu\nu}(\w) &= {\cal S}^{11}_{\mu\nu}(\w)
+{\cal S}^{12}_{\mu\nu} (\w) \\
{\cal S}^{11}_{\mu\nu}(\w) &=
\frac{d}{\pi\w\nu} \int\!\frac{d^dq}{(2\pi)^d}
\int\!\frac{d\nn_1 d\nn_2}{(\W_d)^2} n_{1\mu} n_{1\nu} \\
\Re \big[ & \L^\rho_{11} - \L^g_{11} \big] 
+\Re \left[ \L^\rho_{22} - \L^g_{22} \right]
-2 \Re \left[ \L^\rho_{12} - \L^g_{12} \right] \\
{\cal S}^{12}_{\mu\nu}(\w) &=
\frac{2d}{\pi\w\nu} \int\!\frac{d^dq}{(2\pi)^d}
\int\!\frac{d\nn_1 d\nn_2}{(\W_d)^2} n_{1\mu}n_{2\nu}
\Re \!\left[ \L^\rho_{12} - \L^g_{12} \right]  \\
{\cal E}_{\mu\nu}(\w) &=
-\frac{d\tau}{\pi\w\nu} \int\!\frac{d^dq}{(2\pi)^d}
\int\!\frac{d\nn_1 \ldots d\nn_6}{(\W_d)^6} \gamma^6_{12} \gamma^5_{43} \\
& \left( n_{1\mu}n_{3\nu} - n_{2\mu}n_{3\nu} \right)
\Re \Big\{ \left[ \L^\rho_{56} - \L^g_{56} \right] 
\left[ \L^g_{13} + \L^g_{14} \right]  \Big\} 
\end{split}\ee
where for compactness we kept the singlet channel correction only; inclusion 
of the triplet channel contribution is straightforward\footnote{by the simple 
substitution $\L^\rho - \L^g \to \L^\rho + 3 \L^\sigma - 4 \L^g$ in the 
kernels.}.
We show in appendix \ref{cmp} that our expression for the conductivity 
coincides with the one of Ref.~\onlinecite{ZNA}.
It is natural that the conductivity does not involve any bosonic
distribution function [cf. \req{idu}], because the inelastic electron 
collision with such bosons changes the energy of the electron but not the
direction of its motion.

In contrast, even the electron contribution to the thermal
conductivity tensor $\hat{\kappa}$, such that 
$\j^\e_{tot}=-\hat{\kappa}\nnabla T$,
is sensitive to the bosonic distribution functions.
We represent the total thermal conductivity as
\be\label{ktot}
\hat{\kappa} = \kwf + \delta\hat{\kappa} + 
\hat{\kappa}^\rho - \hat{\kappa}^g 
\ee
The first term in this expression obeys Wiedemann-Franz law even
with the interaction to the conductivity included, i.e. 
$\kwf = \mathrm{L} \hat{\sigma} T$ with the Lorentz number given by
\req{WF}. The second
term  represent the (electronic) correction to the Wiedemann-Franz due
to the energy dependence of the elastic scattering and due to
the inelastic electron scattering on the bosons. Finally,
the third and the fourth term represent the contribution of the
$\rho$ and $g$ bosons to the thermal transport. These additional contributions 
are given by:
{\setlength\arraycolsep{1pt}
\begin{subequations}
\label{delk}
\bea
\delta \hat{\kappa}_{\phantom{el}}
 &=& \delta \hat{\kappa}_{el} + \delta \hat{\kappa}_{in}; \\
\left[ \delta\kappa_{el} \right]_{\mu\nu}&=& \frac{\sd}{e^2T} \int\!d\w 
\Big[ {\cal S}^{el}_{\mu\nu}(\w) -2{\cal E}_{\mu\nu}(\w) 
\Big]\! \left[ \frac{\w^3}{12} \frac{\partial N_P}{\partial \w}\right]
; 
\nonumber \\ \label{dkel} \\
\left[ \delta \kappa_{in}\right]_{\mu\nu} &=&\frac{\sd}{e^2T} \int\!d\w 
\Big[ {\cal S}^{12}_{\mu\nu}(\w) -{\cal S}^{11}_{\mu\nu}(\w) 
\Big] \left[\frac{\w^3}{4} \frac{\partial N_P}
{\partial \w} \right] \nonumber \\
&+& v_F \int\!\frac{d\w}{2\pi} \, \w \int\!\frac{d^dq}{(2\pi)^d}
\int\!\frac{d\nn_1 d\nn_2 d\nn_3}{(\W_d)^3} \, n_{1\mu} \nonumber \\
&\Big\{& \Re \Big[ \L^\rho_{12} \delta_\nu N^\rho_{23} + 
\L^\rho_{32} \delta_\nu N^\rho_{21} - \L^\rho_{12} \delta_\nu N^\rho_{21}
\Big] \nonumber \\
&-&  \Re \Big[ \L^g_{12} \delta_\nu N^g_{23} + 
\L^g_{32} \delta_\nu N^g_{21} - \L^g_{12} \delta_\nu N^g_{21}
\Big] \Big\}; \ \nonumber \\ \label{dkin}
\eea \end{subequations}
\be\label{thcona}
\kappa^\alpha_{\mu\nu}= -\int\!\frac{d\w}{2\pi}\, \w\int\!\frac{d^dq}{(2\pi)^d}
\int\!\frac{d\nn_1 d\nn_2}{(\W_d)^2} 
\left\{ \hat{s}^\alpha_\mu ; \L^\alpha_{12} \delta_\nu N^\alpha_{21} \right\}
\ee
with
\be\label{varder}
\delta_\mu N^\alpha_{ij} = \frac{\delta} {\delta (\nabla_{\mu} T)} 
\big[ \delta N^\alpha (\w,\q; \nn_i,\nn_j ) \big].
\ee}

Equations \rref{elcon}-\rref{thcona} are the complete expressions for the 
electric and thermal transport coefficients. To obtain the explicit result
one has to solve \reqs{boseq} to find the distribution functions $\delta
N^\alpha$. We will do so by
restricting ourselves to the diffusive $T\tau \ll 1$ regime,
except for two-dimensional systems for which we
will consider the arbitrary temperature range
\footnote{The Boltzmann equation description of strictly one-dimensional 
systems is not applicable and considering the quasi one-dimensional
ballistic case within our scheme is meaningless, because of the
effects of boundary scattering. The ballistic regime in three dimensions
also can not be considered within our scheme, because
the main effect on the thermal conductivity is due to the inelastic
scattering processes with momentum transfer of the order of $k_F$. 
}. Moreover for the 
Coulomb interaction we will consider the unitary limit 
(for infrared finite momentum integrals), which will enable us 
to drop all the terms that depend on $\partial_{\q} \hat{F}$.

From now on we will retain only the zeroth harmonic of the Fermi liquid 
constants, which we denote symbolically with $F^\alpha_0$. For the singlet
channel this means
\begin{subequations} \label{f0}
\be
F_0^\rho \equiv \int\!\frac{d\theta}{2\pi} F^\rho (\theta)+ \nu V (\q)
\ee
and for the triplet channel
\be
F_0^\sigma \equiv \int\!\frac{d\theta}{2\pi} F^\sigma (\theta).
\ee
We remind that $F^g=0$ and that for the Coulomb interaction the potential
is:
\be
V(\q) = \left\{ \begin{matrix}
{\displaystyle \frac{4\pi e^2}{q^2}} \, , & \ d=3 \\
{\displaystyle \frac{2\pi e^2}{q}} \, , & \ d=2 \\
e^2 \ln {\displaystyle \frac{1}{(qa)^2}} \, , & \ d=1
\end{matrix}\right.
\ee
where $a$ is a length of the order of the quasi one-dimensional wire width.
\end{subequations}

\subsection{Diffusive regime}

We consider first the distribution function $N^g$; substituting expression
\rref{itsol} into \req{boseq} we obtain at linear order in  $\nnabla T$
\be\begin{split}
- & \frac{\w}{T} \v_1 \!\cdot\!\nnabla T 
\frac{\partial N_P^{12}(\w)}{\partial \w} + 
i\Big[ \v\!\cdot\!\q ; \delta N^g \Big]
 = 2 \left\{ \St_\tau ; \delta N^g \right\} \\ & + 
v_F \frac{\w}{T}\frac{\partial N_P(\w)}{\partial \w} 
\nnabla T \cdot \big( \W_d \dnnn \nn_1 - \nn_1-\nn_2 \big)
\end{split}
\ee
with $N^{ij}_P$ defined in \req{npij}.
The (exact at this order) solution for $\delta N^g(\w;\nn_1,\nn_2)$ is:
\be\label{dngsol}
\delta N^g = \delta N^0 \equiv v_F \tau \nn_1 \cdot \nnabla T 
\frac{\w}{T} \frac{\partial N^{12}_P(\w)}{\partial \w}
\ee

For the distribution function $N^\rho$, the above is only the starting
point for the iterative solution:
\be\label{dnrsol0}
\delta N^\rho = \delta N^0 + \delta N^1.
\ee
The equation for $\delta N^1$ is:
\be\label{dn1eq}
\begin{split}
&\bigg( -\frac{\w}{T}\frac{\partial \hat{N}_P(\w)}{\partial \w}\nnabla T \bigg)
\!\cdot\!\bigg( \w \frac{\partial}{\partial \q} 
\frac{\hat{F}^\rho}{1+\hat{F}^\rho} \bigg) \\ & + 
i\bigg[ \w \frac{\hat{F}^\rho}{1+\hat{F}^\rho} ; \delta \hat{N}^g \bigg] 
+ i\Big[ \hat{H}_{\eh}; \delta \hat{N}^1 \Big]
= 2 \left\{ \St_\tau ; \delta \hat{N}^1 \right\}.
\end{split}
\ee
In the diffusive limit $T\tau \ll 1$ the (first iteration)
solution would be of the form
\[
\delta N^1 \simeq \w\tau \bigg( v_F \tau \frac{\w}{T}\frac{\partial N_P(\w)}
{\partial \w} \nnabla T  \bigg) \cdot \boldsymbol{V}
\]
for a vector $\boldsymbol{V}$ with magnitude of order one. 
However contributions from frequency $\w$ larger than temperature $T$ 
are exponentially suppressed, i.e. $\w\tau\lesssim T\tau \ll 1$;
therefore $\delta N^1$  can be neglected in comparison to
$\delta N^g$. Thus in the diffusive limit
\be\label{d2n}
\delta_\mu N^{\alpha}_{ij} =  v_F \tau \nn_{i\mu}  
\frac{\w}{T} \frac{\partial N^{ij}_P(\w)}{\partial \w}.
\ee

For the propagators $\L^\alpha$ the diffusive approximation
amounts to the substitution
\[
n_\mu n_\nu \to \frac{\delta_{\mu\nu}}{d} \, ,
\]
which leads to the following expression:
\be\label{approxprop}
\begin{split}
\L^\alpha (\nn_1,\nn_2) = & \tau \big( \dnnn -1 \big)+ L_0^\alpha  \\
&+ (n_1+n_2)_\mu L_{1\mu}^\alpha + n_{1\mu}n_{2\nu} L_{2\mu\nu}^\alpha,
\end{split}\ee
where the functions $L_i^\alpha$ depend on $\w, \, \q$ only and are explicitly:
\begin{subequations}
\label{diffL}
\bea
L_{0\phantom{\mu\nu}}^\alpha &=& \frac{1}{\frac{-i\w}{1+F_0^\alpha}+Dq^2} \\
L_{1\mu\phantom{\nu}}^\alpha &=& -i \tau v_F q_\mu L_0^\alpha \\
L_{2\mu\nu}^\alpha &=&  
- d\tau Dq_\mu q_\nu L_0^\alpha.
\eea\end{subequations}
Here $D=\tau v_F^2 / d$ is the diffusion constant. These formulas are valid
for $\w, \, Dq^2 \ll 1/\tau$.

Within this approximation, we have
\[
{\cal S}^{12}_{\mu\nu}(\w) \pm {\cal S}^{11}_{\mu\nu}(\w) 
\propto L^\rho_{2\mu\nu}-L^g_{2\mu\nu}.
\]
This means that in both $\hat{\sigma}$ and $\delta \hat{\kappa}$ --
 see \reqs{elcon} and \rref{delk} --
we can neglect the contributions of the $\hat{\cal S}$ kernels
[note that by inserting 
the solution \rref{d2n}  into \req{dkin}, $\delta \hat{\kappa}_{in}$ 
is given by twice the first line of that equation]: indeed the leading 
contribution is given by the kernel $\hat{\cal E}$ in \req{dkel}.
This kernel has the approximate form
\be
{\cal E}_{\mu\nu}(\w) =
\frac{4}{\pi\nu\tau} \frac{1}{d \w}
\int\!\frac{d^dq}{(2\pi)^d} \Re \Big[ L^g_0 \big(
L^\rho_{2\mu\nu} - L^g_{2\mu\nu} \big) \Big].
\ee
Finally the bosonic contributions [cf. \req{thcona}] 
to the thermal conductivity \rref{ktot} can be written 
as\footnote{We choose to collect the bosonic contributions into a single 
kernel so that the resulting momentum integral is convergent.}
\be\label{krg}
\hat{\kappa}^\rho - \hat{\kappa}^g = \frac{\sd}{e^2 T} \int\!d\w 
\, \hat{\cal B} (\w)
\bigg[ \frac{\w^3}{4}\frac{\partial N_P}{\partial \w} \bigg]
\ee
with
\be
{\cal B}_{\mu\nu} (\w) =- \frac{2\delta_{\mu\nu}}{\pi\w\nu} 
\int\!\frac{d^dq}{(2\pi)^d}
\Re \left[ L^\rho_0 -L^g_0 \right].
\ee

The next step is to evaluate the momentum integrals; first we give the results
for the short range interaction described by the 
(zeroth harmonic of the) Fermi-liquid constant $F_0$ and then 
we indicate the modifications needed to account for the long range part of
the Coulomb interaction in the singlet channel. The triplet channel
contributions are obtained by multiplying the found results by three and 
identifying $F_0$ with $F_0^\sigma$.
 
For the elastic kernel we find:
\be\label{elkerdiff}\begin{split}
{\cal E}_{\mu\nu}(\w) = &
\frac{\delta_{\mu\nu}}{d} \frac{e^2}{\sd} \frac{\W_d}{(2\pi)^d}
\left( \frac{|\w|}{D}\right)^{\frac{d}{2}-1} \frac{1}{\w} \\
& \times \frac{1}{\cos \frac{\pi d}{4}}\left[ \frac{d}{2} - \frac{1}{F_0} 
\left( 1+F_0 -\left( 1+F_0\right)^{1-\frac{d}{2}} \right) \right]
\end{split}\ee
The expression for 
the Coulomb interaction is obtained by taking the 
unitary limit $F_0 \to +\infty$.

The result for the bosonic kernel is:
\be\label{boskerdiff}\begin{split}
{\cal B}_{\mu\nu} (\w) = \frac{\delta_{\mu\nu}}{2} & \frac{e^2}{\sd} 
\frac{\W_d}{(2\pi)^d} 
\left( \frac{|\w|}{D}\right)^{\frac{d}{2}-1} \frac{1}{\w} \\
&\times \frac{1}{\cos \frac{\pi d}{4}}
\left[ 1-\left(1+F_0\right)^{1-\frac{d}{2}} \right] 
\end{split} \ee
In $d=3$ the limit $F_0 \to +\infty$ gives the correct formula
for the long range contribution, but for $d=1$ the limit diverges.
By retaining the full form of the interaction potential however such infrared 
divergence is cut off at the inverse of the screening radius. 
With logarithmic accuracy the result for the Coulomb interaction is found
by substituting $[\ldots ] \to - ak \ln ^{1/2} (Dk^2/|\w |)$, where $a$ is a
length of the order of the wire width and $k^2 = 4\pi e^2 \nu$ is the square 
of the inverse screening radius [in the bulk].

We can now proceed with the calculation of the integrals over $\w$ in 
\reqs{elcon}, \rref{delk} and \rref{krg} using the identity:
\be\label{zgid}
\int\!d\w \, \w^m \partial_\w N_P (\w) = - 2T^m \zeta (m) \Gamma (m+1)
\ee 
Here $\zeta (x)$ is the Riemann zeta function, whose values at the points
relevant for our discussion are:
\[\begin{array}{ll}
\zeta \left(-\frac{1}{2}\right) \approx -0.208 & \, , \quad 
\zeta (0) = -\frac{1}{2} \, , \\ & \\
\zeta \left( \frac{1}{2} \right) \approx -1.460  & \, , \quad 
\zeta \left(\frac{3}{2} \right) \approx 2.612 \, , \\ & \\
\zeta (2) = \frac{\pi^2}{6} \approx 1.645 & \, , \quad 
\zeta \left( \frac{5}{2}\right) \approx 1.341 \, ,
\end{array}\]
and $\Gamma (x)$ is the Euler gamma function, with values
\[\begin{array}{ll}
\Gamma \left(\frac{1}{2}\right) = \sqrt{\pi} & \, , \quad 
\Gamma (1) = 1 \, , \\ & \\
\Gamma \left( \frac{3}{2} \right) = \frac{1}{2} \sqrt{\pi} & \, , \quad 
\Gamma \left( \frac{5}{2} \right) = \frac{3}{4} \sqrt{\pi} \, , \\ & \\
\Gamma (3) = 2 & \, , \quad 
\Gamma \left( \frac{7}{2}\right) = \frac{15}{8}\sqrt{\pi} \, .
\end{array}\]
Performing the final $\w$ integrations we obtain:
\be\begin{split}
\sigma = & \sd + e^2 \frac{\W_d}{(2\pi)^d} \frac{2}{d}
\zeta\left(\frac{d}{2}-1\right) \left( \frac{T}{D} \right)^{\frac{d}{2}-1} 
\Gamma \left( \frac{d}{2} \right) \\
& \times \frac{d-4}{2-d} \frac{1}{\cos \frac{\pi d}{4}}
\left[ \frac{d}{2} - \frac{1}{F_0} 
\left( 1+F_0 -\left( 1+F_0\right)^{1- \frac{d}{2}} \right) \right] 
\end{split}\ee
\bea
&& \delta \kappa =\frac{1}{3d} \frac{\W_d}{(2\pi)^d} 
\left( \frac{T}{D} \right)^{\frac{d}{2}} D \,
\zeta \left(\frac{d}{2} +1 \right) \Gamma \left( \frac{d}{2} +2 \right) 
\nonumber \\
&& \qquad \qquad \times \frac{1}{\cos \frac{\pi d}{4}} 
\left[ \frac{d}{2} - \frac{1}{F_0} 
\left( 1+F_0 -\left( 1+F_0\right)^{1-\frac{d}{2}} \right) \right] 
\nonumber \\ \\
&& \kappa^\rho - \kappa^g = -\frac{1}{4}\frac{\W_d}{(2\pi)^d} 
\left( \frac{T}{D} \right)^{\frac{d}{2}} D \,
\zeta \left(\frac{d}{2} +1 \right) \Gamma \left( \frac{d}{2} +2 \right) 
\nonumber \\
&& \qquad \qquad \times \frac{1}{\cos \frac{\pi d}{4}} 
\left[ 1 -\left(1+F_0\right)^{1-\frac{d}{2}} \right],
\eea
where in the absence of the magnetic field 
$\sigma_{\mu\nu} = \sigma \delta_{\mu\nu}$ and the similar relation holds for
the thermal conductivity. 
According to the previous discussion,
for the Coulomb interaction the correct expressions are given by the 
limit $F_0 \to +\infty$, with the exception of the term
$\kappa^\rho - \kappa^g $ in $d=1$ for which, with logarithmic accuracy,
the result is obtained by substituting 
$[\ldots ] \to - ak \ln ^{1/2} (Dk^2/ T )$. The final answers for the
corrections to the thermal conductivity are given in 
\reqs{k1d}-\rref{k3d}.

We note that for $d=3$ the $\w$ integration in $\hat{\sigma}$ and
$\kwf$ is actually ultraviolet divergent. This divergence can be incorporated
as a renormalization into the Drude results; this 
renormalization however does not invalidate the Wiedemann-Franz law.

\subsection{Two-dimensional system}

In order to evaluate the interaction corrections
 for the whole temperature range, we need the exact form of
the propagators. In two dimensions they are:
\be\begin{split}
& \L^\alpha (\w,\q;\nn_1,\nn_2) = \W_d \dnnn \L_0 (\w,\q;\nn_1) \\ 
& \, + \L_0 (\w,\q;\nn_1) \L_0 (\w,\q;\nn_2)
\frac{\left( -i\w \frac{F^\alpha_0}{1+F_0^\alpha} 
+\frac{1}{\tau} \right) {\cal C}}
{{\cal C} - \left( -i\w \frac{F_0^\alpha}{1+F_0^\alpha}+\frac{1}{\tau} \right)}
\end{split}\ee
where
\be\label{somedef}
\begin{split}
& \L_0 (\w,\q;\nn)=\frac{1}{-i\w+i\v\cdot\q + 1/\tau},
\\
& {\cal C} = \sqrt{\left( -i\w +1/\tau \right)^2 + 
\left( v_Fq \right)^2 }.
\end{split}\ee
Note that now the variables $\w$ and $v_F q$ are limited only by the Fermi
energy $E_F$.

As before, we need to find the non-equilibrium corrections $\delta N^{\rho,g}$
to the bosonic distribution functions. These are given again by 
\reqs{dngsol}-\rref{dnrsol0}, but to obtain the thermal
conductivity on the arbitrary temperature range, we calculate exactly 
(at linear order in $\nnabla T$) the solution to \req{dn1eq}:
\be\begin{split}
\delta N^1 = \tilde{N} \bigg\{ \frac{ \q \cdot \nnabla T}{v_F q^2} 
 + \frac{i}{iv_F \q \cdot (\nn_1 - \nn_2 ) +2/\tau} \\
\times \bigg[ \bar{\lambda} (\nn_1 ) \nn_1^\perp   
- \lambda (\nn_2 ) \nn_2^\perp  \bigg] \cdot \nnabla T \bigg\},
\end{split}\ee
where the bar indicates complex conjugation and we introduced the following
quantities:
{\setlength\arraycolsep{0pt}
\bea
&& \tilde{N} = v_F \tau \frac{\w^2}{T} \frac{\partial N_P}{\partial \w} 
\frac{F_0}{1+F_0}
\, , \ \nn^\perp = \nn - \frac{(\nn\cdot\q) \q}{q^2} \, , \nonumber
\\ && \lambda (\nn) = \frac{a(\nn)} { a(\nn)  - b }  \, , \qquad
 b = \frac{-i\w \, F_0}{1+F_0}+ \frac{1}{\tau}
\label{somedef2} \\
&& a(\nn)=\sqrt{(v_F q)^2 + \left (-i \q\!\cdot\!\v +\frac{2}{\tau} \right)^2 }
\, . \nonumber
\eea}

We will not calculate the corrections to the electrical conductivity, which
would reproduce the results of Ref.~\onlinecite{ZNA}, as shown in Appendix
\ref{cmp} [see also Ref.~\onlinecite{Mirlin} for the generalization
to the arbitrary disorder]. 
For convenience in the calculations, we separate in $\delta \hat{\kappa}_{in}$
and $\hat{\kappa}^\rho$ the contributions due to $\delta N^0$ and $\delta N^1$
[cf. \reqs{dkin} and \rref{krg}]:
\begin{subequations}
\bea
\delta \hat{\kappa}_{in} \ &=& \delta \hat{\kappa}_{in}^0 + 
\delta \hat{\kappa}_{in}^1 \\
\delta \hat{\kappa}_{in}^0 \ &=& \frac{2\sd}{e^2T} \!\int\!d\w 
\Big[ \hat{\cal S}^{12}(\w) -\hat{\cal S}^{11}(\w) 
\Big] \! \left[\frac{\w^3}{4} \frac{\partial N_P}
{\partial \w} \right] \\
\label{dkin1} \left[\delta \kappa_{in}^1\right]_{\mu\nu} &=& \,
v_F \int\!\frac{d\w}{2\pi} \, \w \int\!\frac{d^2 q}{(2\pi)^2}
\int\!\frac{d\nn_1 d\nn_2 d\nn_3}{(\W_2)^3} \\
&& n_{1\mu} \Re \Big\{ \L^\rho_{12} \delta_\nu N^1_{23} + 
\L^\rho_{32} \delta_\nu N^1_{21} - \L^\rho_{12} \delta_\nu N^1_{21}
\Big\} \quad \nonumber 
\eea\end{subequations}
\begin{subequations}
\bea
\hat{\kappa}^\rho &=& \hat{\kappa}^\rho_0 + \hat{\kappa}^\rho_1 \\
\label{thcor0g}
\hat{\kappa}^\rho_0 &-& \hat{\kappa}^g = \frac{\sd}{e^2 T} \int\!d\w 
\, \hat{\cal B}^0 (\w)
\bigg[ \frac{\w^3}{4}\frac{\partial N_P}{\partial \w} \bigg]
\\
\label{thcor1}
\hat{\kappa}^\rho_1 &=& -\int\!\frac{d\w}{2\pi}\, \w\int\!\frac{d^2q}{(2\pi)^2}
\int\!\frac{d\nn_1 d\nn_2}{(\W_2)^2} 
\left\{ \hat{s}_\mu ; \L^\rho_{12} \delta_\nu N^1_{21} \right\}, \nonumber \\
\eea
where, as in \req{varder}, we indicate with $\delta_\nu N^1$ the variational 
derivative with respect to the temperature gradient and
\be\label{b01}
{\cal B}^0_{\mu\nu} (\w) =- \frac{4}{\pi\w\nu} 
\int\!\frac{d^2q}{(2\pi)^2} \int\!\frac{d\nn_1}{\W_2}
n_{1\mu}n_{1\nu} \Re \left[ \L^\rho_{11} - \L^g_{11} \right].
\ee\end{subequations}
Expressions \rref{elcon} for the electrical conductivity and \rref{dkel} for 
the elastic correction to the thermal conductivity and the definition 
\rref{kernels} for the kernels remain unchanged. The momentum and
angular integrals in the latter can be exactly calculated and
for the singlet channel in the unitary limit we find:
{\setlength\arraycolsep{0pt}
\begin{subequations}
\bea
{\cal E}_{\mu\nu}(\w)  = &&
- \frac{e^2 \delta_{\mu\nu}}{\sd 2 \pi^2} \frac{1}{2\w}
\bigg[ 2 - 2 \w\tau H(\w\tau) \arctan \frac{1}{\w\tau} \nonumber \\
&& + \frac{(\w\tau)^2}{2} \left( \frac{1}{2} - H(\w\tau) \right)
\ln \left( 1+\frac{1}{(\w\tau)^2} \right) \\ 
&& \quad\quad -(\w\tau)^2 H(\w\tau) \ln 2 \bigg] \nonumber \\
 {\cal S}^{11}_{\mu\nu}(\w) = &&
- \frac{e^2 \delta_{\mu\nu}}{\sd 2 \pi^2} \, \tau \, 
\frac{\pi}{2} \sgn \, \w \\
{\cal S}^{12}_{\mu\nu}(\w) = &&
- \frac{e^2 \delta_{\mu\nu}}{\sd 2 \pi^2} \, \tau
\bigg\{ \Big( 2H(\w\tau) - 1 \Big) \arctan \frac{1}{\w\tau} \\
+ \frac{\pi}{2} && \sgn \,  \w + \w\tau H(\w\tau) \left[ 
\frac{1}{2} \ln \left( 1+\frac{1}{(\w\tau)^2} \right) + \ln 2 \right]
\!\bigg\} \nonumber
\eea
with the function $H$ defined as:
\be
H(x) = \frac{1}{4+x^2}
\ee
To perform the momentum integral in \req{b01} we must keep the full form
of the propagator in order to avoid the infrared divergence 
that we would obtain in the unitary limit:
\be\begin{split}
{\cal B}^0_{\mu\nu}&(\w) = 
-\frac{e^2 \delta_{\mu\nu}}{\sd 2\pi^2} \frac{1}{\w} \bigg\{
\ln \left( \frac{D k^2}{|\w |} \right) +\pi |\w | \tau \\ 
&  - \frac{1}{2} \ln \left[1+ (\w\tau)^2 \right]
 - \w \tau \arctan \left( \frac{1}{\w\tau } \right) \bigg\}
\end{split}\ee
\end{subequations}
where $k=2\pi e^2 \nu$ is the inverse screening radius.}
Next, we calculate the angular integrals in \reqs{dkin1} and \rref{thcor1} 
as well as the angular part of the momentum integrals and obtain:
\begin{subequations}\be\begin{split}
& \int\!\frac{d\theta_q}{2\pi}
\int\!\frac{d\nn_1 d\nn_2 d\nn_3}{(\W_2)^3} \, n_{1\mu} 
 \Re \Big\{ \L^\rho_{12} \delta_\nu N^1_{23} + 
\L^\rho_{32} \delta_\nu N^1_{21} \Big\}, \\
& = \frac{i}{4} \tilde{N} \delta_{\mu\nu}\bigg\{
\frac{1}{(vq)^2}\frac{1}{{\cal C} \bar{\cal C}} \left[ 
\frac{{\cal C} \bar{b}'}{{\cal C} - b} + 
\frac{\bar{\cal C} b'}{\bar{\cal C} - \bar{b}}\right]
\left[ \bar{\cal C} - \bar{b}' - {\cal C} + b' \right]
\\ & \quad +\frac{1}{(vq)^2}\left[ \frac{{\cal C} - b'}{{\cal C}} - 
 \frac{\bar{\cal C} - \bar{b}'}{\bar{\cal C}}\right]
- \frac{1}{{\cal C}\bar{\cal C}} \left[
\frac{{\cal C}}{{\cal C} - b} - 
\frac{\bar{\cal C}}{\bar{\cal C} - \bar{b}}\right] \bigg\}
\end{split}\ee
\be\begin{split}
& \int\!\frac{d\theta_q}{2\pi}
\int\!\frac{d\nn_1 d\nn_2}{(\W_2)^2} \, n_{1\mu} 
 \Re \Big\{ \L^\rho_{12} \delta_\nu N^1_{21} \Big\} \\
& \ = \frac{i}{4} \tilde{N} \delta_{\mu\nu} \frac{\tau}{2}  
\bigg\{ \left[ \frac{\bar{\cal C}}{\bar{\cal C} - \bar{b}} - 
\frac{{\cal C}}{{\cal C} - b}\right] \left[\frac{1}{{\cal C}} +
\frac{1}{\bar{\cal C}} \right] \\
& - \frac{1}{(vq)^2} \left[ 
\frac{{\cal C} \bar{b}'}{{\cal C} - b} + 
\frac{\bar{\cal C} b'}{\bar{\cal C} - \bar{b}}\right]
\left[  \frac{{\cal C} - b'}{{\cal C}} - 
 \frac{\bar{\cal C} - \bar{b}'}{\bar{\cal C}}\right] .
\bigg\}
\end{split}\ee\end{subequations}
The function ${\cal C}$ was defined in \req{somedef} and $b'$ is given by 
the limit $F_0 \to +\infty$ of $b$ in \req{somedef2}, where $\tilde{N}$ 
is also defined.
The remaining integrals over the magnitude of the momentum can 
be evaluated approximately; the result can be written as:
\begin{subequations}\be
\delta \hat{\kappa}_{in}^1 + \hat{\kappa}_1^\rho = \frac{\sd}{e^2 T}
\int\!d\w \, \hat{\cal B}^1(\w) 
\bigg[ \frac{\w^3}{4}\frac{\partial N_P}{\partial \w} \bigg] 
\ee
with
\be\begin{split}
{\cal B}^1_{\mu\nu}(\w)
= \frac{e^2 \delta_{\mu\nu}}{\sd 2\pi^2} \, \tau \bigg\{
\frac{2\w\tau}{1+(\w\tau)^2} \ln \left( \frac{v_F k}{2 |\w| } \right) 
\\ - \arctan \w\tau
-\w\tau \ln \left( \frac{E_F}{\sqrt{\w^2 + \tau^{-2}}} \right).
\bigg\}
\end{split}\ee\end{subequations}
In the above kernel the first term in curly bracket originates from
$\hat{\kappa}_1^\rho$ only -- as discussed in Sec.~\ref{sec:gaugetr} no long
range terms can be present in the electronic contribution to the 
thermal conductivity.
Note that the second term in the above expression is beyond the
logarithmic accuracy of our approximate calculation and must be dropped. 
Similarly, most of the terms in the other kernels can be
neglected and collecting the logarithmic contributions:
\be\begin{split}
\Delta\kappa_s & = -\frac{1}{2\pi^2 T} \int\!d\w
\bigg[ \frac{\w^2}{4}\frac{\partial N_P}{\partial \w} \bigg]
\\ &\times \bigg\{(\w\tau)^2 \ln \left(\frac{E_F}{|\w|}\right) 
+ \frac{2}{1+(\w\tau)^2} \ln \left( \frac{v_F k}{|\w|}\right) 
\\ &+\left[ \frac{7}{12}(\w\tau)^2 -\frac{5}{6} + 
\frac{16}{3} \frac{1}{4+(\w\tau)^2} \right] 
\ln \left( 1+\frac{1}{(\w\tau)^2}\right)\bigg\},
\end{split}\ee
where we defined the singlet and triplet channel corrections as:
\[
\left[ \kappa -\kappa_{\scriptscriptstyle W\!F} \right]_{\mu\nu}
= \left( \Delta\kappa_s +3\Delta\kappa_t\right) \delta_{\mu\nu}
\]
The final integration can now be performed within the logarithmic accuracy; 
we find [cf. \req{k2ds}]:
\be\label{Dkappa}\begin{split}
&\Delta\kappa_s = -\frac{\pi^2}{15}T(T\tau)^2 \ln\left(\frac{E_F}{T}\right)
\\ & + \frac{T}{6} g_1(2\pi T\tau) \ln \left( \frac{v_F k}{T} \right) 
- \frac{T}{24} g_2(\pi T\tau) \ln \left( 1+ \frac{1}{(T\tau)^2} \right)
\end{split}\ee
where
\begin{subequations} \label{gsdef}
\bea
g_1 (x) &=& 
\frac{3}{x^2} \left\{ \frac{1}{x} \left[ 2\psi' \left(\frac{1}{x}\right)
- x^2 \right] -2 \right\} \\
g_2 (x) &=& \frac{14}{15} x^2 +\frac{8}{3} g_1(x) - \frac{5}{3},
\eea
and $\psi'$ is the derivative of the digamma function.
Since the asymptotic behaviour of $g_1(x)$ is
\be
g_1 (x) = \left\{
\begin{matrix}
1 - \frac{1}{5} x^2 + \frac{1}{7} x^4 + \ldots & x \ll 1 \\
\frac{3}{x}-\frac{6}{x^2} +\frac{\pi^2}{x^3} + \ldots & x \gg 1
\end{matrix}\right.,
\ee\end{subequations}
both these functions tend to 1 as $T\tau \to 0$; therefore in 
the diffusive limit the main contribution is $T \ln (Dk^2/T) /12$. On the
other hand, for $T\tau \gg 1$ the first term in \req{Dkappa} is the 
dominant one.

Turning to the triplet channel, for simplicity
we will restrict ourselves to the limiting
diffusive and quasiballistic cases, although one can extend the calculation
to the whole temperature range, as done in Ref.~\onlinecite{ZNA} for the 
electrical conductivity.

In the diffusive limit $T\tau \ll 1$ we know from our previous analysis
that we can discard the ${\cal B}^1$ term as well as the ${\cal S}$ terms.
The relevant kernels are then\footnote{They can be obtained from 
\reqs{elkerdiff}-\rref{boskerdiff} in the limit $d \to 2$.}:
\be\label{diffkere}
{\cal E}_{\mu\nu}(\w) = 
- \frac{e^2 \delta_{\mu\nu}}{\sd 2 \pi^2 \w} \left[ 1- \frac{1}{F_0^\sigma} 
\ln \left( 1+F_0^\sigma \right) \right]
\ee
\be\label{diffkerb}
{\cal B}^0_{\mu\nu} (\w) =-\frac{e^2 \delta_{\mu\nu}}{\sd 2\pi^2 \w}
 \ln \left( 1+F_0^\sigma \right) 
\ee
which substituted into \reqs{delk}-\rref{thcor0g} give
\be
\Delta\kappa_t = -\frac{T}{18} \left[ 1- \frac{1}{F_0^\sigma} 
\ln \left( 1+F_0^\sigma \right) \right] + \frac{T}{12} \ln \left( 
1+F_0^\sigma \right).
\ee
In the opposite limit $T\tau \gg 1$, the main contribution comes, as for the
singlet channel, from the logarithmic divergence at large momentum in the
kernel ${\cal B}^1$:
\be
{\cal B}^1_{\mu\nu}(\w)
= -\frac{e^2 \delta_{\mu\nu}}{\sd 2\pi^2} \, \tau \bigg\{
\w\tau \ln \left( \frac{E_F}{\sqrt{\w^2 + \tau^{-2}}} \right)\!
\bigg\} \left( \frac{F_0^\sigma}{1+F_0^\sigma}\right)^2.
\ee
Then, the correction to the thermal conductivity is:
\be\label{dkinrate}
\Delta\kappa_t = -\frac{\pi^2}{15}T(T\tau)^2 \ln\left(\frac{E_F}{T}\right) 
\left( \frac{F_0^\sigma}{1+F_0^\sigma}\right)^2
\ee
which concludes the derivation of \req{k2dt}.

This correction to the thermal conductivity (and the corresponding one 
in the singlet channel) is the inelastic processes contribution to the 
energy relaxation rate\footnote{A similar argument is presented
in Ref.~\onlinecite{Langer}.}. 
In the clean system, such inelastic processes 
cannot relax momentum (because of Galilean invariance) 
and hence they do not affect the electrical conductivity, but they
can contribute to the energy relaxation rate $\Gamma_\e$. 
In the kinetic theory, the thermal conductivity can be written, 
up to a numerical coefficient, as:
\[ 
\kappa \propto TE_F/\Gamma_\e
\]
and the rate is given by the sum of the rates for the relevant processes,
namely the electron-impurity and electron-electron scattering rates:
\[
\Gamma_\e = \Gamma_{imp}+ \Gamma_{\da{$e$}{$e$}}
\]
with $\Gamma_{imp} = 1/\tau$ and
\[
\Gamma_{\da{$e$}{$e$}} =a \frac{T^2}{E_F} \ln \left( \frac{E_F}{T} \right)
\]
Here $a$ is a constant whose exact value is irrelevant for our 
argument. In the limit $(T^2/E_F) \ln (E_F /T) \ll 1/\tau$ we can expand the 
expression for the total rate, substitute the result into the above formula
for $\kappa$ and obtain:
\[
\kappa \propto T\tau E_F - a T (T\tau )^2 \ln \left( \frac{E_F}{T} \right)
\]
The first term on the right hand side is the usual Drude result for the thermal
conductivity and the second term has the form of the correction 
\rref{dkinrate}. Note that
in the opposite limit (clean system) one would find:
\[
\kappa \propto \frac{E_F^2}{T \ln \left( \frac{E_F}{T} \right)}
\]
in agreement with the result of Ref.~\onlinecite{mish}.

\subsection{Specific heat}
 
Recalling our discussion on the structure of the kinetic equation in Section 
\ref{Sec.2}, we write the total specific heat as the sum of the electronic 
 and bosonic contributions:

\begin{subequations}
\bea
\cV &=& \cV^0 + \delta \cV,
\\
\cV^0 &=& \frac{\partial u_e}{\partial T} = \frac{\pi^2}{3} \nu T,
\\ \label{dcv2}
\delta \cV &=& \frac{\partial}{\partial T} \Big( u_\rho - u_g \Big), 
\eea
\end{subequations}
where, according to \req{bos1c}, the bosonic energy densities 
[in the equilibrium \rref{thermeq}] are
\be
u_{\alpha} = \int\!d\w \, \w \, b^{\alpha}(\w ) N_P (\w)
\ee
with\footnote{This definition of the density of states is one half
of the one given in \reqs{cVDoS}, because of the different limits for the $\w$
integration in the energy density and the specific heat.}
\be
b^\alpha (\w) = \frac{\Re}{2\pi}\!\int\!\frac{d^d q}{(2\pi)^d}
\Tr_{\nn}  \bigg\{ \frac{1}{1+\hat{F}^\alpha}; \L^\alpha (\w,\q) \bigg\}.
\ee
As before, we will consider explicitly the singlet channel, short range 
interaction in the zeroth harmonic approximation for the Fermi liquid constant,
which we denote with $F_0$. The results for the long range interaction in the 
unitary limit are obtained by letting $F_0^\rho \to +\infty$. For the triplet 
channel one must substitute $F_0^\rho$ with $F_0^\sigma$ and multiply by an 
overall factor of 3. The final answer with the correct coefficients is 
given in Sec.~\ref{cvsumm}.

In the diffusive limit, to which we restrict our attention for $d=1,3$, 
we have:
\be
b^\rho (\w) - b^g(\w)= 
\frac{\Re}{2\pi}\!\int\!\frac{d^d q}{(2\pi)^d} \left[
\frac{1}{1+F_0^\rho} L_0^\rho (\w,\q) - L_0^g (\w,\q) \right]
\ee
and the functions $L_0^\alpha$ were defined in \reqs{approxprop} and 
\rref{diffL}.
After the integration over momentum we find:
\be\begin{split}
b^\rho(\w ) -b^g(\w) &
=  \frac{\W_d}{(2\pi)^d} \left( \frac{|\w |}{D} \right)^{\frac{d}{2}} 
\frac{1}{4\w}
\\ \times & \frac{\cos \frac{\pi}{4} (d-2) }{\sin \frac{\pi}{2}(d-2) }
\left[ 1 - \frac{1}{(1+F_0^\rho)^{\frac{d}{2}}} \right]
\end{split}\ee
which inserted into \req{dcv2} gives:
\be\begin{split}
\delta \cV = &\frac{1}{2} \frac{\W_d}{(2\pi)^d}
\left(\frac{T}{D}\right)^{\frac{d}{2}} 
\zeta\left(\frac{d}{2}+1\right) \Gamma \left( \frac{d}{2} + 2\right) \\
&\times \frac{\cos \frac{\pi}{4} (d-2) }{\sin \frac{\pi}{2}(d-2) }
\left[ 1 - \frac{1}{(1+F_0^\rho)^{\frac{d}{2}}} \right].
\end{split}\ee
The relevant numerical values for  the zeta and gamma functions are given
after \req{zgid}, which has been used to perform the $\w$ integral.

For $d=2$ we can keep the full form of the propagators
to find the singular contribution to the specific heat
at arbitrary value of $T\tau$:
\be\begin{split}
b^\rho(\w) - & b^g(\w) =  -\frac{\Re}{2\pi} \int\!\frac{d^2
  q}{(2\pi)^2} 
\bigg[
\frac{F_0^\rho}{1+F_0^\rho}\frac{1}{{\cal C} - b }
\\ &-\frac{\left(-i\w + 1/\tau \right)}{{\cal C}} 
\left( \frac{1}{{\cal C} - b } 
- \frac{1}{{\cal C} - 1/\tau} \right) \bigg],
\end{split}\ee
with ${\cal C}$ and $b$ defined respectively in \reqs{somedef} and 
\rref{somedef2}. 
The first term in the integral is formally divergent
at $|q| \to \infty$ and this divergence gives a linear in $T$
contribution to the specific heat which does not depend on disorder. 
This term must be disregarded as all the linear terms are included in the
definition of the effective electron mass -- taking it into account
here would be a double counting. To regularize the integral, we
replace $\frac{1}{{\cal C} - b } \to \frac{1}{{\cal C} - b
}-\frac{1}{\cal C}$ in the first line.

Performing the momentum integral we obtain:
\be\begin{split}
b^\rho(\w) &- b^g(\w) =-\frac{1}{8\pi^2 D} \bigg[ \frac{F_0^\rho}{1+F_0^\rho} 
\ln \left(\frac{E_F}{|\w |}\right) \\
&+\left(\frac{F_0^\rho}{1+F_0^\rho}\right)^2 \frac{\pi}{2} \tau |\w | 
-\frac{1}{1+F_0^\rho} \ln (1+F_0^\rho) \bigg].
\end{split}\ee
The final answer for the correction to the specific heat is then:
\be\begin{split}
\delta \cV = & - \frac{1}{12} \frac{T}{D}  \frac{1}{1+F_0^\rho} \left[ F_0^\rho
\ln \left(\frac{E_F}{T} \right) - \ln (1+F_0^\rho) \right] \\
& - \frac{1}{4\pi^2}\Big[ (2\gamma -3) \zeta (2) -2 \zeta '(2) \Big] 
\frac{T}{D} \frac{F_0^\rho}{1+F_0^\rho} \\
& -\frac{3}{4\pi}\zeta (3)  \frac{T}{D}\left( T\tau \right) 
\left( \frac{F_0^\rho}{1+F_0^\rho} \right)^2  
\end{split}\ee
where $\zeta (2) \approx 1.645$, $\zeta ' (2) \approx -0.938$ and 
$\zeta (3) \approx 1.202 $.
In the quasi ballistic limit $\tau \to +\infty$ only the last line is relevant:
\[
\delta \cV = -\frac{3}{2\pi} \zeta (3) 
\left( \frac{F_0^\rho}{1+F_0^\rho} \right)^2 \frac{T^2}{v_F^2}.
\]
This $T^2$ correction to the specific heat has the same form found
for two-dimensional Fermi liquids\cite{millisetc}. 

As discussed before, the long-range interaction 
can be accounted for by taking
the limit $F^\rho_0 \to +\infty$, while the triplet channel contribution
is three times larger -- see also Sec.~\ref{cvsumm}.

\section{Conclusions}

Locality on the scale determined by the temperature and the validity
of the conservation laws are two main requirements for a proper
kinetic description of any system.
In the present paper we derived such a description for the 
interaction effects in disordered metals [assuming
that the clean counterpart of the system is a stable Fermi liquid]. 

We showed that this description requires the introduction, along with
the usual fermionic quasiparticle distribution function, of 
additional bosonic distribution functions. These neutral bosons
are of two types: (i) the ones describing the oscillation in
charge density (singlet) or spin density (triplet) and
(ii) fictitious (ghost) bosons which prevent over-counting
the degrees of freedoms (electron-hole pairs) 
already included in the fermionic part. The obtained conservation
laws together with gauge invariance allowed for
the unambiguous definition of the corresponding electric and energy
currents.

For the electric transport the neutral bosons are not important and
our description reproduces the known results for the correction to the
conductivity obtained in Ref.~\onlinecite{AAL} for the diffusive regime
and in Ref.~\onlinecite{ZNA} in the ballistic and crossover regime.

The neutral bosons, however, are crucial for the thermal
properties of the system.
Namely, their contributions to the energy density
are responsible for the non-analytic corrections to the specific
heat, see \reqs{dcvdiff} and \rref{dcv2d}. 
Our kinetic equation approach reproduces the results for
the interaction corrections to the specific heat previously calculated 
within the equilibrium diagrammatic technique\cite{AA85}.
Moreover the neutral bosons 
contributions to the energy current cause the violation of the 
Wiedemann-Franz law, see \req{wfvio} and the discussion that follows it. 
The violation is stronger for low 
dimensionality systems ($d=1,2$) in the diffusive regime, see \reqs{k1d} and
\rref{k2ds}. Other effects contributing to the
violation of the Wiedemann-Franz law
are the energy dependence of the electron elastic scattering
and the inelastic scattering of the electrons on the neutral bosons. The
latter effect was found to be relevant in the quasi-ballistic regime 
$T\tau \gg \hbar$ for two-dimensional systems, see \reqs{k2d}.

The violation of the Wiedemann-Franz law was investigated before in
the diffusive regime in Refs.~\onlinecite{Livanov} and \onlinecite{Raimondi}
within the ``quantum kinetic equation'' approach and by Kubo formula in 
Ref.~\onlinecite{Smith}.  Ironically, even though the forms of the 
energy current operator used in those references are wrong, the final 
results for the thermal conductivity are consistent with our 
\reqs{k1d}-\rref{k2d}. We think that this agreement is accidental.

\acknowledgments

We are grateful to B.L. Altshuler for initiating this work.
Interesting conversations with M.Yu. Reizer are
acknowledged. We would like to 
thank A.I. Larkin, A.J. Millis and B.N. Narozhny for discussions of
our results. 
I.A. was supported by Packard foundation.

\appendix

\section{On the correction to the thermodynamic potential}
\label{app1}

Standard analytic continuation of \req{dom} gives:
\bea
\delta \W &=& \int\!\frac{d\w}{4\pi}  \coth \left( \frac{\w}{2T} \right) 
\int\!\frac{d^d q}{(2\pi)^d} \Im \, \ln \left( 1 + \frac{F}{\nu} 
\Pi^R (\w ,\q) \right) \nonumber 
\\ & =& \int\!\frac{d\w}{2\pi} \frac{1}{2}\coth \left( \frac{\w}{2T} \right)
\Im \,\Tr \, \ln \left( 1 + \frac{\hat{F}}{\nu} \hat{\Pi}^R \right)
\label{ancont}
\eea
In the second line we used the operator notation -- see \reqs{operator}
and \rref{trace1} -- which gives the correct generalization for the 
momentum dependent Fermi liquid parameter.
Substituting the transform of the explicit expression \rref{pira} for the 
polarization operator, the argument of the logarithm can be rewritten as:
\[
\left( 1+ \hat{F} \right) \left[ \left(\hat{\L}^g\right)^{-1} + 
\frac{\hat{F}}{1+\hat{F}} i \w \right] \hat{\L}^g
\]
According to the definition \rref{Ls}, the term in square brackets is 
$\left(\hat{\L}^\rho\right)^{-1}$. Using the property
\[
\Tr \ln \left( \hat{A} \hat{B} \right) = \Tr \ln \hat{A} + \Tr \ln \hat{B} 
\]
and since $\ln \left( 1 + \hat{F} \right) $ does not contribute to the 
imaginary part, we conclude that
\[
\Im \, \Tr \, \ln \left( 1 + \frac{\hat{F}}{\nu} \hat{\Pi}^R \right) =
- \Im \, \Tr \, \left[ \ln \hat{\L}^\rho - \ln \hat{\L}^g \right]
\]
Substituting this identity into \req{ancont} we finally obtain \req{dop}.

\section{On the microscopic form of the energy current operator}
\label{microcurr}

The action entering into the partition function that 
describes the electron gas in the presence of an external electric field is
\be
S = 
\int dt d^dr \Big[
i \psi^\dag \pt \psi - \psi^\dag \hat{H}_{gi} \psi 
- \psi^\dag \varphi \psi \Big]
\ee
with the condition: $\nnabla \times \A =0$,
which ensures the absence of the magnetic field. The variables $(t ,\r)$ on
which all the fields depend have been suppressed. 
The gauge invariant part of the Hamiltonian for the non-interacting 
system has the usual form:
\be\begin{split}
\psi^\dag \hat{H}_{gi} \psi = & \frac{1}{2m} 
\left( i\nnabla + \A \right) \psi^\dag \left( -i\nnabla + \A \right) \psi 
\\ & + \psi^\dag V_{imp} \psi 
\end{split}\ee
where $V_{imp}$ is the impurity potential and the potentials 
$\varphi$ and $\A$ describe the external electric field:
\[
e\E_{ext} = -\nnabla \varphi + \pt \A
\]
As usual, the charge conservation law
\be\label{coneq}
\pt \rho + \nnabla \cdot \j = 0
\ee
follows from the requirement of gauge invariance,
with the  charge and current densities given by:
\be\label{elcdef}\begin{split}
\rho &= e \psi^\dag \psi \\
\j &= \frac{e}{2m} \left[ \psi^\dag \left( -i\nnabla + \A \right) \psi + 
 \left( i\nnabla + \A \right) \psi^\dag \psi \right]
\end{split}\ee

The invariance of the action upon the replacement
\[
\psi ( t,\r) \to \psi (t +\alpha(t,\r) ,\r) 
\]
[and the similar replacement for $\psi^\dag$] is the basis for the
derivation of the energy conservation law. A straightforward calculation gives
\be\begin{split}
& \frac{\delta S}{\delta \alpha} = 0 =  \partial^\prime_t 
( \psi^\dag \hat{H}_{gi} \psi ) + \varphi \pt (\psi^\dag \psi ) \\
& -\frac{i}{2m} \nnabla \cdot \left[ \pt \psi^\dag (-i\nnabla + \A ) 
\psi - (i\nnabla + \A)\psi^\dag \pt \psi \right]
\end{split}\ee
where the prime means that the derivative acts on $\psi ,\, \psi^\dag$ only.
By adding and subtracting terms proportional to $\varphi$ in the last bracket
and to $\pt A$ in the first term, we find the energy conservation law:
\be\label{ecl}
\pt u_0 + \nnabla \cdot \j_\e^0 = \j\cdot\E_{ext} -\frac{1}{e} \varphi 
\left[ \pt \rho + \nnabla \cdot \j \right]
\ee
where
\begin{subequations}
\label{apu0}
\bea
u_{0} &=& \psi^\dag \hat{H}_{gi} \psi  \\
\j_\e^0 &=& -\frac{1}{2m} \Big[ ( i\pt +\varphi ) \psi^\dag 
(-i\nnabla + \A ) \psi \label{mec}
\\ && \quad -(i\nnabla + \A )\psi^\dag (i\pt - \varphi) \psi \Big]
\nonumber
\eea
The first term on the right hand side of \req{ecl} is the usual
Joule heat; the last term in that equation is not gauge invariant, but
it vanishes because it is proportional to the continuity equation \rref{coneq}.
\end{subequations}

We now consider the generalization to the interacting case. The Hamiltonian
now contain an additional term:
\[
\frac{1}{2}\int d^d r_1 \psi^\dag \psi (\r) V (\r -\r_1) \psi^\dag \psi(\r_1)
\]
where $V(\r) = e^2/|\r|$ describes the density-density Coulomb interaction,
which can be decoupled by the Hubbard-Stratonovich transformation. This amount 
to the introduction of the quantum fields $\phi$ and $\cA$ in the action
by adding the term
\be\label{efldef}
-\psi^\dag \phi\psi+\frac{1}{2} \E_{fl}^2 \, , 
\quad e\E_{fl} = -\nnabla \phi + \pt \cA
\ee
and redefining the vector potential as the sum of the external and 
fluctuating ones:
\be\label{subst}
\A \to  \A_{ext} + \cA
\ee
The variation of the action
with respect to the fluctuating potentials results in the first and 
fourth Maxwell equations relating the fluctuating electric field to
the charge and current densities:
\be\label{maxeq}
 \nnabla \cdot \E_{fl} = \rho \, , \quad 0 = \j + \pt \E_{fl} 
\ee
where the electric current is defined in \req{elcdef}, but the substitution
\rref{subst} must be performed.

To obtain the energy conservation law we must consider the further 
transformation
\[
\phi(t,\r) \to \phi (t+\alpha(t,\r) ,\r)
\]
and the similar one for $\cA$. Proceeding as before we find the conservation
law:
 \be\label{step1}
 \begin{split}
 &
 \frac{\partial u^?}{\partial t}
 + \nnabla \cdot \j_\e^?
 = \j \cdot \E_{ext}\\
 & u^?= u_0 - \frac{1}{2} \E_{fl}^2
 + \frac{1}{e}\left[ \rho \phi 
 + \frac{\partial \cA}{\partial t} \cdot \E_{fl} \right]
 \\
 & \j_\e^? = \j_\e^0 - \frac{1}{e}\phi\j +
\frac{1}{e}
\left[\phi\j - \frac{\partial \phi}{\partial t} \E_{fl}  \right] 
\end{split}
\ee
where $u_0,\ \j_\e^0$ are defined in \req{apu0} [with the substitution 
\rref{subst}].
Given the form of \req{step1}, 
one might be tempted to call $u^?$ and
$\j_\e^?$  the energy and energy current densities so that the conservation
law takes exactly the same form as in the non-interacting case. 
However, such a redefinition would result in gauge dependent expressions
for the densities, since the terms  the square brackets
taken separately are not gauge invariant. Hence this naive
redefinition of the conserved quantities is unphysical, 
since any physical perturbation  can be coupled only to gauge 
invariant quantities. To find the gauge invariant definitions, we 
rewrite the contribution of those non-gauge invariant terms as:
\[\begin{split}
&\frac{\partial}{\partial t}
\left[ \rho \phi 
 + \frac{\partial \cA}{\partial t} \cdot \E_{fl} \right]
+
\nnabla
\left[ \phi\j - \frac{\partial \phi}{\partial t} \E_{fl}  \right]
\\
& = \frac{1}{e}\phi \left[ \pt\rho + \nnabla\cdot\j \right] + \frac{1}{e}
\frac{\partial \phi}{\partial t} \left[ \rho - \nnabla \cdot \E_{fl} \right]
\\& \ 
+\frac{1}{e}\frac{\partial}{\partial t}\left( \pt \cA -\nnabla \phi \right)
\cdot \E_{fl} + \frac{1}{e}\left[ \nnabla \phi \cdot \j +\pt \cA 
\, \pt \E_{fl} \right]
\end{split}\]
Here the first line vanishes because of charge conservation, \req{coneq},
and because of the first of the Maxwell equations \rref{maxeq}. 
As for the second line, we use the second
Maxwell equation to eliminate the current; in the result we substitute 
the definition for the fluctuating field given in \req{efldef} and
we obtain that the second line of \req{step1} equals $\pt \E_{fl}^2$. 
This enables us to conclude that the correct, gauge invariant expressions 
for the energy and energy current densities are:
\bea
u \,  &=& \psi^\dag \hat{H}_{gi} \psi + \frac{1}{2} \E_{fl}^2 \\
\j_\e &=& -\frac{1}{2m} \Big[ ( i\pt + \varphi ) \psi^\dag 
(-i\nnabla + \A ) \psi 
\\ && \quad -(i\nnabla + \A )\psi^\dag (i\pt - \varphi) 
\psi \Big] \nonumber
\eea
where the potentials are the total ones:
\[
\A = \A_{ext} + \cA \, , \quad \varphi  = \varphi_{ext} + \phi
\]
We note that these expressions are gauge invariant with respect to gauge
transformations of both the external and fluctuating potentials. 
We believe, that only such quantities can be coupled to the
``gravitational field'' in the Luttinger scheme for the calculation of
the thermal conductivity \cite{Luttinger}.

\begin{figure}[!t]
\includegraphics[width=0.4\textwidth]{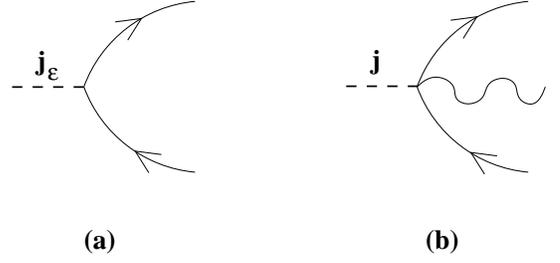}
\caption{(a) the energy current
vertex for the non-interacting case; (b) the additional vertex
induced by the interaction. The solid lines with arrows are electron Green's
functions, the wavy line is the interaction propagator, the dashed lines
are the ``standard'' (non-interacting case)
energy and electric current operators defined respectively
in \reqs{mec} and \rref{elcdef}.}
\label{figcurrver}
\end{figure}

The same final answer is obtained if the interaction is decoupled in the
``gauge-fixed'' form $\cA =0$. In this case, which is the most widely used
in the literature, there are two contributions to the energy
current vertex in the diagrammatic approach, see Fig.~\ref{figcurrver}: 
beside the usual vertex of the non-interacting case, which 
arises from the terms $\partial_\tau \psi^\dag \nnabla \psi$, there is an 
additional vertex from the $\phi \psi^\dag \nnabla \psi$ terms.
These vertices were not taken into account in 
Refs.~\onlinecite{Livanov,Smith,Raimondi}. However, analogous vertices 
were previuosly considered in the calculations of the thermoelectric
coefficient with the inclusion of the electron-electron interaction in
the particle-hole channel\cite{veree,Reyzerprivate} 
and in the Cooper channel\cite{vercoop}
and for the electron-phonon interaction\cite{vereph}.

\section{Alternative parametrization}
\label{alter}

The operator $\hat{H}_{\eh}$ defined in \req{LR} is  
clearly not a standard Hamiltonian. However, one can
introduce a different definition for
the propagator $\L^\rho$
\be
\label{LRalt}
\left[\frac{\partial}{\partial t_1} + i\hat{H}_{\eh}(-i\nnabla_1)  
- \St_{s}^{out}   \right]
\L^\rho
=\delta_{12},
\ee
such that the (new) $\hat{H}_\eh$ operator is indeed a Hamiltonian:
\be
\begin{split}
\hat{H}_{\eh}(\q) \equiv \sqF\,\v\cdot & \q\, \sqF \, , \quad 
\St_{s}^{out}  \equiv \sqF\,\St_{\tau}\sqF \, ,
\\
& \sqF  \equiv \left(1+\hat{F}\right)^{1/2}
\end{split}
\label{LRdef}
\ee  
[the action of the operator $\hat{F}$ was defined in \req{Fdef}].
Proceeding as in Section \ref{abf}, one obtains the following expressions
for the $\K$ propagators:
\begin{subequations}
\be \label{krfinalt}
\begin{split}
&{\nu}\hat{\K}^R =  \left( \pt \right)^{-1}\Big[  
\hat\L^g - \sqF \hat{\L}^\rho \sqF \Big] 
\\
&{\nu}\hat{\K}^A = - \Big[  
\hat{\bar{\L^g}} - \sqF \hat{\bar{\L}}^\rho \sqF \Big] \left( \pt \right)^{-1}
\end{split}
\ee
\bea
\label{kknalt}
&& \nu \hat{\K}^K = -i \left( \pt \right)^{-1} 
\Big[ \hat\L^g \hat{\cal N}^g +  \hat{\cal N}^g
\hat{\bar{\L^g}} \Big] 
\left( \pt \right)^{-1} \\
&& +i \left( \pt \right)^{-1} \sqF
\Big[ \hat{\L}^\rho  \hat{\cal N}^\rho   + 
\hat{\cal N}^\rho \bar{\hat{\L}}^\rho \Big]
\sqF \left( \pt \right)^{-1} \! . \nonumber 
\eea
\end{subequations}
The ``kinetic equation'' for ${\cal N}^g$, \req{kineq0}, remains unchanged, 
while ${\cal N}^\rho$  now obeys:
\be
\Big[\pt + \hat{H}_{\eh}(-i\nnabla) ;  \hat{\cal N}^\rho\Big]
= \St^{\rho}\left\{{\cal N}^\rho , g^K\right\}
\ee
with
\be
\St^{\rho}\left\{{\cal N}^\rho, g^K\right\}\equiv
2\Big\{\St_s^{out} ; \hat{\cal N}^\rho \Big\} +
2\sqF\hat{M}\sqF 
\ee
or, after the Wigner transforms \rref{Nb}-\rref{wigtr2}:
\be
\w \biggl[
\pt {\hat N}^\rho + \left\{\hat{\s}, \nnabla  {\hat N}^\rho\right\} 
+ i\left[\hat{H}_{\eh}(\q), {\hat N}^\rho \right]
\biggr]= \St^\rho\left\{N^\rho, f\right\}
\ee
\be\begin{split}
& \St^\rho\left\{N^\rho, f\right\} (\w ,\nn_1,\nn_2) = \int\!d\e 
\int\!\frac{d\nn_3\ldots d\nn_6}{\W_d^4} 
\\ & \Big[ {\cal F}_{16} \gamma^6_{34}
\Upsilon^\rho_{52;43}(\e,\w) {\cal F}_{45}  + {\cal F}_{56} \gamma^6_{34}
\Upsilon^\rho_{15;34}(\e,\w) {\cal F}_{42} \Big]
\end{split}\ee
with the definitions \rref{gamma1} and \rref{Delta} for 
$\gamma$ and $\Upsilon$.

We can then proceed as in Sec. \ref{conlaw} and obtain the conservation laws 
\rref{rhocont1} and \rref{bal}; the only formal difference 
is in the definition of the bosonic energy density, which now is:
\[
u_\alpha (t,\r) = \frac{1}{2} \Tr_{\nn} \left[ \hat{\L}^\alpha 
\hat{\cal N}^\alpha \right]
\] 

In the alternative parametrization the formalism can be developed with no 
more difficulties than in the original one. However the evaluation of
the thermal conductivity become cumbersome. Also in the original
parametrization it is easier to include (at least perturbatively) the
effects due to higher harmonics of the Fermi liquid parameters.

\section{Derivation of the electrons collision integral}
\label{app2}

The calculation of the matrix collision integral is simplified by
the introduction of two functions $A(t,\r,\nn,\tilde{\nn})$ and 
$B (t,\r,\nn,\tilde{\nn})$ such that 
\be\label{abdef}
e^{i\hat{K}(t,\nn,\r)}e^{-i\hat{K}(t,\tilde{\nn},\r)} = \left( A 
\hat{\openone}_K + B\hat{\sigma}^x_K \right)_{\nn,\tilde{\nn}}
\ee
We remind that $\hat{K} = K_+ \hat{\openone}_K + K_- \hat{\sigma}^x_K$ and
\[
\hat{\sigma}^x_K = \left( \begin{matrix}
0 & \ 1 \cr
1 & \ 0 \end{matrix} \right)
\]

The collision integral -- right hand side of \req{eil2} -- in the
matrix notation is then:
\be\begin{split}
\frac{1}{2\tau} \langle \Big[ \hat{g}(\nn) \con \big( A\hat{\openone}_K + 
B\hat{\sigma}^x_K \big)_{\nn,\tilde{\nn}} \hat{g}(\tilde{\nn})
\big( A\hat{\openone}_K + B\hat{\sigma}^x_K \big)_{\tilde{\nn},{\nn}} \\
- \big( A\hat{\openone}_K + B\hat{\sigma}^x_K \big)_{\nn,\tilde{\nn}} 
\hat{g}(\tilde{\nn})
\big( A\hat{\openone}_K + B\hat{\sigma}^x_K \big)_{\tilde{\nn},{\nn}}
\con \hat{g}(\nn) \Big] \rangle_{\tilde{\nn}}
\end{split}\ee
where the open dot indicates the time convolution [cf. \req{tcon}] and
the time argument of the functions $A$ and $B$ is 
the first (second) time argument of the Green's function on their right 
(left), e.g.:
\[
B g B \con g \equiv \int dt_3 B(t_1) g^K(t_1,t_3) B(t_3) g^K(t_3,t_2) 
\]

Substituting the matrix Green's function of the form \rref{gzero}, we find 
that the collision integral becomes:
\be
\left\langle \left( \begin{matrix}
\St^R & \St^K \cr
\St^Z & \St^A
\end{matrix} \right) \right\rangle_{\tilde{\nn}}
\ee
The explicit expressions for the retarded, advanced and `Z' components are:
{\setlength\arraycolsep{1pt}
\bea
\St^R &=& g^K(\nn) \big[ B ; A \big]
+ g^K(\nn) \con 
 B(\nn,\tilde{\nn})g^K(\tilde{\nn}) B(\tilde{\nn},\nn) 
\nonumber \\
\St^A &=&  \big[ A ; B \big] g^K(\nn)
- B(\nn,\tilde{\nn})g^K(\tilde{\nn}) B(\tilde{\nn},\nn) 
\con g^K(\nn)
\nonumber \\
\St^Z &=& 2 \delta (t_1-t_2 ) \big[ A ; B \big] 
-2 B(\nn,\tilde{\nn})g^K(\tilde{\nn}) B(\tilde{\nn},\nn) \nonumber
\eea
where the (equal time) commutator is 
\[
\big[ A ; B \big] \equiv A(\nn,\tilde{\nn}) B(\tilde{\nn},\nn) - 
 B(\nn,\tilde{\nn}) A(\tilde{\nn},\nn) 
\] }
The calculations performed so far are exact. However at one loop we are 
interested in terms up to the second order in the fluctuating fields.
Then the expansion of the exponentials in \req{abdef} shows that any product 
of two functions $B$ is proportional to terms of
the form $K_-K_-$, which vanish after averaging over the fluctuating fields;
accordingly, we drop such terms. 
The remaining terms are all commutators, whose
explicit (approximate) form is:
\be
\big[ A ; B \big] = 2i \Big( K_-(\tilde{\nn}) - K_-(\nn) \Big)
\ee
Note that the second order terms cancel each other exactly.
The surviving first order terms lead to \req{eqKminus} for $K_-$.

For reference, we present the expression for $\St^K$ from which
\req{i1} and \req{st2} are derived [with the exception of the last line
of \req{stm}, which is a consequence of the requirement \rref{kpeq} for $K_+$]:
\bea
\St^K &=& 2 \delta (t_1-t_2 ) \big[ A ; B \big]
 +2 A (\nn,\tilde{\nn}) g^K (\tilde{\nn}) A (\tilde{\nn},\nn) \nonumber \\
&& - g^K(\nn) A (\nn,\tilde{\nn}) A (\tilde{\nn},\nn)
- A (\nn,\tilde{\nn}) A (\tilde{\nn},\nn) g^K(\nn) 
\nonumber \\
&& +g^K (\nn) B(\nn,\tilde{\nn})B(\tilde{\nn},\nn) 
+  B(\nn,\tilde{\nn})B(\tilde{\nn},\nn) g^K(\nn) \nonumber \\
&&+ g^K(\nn) \con B(\nn,\tilde{\nn}) g^K(\tilde{\nn}) A(\tilde{\nn},\nn)
\nonumber \\
&&- A(\nn,\tilde{\nn}) g^K(\tilde{\nn}) B(\tilde{\nn},\nn) \con g^K(\nn) 
\eea

\section{Derivation of \reqs{stin1b} and \rref{stin2b}}
\label{deriv1}

We start the derivation by separating the Keldysh and retarded/advanced
propagators contributions to $\St_1^{\mathrm{in}}$ in \req{i1split2}: 
{\setlength\arraycolsep{0pt}
\bea \label{st1k}
\St_{1}^K =&& -\frac{i}{16} \!\int\!\frac{d\nn_2 d\nn_3}{(\W_d)^2}
\, \gamma_{12}^{3} \: g(\bar{t},\tta, \nn_2) \\
&& \big[ 
2\K^K(\bar{t},\tta,\nn_3,\nn_2)
- \K^K(\bar{t}+\tta/2,0,\nn_3,\nn_2) \nonumber \\ 
&& \; - \K^K(\bar{t}-\tta/2,0,\nn_3,\nn_2) 
 - (\nn_2 \to \nn_1) \big]
\nonumber \\
\label{st1ra}
\St_{1}^{RA} =&& -\frac{i}{32} \!\int\!\!dt_3
 \int\!\frac{d\nn_2 d\nn_3}{(\W_d)^2} \, \gamma_{12}^{3} \\
 \big[\K^A&&(t_3,t_2,\nn_3,\nn_2) - 
\K^R(t_1,t_3,\nn_2,\nn_3) - (\nn_2 \to \nn_1)
\big] \nonumber \\
\times  \big[g&&(t_1,t_3,\nn_1) 
g(t_3,t_2,\nn_2) 
+ g(t_1,t_3,\nn_2) g(t_3,t_2,\nn_1) \big]
\nonumber
\eea
For our convenience, we rewrote the Keldysh part in terms of the
new time variables $\bar{t},\tta$:
\bea
&& \bar{t} = \frac{t_1+t_2}{2} \, , \quad \tta = t_1 - t_2 \, , \\
&& g(t_1,t_2) \to g(\bar{t},\tta), \
\K^K(t_1,t_1) \to \K^K(\bar{t}+\tta/2,0), \dots \nonumber
\eea}

Let us consider the limit $t_2 \to t_1$ of \req{st1k}; 
clearly when $\tta \to 0$, the square bracket vanishes. 
However we know that in this 
limit $g \to -2i/\pi\tta$ [cf. \req{gklim}]: 
in principle there could be a non-vanishing 
contribution from the first order expansion of the propagators in $\tta$. 
The last two Keldysh propagators depend on $\tta$ in their first 
variable, but with opposite signs, and so their respective first order 
terms cancel each other. As for the first propagator, the property 
$\K^K(1,2) = \K^K(2,1)$ translates into $\K^K(t,\tta) = \K^K(t,-\tta)$, which
ensures the absence of first order terms. We conclude that in the limit 
$\tta \to 0$ the Keldysh propagator terms vanish.
Similarly from the property $\K^A(1,2) = \K^R(2,1)$ it follows that
\req{st1ra} vanish for $t_2 = t_1$; this concludes the proof of \req{stin1b}.

Let us turn to \req{stin2b}. Since  
$\partial_{t_{1}} - \partial_{t_{2}} = 2\partial_{\tta}$,
we must expand
the Keldysh propagators to the second order in $\tta$. At this order the square 
bracket in \req{st1k} is (up to the proper combination of angular variables):
\be
\tta^2 \left( \partial_{\tta}^2 - \frac{1}{4} \partial_{\bar{t}}^2 \right) 
\K^K(\bar{t},0) =
- \tta^2 \lim_{t_{2}\to t_{1}} \partial_{t_{1}}\partial_{t_{2}} 
\K^K(t_1,t_2)
\nonumber \ee
where in the last equality we restored the original 
time variables. In the operator notation this is:
\be
\tta^2 \Bigl[ \partial_t \, \K^K \partial_t \Bigr]_{t_{2}=t_{1}}
\nonumber \ee
Therefore
\[ \begin{split}
& \lim_{t_{2} \to t_{1}} (\partial_{t_{1}}- \partial_{t_{2}})
 \, \St_{1}^K (t_1,t_2)  =
\frac{1}{4\pi} \int\!\frac{d\nn_2 d\nn_3}{\W_d^2} \gamma_{12}^{3} \\
& \quad \Big\{ \Bigl[ \pt\, \K^K \pt \Bigr] (t_1,t_1,\nn_3,\nn_1)
- \Bigl[ \pt\, \K^K \pt \Bigr] (t_1,t_1,\nn_3,\nn_2) \Big\}
\end{split} \]
which proves the first part of \req{stin2b}.

As for \req{st1ra}, using again the analytic property
$\K^A(1,2)=\K^R(2,1)$ we conclude that when the derivatives $\partial_{t_{1}}$,
 $\partial_{t_{2}}$
act on the distribution functions $g$, the terms in the second line
cancel each other.
However there are non-vanishing contributions when a derivative acts on the
propagators, such as:
\[\begin{split}
&\int\!dt_3 \int\!\frac{d\nn_2 d\nn_3}{\W_d^2} \, \gamma_{12}^{3}
 \, g(t_1,t_3,\nn_1) g(t_3,t_1,\nn_2) 
\\ &\times  \Big\{ \Bigl[ \partial_t \K^R \Bigr] (t_1,t_3,\nn_1,\nn_3)
- \Bigl[ \partial_t \K^R \Bigr] (t_1,t_3,\nn_2,\nn_3) \Big\}
\end{split}\]
Collecting all the terms we arrive at:
{\setlength\arraycolsep{0pt}
\bea
&& \lim_{t_{2} \to t_{1}} (\partial_{t_{1}}- \partial_{t_{2}})
 \, \St_{1}^{RA} (t_1,t_2)  = 
\frac{i}{16} \!\int\!\!dt_3 
\int\!\frac{d\nn_3 d\nn_2}{\W_d^2} \, \gamma_{12}^{3}
\nonumber \\ &&
\Big\{ \Bigl[ \pt \K^R \Bigr] 
(t_1,t_3,\nn_2,\nn_3) - 
\Bigl[ \pt \K^R \Bigr] (t_1,t_3,\nn_1,\nn_3) \Big\} \nonumber \\
&& \times  \big[g(t_1,t_3,\nn_1) g(t_3,t_1,\nn_2) 
+ g(t_1,t_3,\nn_2) g(t_3,t_1,\nn_1) \big]
\nonumber
\eea
which concludes the derivation of \req{stin2b}.}

\section{Elastic kernels in terms of the interaction propagator $D$}
\label{cmp}

To compare the kernels in \req{kernels} to the corresponding expressions in
Ref.~\onlinecite{ZNA}, we use the Fourier transforms of \reqs{Ks} and 
\rref{krfin} to obtain
\be\label{LDrel}
\Re \Big[ \hat{\L}^\rho - \hat{\L}^g \Big] = -\nu\w\,\Im \Big[ \hat{\L}^g 
 \hat{D}^R \hat{\L}^g\Big]
\ee
If we assume, as is done in Ref.~\onlinecite{ZNA}, that the Fermi liquid 
parameters do not depend on the momentum direction, then the 
interaction propagators $D^{R,A}$ also do not depend on it and the above
equation becomes
\be\label{LD2}
\Re \Big[ \L^\rho - \L^g \Big] = -\nu\w\,\Im \Big[ \L^g \rangle
 D^R \langle \L^g\Big]
\ee
where we generalized the angular integral notation so that
\[
\langle \L^g = \int\!\frac{d\nn_1}{\W_d} \L^g (\nn_1,\nn_2) , \quad
\L^g \rangle = \int\!\frac{d\nn_2}{\W_d} \L^g (\nn_1,\nn_2)
\]
We remind that our ghost propagator $\L^g$ coincides with the diffuson
propagator $D$ of Ref.~\onlinecite{ZNA}.

By substituting \req{LD2}, we rewrite the kernels 
\rref{kernels} as:
\bea
{\cal S}^{11}_{\mu\nu}(\w) &=& \frac{2}{\pi}\delta_{\mu\nu}
\int\!\frac{d^dq}{(2\pi)^d} \Big(
\langle \L^g\rangle\langle \L^g \rangle
- \langle \L^g \L^g \rangle \Big) D^R \quad
\\
{\cal S}^{12}_{\mu\nu}(\w) &=& -\frac{2d}{\pi}
\int\!\frac{d^dq}{(2\pi)^d} 
\langle n_{\alpha} \L^g \rangle \langle \L^g n_{\beta}\rangle D^R
\eea
\be\begin{split}
&{\cal E}_{\mu\nu}(\w) = \frac{d}{\pi\tau}\, \Im 
\int\!\frac{d^dq}{(2\pi)^d} \, 
D^R \Big[ \langle \L^g n_\alpha \L^g n_\beta \rangle 
\langle \L^g \rangle \\ &- \langle \L^g n_\alpha \L^g \rangle
\langle n_\beta \L^g \rangle 
+ \langle \L^g n_\alpha \rangle 
\langle \L^g n_\beta \rangle \langle \L^g \rangle  \\ &-
\langle \L^g \rangle \langle n_\alpha \L^g n_\beta \rangle
\langle \L^g \rangle \Big]
+ D^R\Big[ \langle \L^g \rangle \langle n_\alpha \L^g 
n_\beta \L^g \rangle 
\\& - \langle \L^g n_\alpha \rangle
\langle \L^g n_\beta \L^g \rangle - \langle \L^g
n_\alpha \L^g n_\beta \L^g \rangle \Big] \\ 
\end{split}
\ee
The first square bracket in the kernel ${\cal E}$ can be expressed as
\be
\tau \bigg[ 
\langle \L^g n_\alpha \L^g \Big[ n_\beta ; \St_\tau \Big] \L^g
\rangle + \langle \L^g \Big[ n_\alpha ; \St_\tau \Big] \L^g n_\beta
\rangle \langle \L^g \rangle \bigg]
\ee
and using the identity:
\be
\L^g \Big[ \nn ; \St_\tau \Big] \L^g = \Big[ \nn ; \L^g \Big]
\ee
we recast it as follows:
\be \begin{split}
\tau \bigg[ & \frac{\delta_{\mu\nu}}{d}\Big( \langle \L^g \L^g \rangle - 
\langle \L^g \rangle \langle \L^g \rangle \Big) 
\\& +\langle \L^g \rangle \langle n_\alpha \L^g n_\beta \rangle 
- \langle \L^g n_\alpha \L^g n_\beta \rangle \bigg]
\end{split}\ee
In the second square bracket we use the identity
\be
\L^g \nn \L^g = \frac{i}{v_F} \partial_{\q} \L^g
\ee
to obtain:
\be
\frac{i}{v_F} \Big[ \langle\L^g \rangle 
\partial_{q_{\beta}} \langle n_\alpha \L^g \rangle
- \langle \L^g n_\alpha \rangle \partial_{q_{\beta}}\langle\L^g\rangle
- \langle \L^g n_\alpha \partial_{q_{\beta}} \L^g \rangle \Big]
\ee
Finally the identity 
\be
\L^g \L^g = -i \partial_\w \L^g
\ee
enables us to conclude that the sum of the three kernels 
\[
{\cal S}^{11} +{\cal S}^{12} + {\cal E} 
\]
that determines the correction to the conductivity -- \req{elcon} --  
coincides with the combination $(K_0 - K_1 - L_0/v_F \tau)$ in the expression
for the conductivity of Ref.~\onlinecite{ZNA}.


\section{The inelastic kernel for the phase relaxation time}
\label{deph}

Let us consider a uniform system in which the bosons are assumed to be in
equilibrium with the electrons. In other words, the distribution function $f$
is independent of $\r, \nn$ and the boson-electron collision integral 
\rref{boselci} must vanish. The latter condition enables us to express
the bosonic distributions $N^\alpha$ in terms of $f$ and to obtain
\be
\Upsilon_{ij} (\e,\w) = -\W_d\delta(\widehat{\nn_i\nn_j}) \frac{1}{\w}
\int d\e_1 \, \Psi (\e,\e_1;\w)
\ee
[from now on irrelevant angular and momentum variables are omitted; 
all relevant definitions can be found in Sec. \ref{final}].
The former condition implies that the collision integrals \rref{stnonloc},
 \rref{steeel} and \rref{steel} vanish and therefore the kinetic equation 
for (the zeroth harmonic of) $f$ reduces to
\be
\pt f(\e;t) = \int\!d\w\int\!d\e_1 \, A(\w) \Psi (\e,\e_1;\w) 
\ee
where
\be\begin{split}
A(\w) =& \frac{-2}{\nu\pi\w^2} \Re \,
\Tr \Big\{\!\left[2\St_\tau \hat{\bar{\L}}^g +\hat{1} \right] 
\St_\tau \! \left[
\hat{\L}^\rho + 3\hat{\L}^\sigma \! - 4 \hat{\L}^g\right] \!\Big\}
\end{split}\ee
We substitute \req{LDrel} and the similar relation for the triplet channel
[$D^R_T$ being the triplet channel propagator] 
into the expression for $A(\w)$; 
then we use the identity \rref{diffid} and obtain
\be
A(\w) = -\frac{2}{\pi\w} \Im \, \Tr \left[ \St_\tau \hat{\bar{\L}}^g
\left( \hat{D}^R + \hat{D}^R_T \right) \hat{\L}^g \right]
\ee
Using again \req{diffid}, we recover immediately the form of the inelastic
kernel $A(\w)$ given in Ref.~\onlinecite{NZA}.



\end{document}